\RequirePackage{amsmath}
\documentclass{lmcs}
\pdfoutput=1
\usepackage[utf8]{inputenc}

\usepackage{lastpage}
\lmcsdoi{20}{1}{13}
\lmcsheading{}{\pageref{LastPage}}{}{}%
{Mar.~03,~2023}{Feb.~12,~2024}{}

\usepackage{amssymb}
\usepackage{booktabs}
\usepackage{hyperref}
\usepackage{mathpartir}
\usepackage{stmaryrd}
\usepackage{subcaption}
\usepackage{tikz,tikz-cd}
\usepackage{xifthen}

\newcommand\phantomprimel[1]{\mathrlap{#1}\phantom{#1'}}

\newcommand\syntax[1]{\mathbf{#1}}
\newcommand\syntaxtype[1]{\syntax{#1}}
\newcommand\syntaxarg[2][\,]{\ifthenelse{\isempty{#2}}{}{#1 #2}}
\newcommand\comp[1]{\underline{#1}}

\newcommand\unittype{\syntaxtype{unit}}
\newcommand\booltype{\syntaxtype{bool}}

\newcommand\thunktype[1]{\syntaxtype{U} \syntaxarg[\mspace{2mu}]{#1}}
\newcommand\freetype[1]{\syntaxtype{F} \syntaxarg[\mspace{2mu}]{#1}}
\newcommand\fakecomp[3]{%
  \smash{\phantom{\underline{#3}}}
  \mathllap{#3}
  \mathllap{\raisebox{#1}{$
    \mspace{#2}
    \underline{\mspace{-#2}\phantom{#3}\mspace{-#2}}
    \mspace{#2}$}}}
\newcommand\compprod{\mathbin{\fakecomp{0.2ex}{2mu}{\times}}}

\newcommand\unitterm{{()}}
\newcommand\pairterm[2]{(#1,#2)}
\newcommand\fstkeyword{\syntax{fst}}
\newcommand\sndkeyword{\syntax{snd}}
\newcommand\fstterm[1]{\fstkeyword \syntaxarg{#1}}
\newcommand\sndterm[1]{\sndkeyword \syntaxarg{#1}}
\newcommand\matchpairterm[4]{%
  \syntax{match}~#1~\syntax{with}~(#2, #3).\,#4}
\NewDocumentCommand{\absterm}{ m o m }
  {\lambda #1 \IfNoValueF{#2}{\!:\!#2}.\,#3}
\newcommand\returntermkeyword{\syntax{return}}
\newcommand\returnterm[1]{\returntermkeyword\syntaxarg{#1}}
\newcommand\trueterm{\syntax{true}}
\newcommand\falseterm{\syntax{false}}
\newcommand\iftermkeyword{\syntax{if}}
\newcommand\ifterm[3]{%
  \iftermkeyword~#1~\syntax{then}~#2~\syntax{else}~#3}
\newcommand\totermkeyword{\syntax{to}}
\newcommand\toterm[3]{#1~\totermkeyword~#2.\,#3}
\newcommand\thunktermkeyword{\syntax{thunk}}
\newcommand\forcetermkeyword{\syntax{force}}
\newcommand\thunkterm[1]{\thunktermkeyword \syntaxarg{#1}}
\newcommand\forceterm[1]{\forcetermkeyword \syntaxarg{#1}}
\NewDocumentCommand\comppairterm{ s O{} m m }{%
  \IfBooleanTF{#1}
    {\begin{aligned}\lambda\{&1.\,#3,\\#2 &2.\,#4\}\end{aligned}}
    {\lambda\{1.\,#3,#2 2.\,#4\}}}
\newcommand\pushterm[2]{#1\,\syntax{`}\,#2}
\newcommand\pushtermtight[2]{#1\syntax{`}#2}

\newcommand{\truevalue}{\mathit{true}}
\newcommand{\falsevalue}{\mathit{false}}

\NewDocumentCommand{\cbpvrecterm}{ m o m }
  {\syntax{rec}\,#1 \IfNoValueF{#2}{\!:\!\thunktype{#2}}.\,#3}
\makeatletter
\NewDocumentCommand{\valuerecterm}{ m o o m m }
  {\syntax{rec}\,#1
    \IfNoValueF{#2}{
      \IfNoValueT{#3}{\@latex@error{Missing type}}
      \!:\!#2 \to #3}
    .\,\absterm{#4}{#5}}
\makeatother
\newcommand\failconstant[1][]{\syntax{fail}_{#1}}
\newcommand\getconstant{\syntax{get}}
\newcommand\orconstant{\syntax{or}}
\newcommand\orterm[2]{#1 \mathbin{\orconstant} #2}

\newcommand\emptycontext\diamond
\newcommand\substitute[2]{#1 [#2]}
\newcommand\reducesto\Downarrow
\newcommand\syntaxleq\preccurlyeq
\newcommand\syntaxgeq\succcurlyeq
\newcommand\ctxleq[1][]{\syntaxleq^{#1}_\mathrm{ctx}}
\newcommand\ctxgeq[1][]{\syntaxgeq^{#1}_\mathrm{ctx}}
\newcommand\ctxeq[1][]{\cong^{#1}_\mathrm{ctx}}
\newcommand\welltyped[3]{#1 \vdash #2 : #3}
\newcommand\valuetyped[3]{#1 \vdash #2 : #3}
\newcommand\comptyped[3]{#1 \mathbin{\vdash_{\mspace{-0.5mu}c}} #2 : #3}
\newcommand\hole\square

\newcommand\translate[2]{%
  #2
  {\llparenthesis\mspace{1mu}
  {}#1{}
  \mspace{1mu}\rrparenthesis}}
\newcommand{\translatevalue}[1]{\translate{#1}{\mathcal{V}}}

\newcommand{\translatename}[1]{\translate{#1}{\mathcal{N}}}
\NewDocumentCommand{\semrel}{m o}{\sem[\mathcal{R}]{#1}_{\IfNoValueF{#2}{#2}}}
\NewDocumentCommand{\semrels}{m}{\sem[\mathcal{R}']{#1}}
\NewDocumentCommand{\vnrel}{m m}{\ltimes^{#1}_{#2}}

\newcommand{\semvalue}[1]{\sem[\mathcal{V}]{#1}}

\newcommand{\semname}[1]{\sem[\mathcal{N}]{#1}}
\newcommand{\toname}[1]{\phi_{\cramped{#1}}}
\newcommand{\tonamehat}[1]{\hat{\phi}_{\cramped{#1}}}
\newcommand{\tovalue}[1]{\psi_{\cramped{#1}}}
\newcommand{\ToName}[1]{\Phi_{\cramped{#1}}}
\newcommand{\ToNameHat}[1]{\hat{\Phi}_{\cramped{#1}}}
\newcommand{\ToValue}[1]{\Psi_{\cramped{#1}}}

\newcommand\natset{\mathbb{N}}
\newcommand\poset[1]{#1}
\newcommand\fix[1]{\mathop{\mathit{fix}} #1}
\newcommand\join\sqcup
\newcommand\Lub\bigsqcup
\newcommand\wcpo{$\omega$cpo}

\newcommand\downclose[1]{{\downarrow} #1}
\newcommand\finpowerset[1]{\mathcal{P}_\mathrm{fin} #1}

\newcommand\xto[1]{\xrightarrow{#1}}
\newcommand\ob[1]{|#1|}
\newcommand\Hom[3]{#1 (#2, #3)}
\newcommand\id{\mathit{id}}

\newcommand\compose\circ
\newcommand\iso\cong
\newcommand\pleq\sqsubseteq
\newcommand\pgeq\sqsupseteq
\newcommand\tuplemorphism[1]{\langle #1 \rangle}
\newcommand\vtuplemorphism[1]{\langle\!\langle #1 \rangle\!\rangle}
\newcommand\terminalmorphism[1]{\tuplemorphism{}_{#1}}
\newcommand\proj[1]{\pi_{#1}}
\newcommand\pairmorphism[2]{\tuplemorphism{#1, #2}}
\newcommand\vpairmorphism[2]{\vtuplemorphism{#1, #2}}
\newcommand\cotuplemorphism[1]{[#1]}

\newcommand\inl{\mathit{inl}}
\newcommand\inr{\mathit{inr}}
\newcommand\copairmorphism[2]{\cotuplemorphism{#1, #2}}
\renewcommand\exp[2]{#1 \Rightarrow #2}
\newcommand\compexp[2]{#1 \Rightarrow #2}
\newcommand\curry\Lambda
\newcommand\uncurry{\curry^{-1}}
\newcommand\ev{\mathit{ev}}
\newcommand\dist{\mathit{dist}}
\newcommand\assoc{\mathit{assoc}}

\newcommand\seql{\mathit{seq}^{\mathrm{L}}}
\newcommand\seqr{\mathit{seq}^{\mathrm{R}}}
\newcommand\unit\eta

\newcommand\catname[1]{\mathbf{#1}}
\newcommand\monad[1]{\mathsf{#1}}
\newcommand\algebra[1]{\mathsf{#1}}
\newcommand\C{\catname{C}}
\newcommand\M{\mathcal{M}}
\newcommand\Z{\algebra{Z}}
\newcommand\T{\monad{T}}

\newcommand\forget[1]{U_{#1}}
\newcommand\free[1]{F_{#1}}

\newcommand\Catname[1]{\mathbf{#1}}
\newcommand\Set{\Catname{Set}}
\newcommand\Poset{\Catname{Poset}}
\newcommand\wCpo{\Catname{\omega Cpo}}

\newcommand\sem[2][]{#1{\left\llbracket #2 \right\rrbracket}}

\newcommand\extend[1]{{#1}^\dagger}
\NewDocumentCommand{\extendr}{ o m }{{#2}^{\dagger^{\IfNoValueF{#1}{#1{\times}\hole}}}}
\NewDocumentCommand{\extendl}{ o m }{{#2}^{\dagger^{\IfNoValueF{#1}{\hole{\times}#1}}}}
\newcommand\algextend[1]{{#1}^\ddagger}
\NewDocumentCommand{\algextendr}{ o m }{{#2}^{\ddagger^{\IfNoValueF{#1}{#1{\times}\hole}}}}
\NewDocumentCommand{\algextendl}{ o m }{{#2}^{\ddagger^{\IfNoValueF{#1}{\hole{\times}#1}}}}


\begin{document}

\title{Galois connecting call-by-value and call-by-name}
\titlecomment{{\lsuper*}This article is an extended version of
\cite{mcdermott2022galois}.}

\author[D.~McDermott]{Dylan McDermott\lmcsorcid{0000-0002-6705-1449}}[a]
\author[A.~Mycroft]{Alan Mycroft\lmcsorcid{0000-0001-7013-8572}}[b]

\address{Reykjavik University, Iceland}
\email{dylan@dylanm.org}

\address{University of Cambridge, UK}
\email{Alan.Mycroft@cl.cam.ac.uk}

\begin{abstract}
  We establish a general framework for reasoning about the relationship
  between call-by-value and call-by-name.

  In languages with computational effects, call-by-value and call-by-name
  executions of programs often have different, but related, observable
  behaviours. For example, if a program might diverge but otherwise has
  no effects, then whenever it terminates under call-by-value,
  it terminates with the same result under call-by-name.
  We propose a technique for stating and proving properties like these. The key
  ingredient is Levy's call-by-push-value calculus, which we use as a
  framework for reasoning about evaluation orders.
  We show that the call-by-value and call-by-name translations of
  expressions into call-by-push-value have related observable behaviour
  under certain conditions on computational effects, which we identify.
  We then use this fact to construct maps
  between the call-by-value and call-by-name interpretations of types,
  and identify further properties of effects that imply these
  maps form a Galois connection.
  These properties hold for some computational effects (such as divergence), but not
  others (such as mutable state).
  This gives rise to a general reasoning principle that relates call-by-value and
  call-by-name. We apply the reasoning principle to example
  computational effects including
  divergence and nondeterminism.
\end{abstract}

\maketitle

\section{Introduction}
Suppose that we have a language in which terms can be statically
tagged either as using call-by-value evaluation or as using call-by-name
evaluation. Each program in this language would therefore use a mix of
call-by-value and call-by-name at runtime. Given any such program $M$,
we can construct a new program $M'$ by changing call-by-value to
call-by-name for some subterm. The question we consider in this paper
is: what is the relationship between the observable behaviour of $M$ and
the observable behaviour of $M'$?

For a language with computational effects (such as divergence), changing the
evaluation order in this way will in general change the behaviour of the
program, but for some effects we can often say something about how
we expect the behaviour to change:
\begin{itemize}
  \item If there are no effects at all (in particular, programs are
    normalizing), the choice of evaluation order is irrelevant:
    $M$ and $M'$ terminate with the same result.
  \item If there are diverging terms (for instance, via recursion), then
    the behaviour may change: a program might diverge under
    call-by-value and return a result under call-by-name. However, we
    can say something about how the behaviour changes: if $M$ terminates
    with some result, then $M'$ terminates with the same result.
  \item If nondeterminism is the only effect, every result of $M$
    is a possible result of $M'$.
\end{itemize}
These three instances of the problem are intuitively obvious, and each can
be proved separately.
We develop a \emph{general} technique for proving these properties.

The idea is to use a calculus that captures both call-by-value and
call-by-name, as a setting in which we can reason about both evaluation
orders (this is where $M$ and $M'$ live).
The calculus we use is Levy's
\emph{call-by-push-value} (CBPV)~\cite{levy:call-by-push-value}.
Levy describes how to translate (possibly open) expressions $e$ into
CBPV terms $\translatevalue{e}$ and $\translatename{e}$, which
respectively correspond to call-by-value and call-by-name.
We study the relationship between the behaviour of $\translatevalue{e}$
and the behaviour of $\translatename{e}$ in
a given program context.

The main obstacle is that $\translatevalue{e}$ and
$\translatename{e}$ have different types. The former has a
``call-by-value type'' $\freetype(\translatevalue{\tau})$ and the latter a
``call-by-name type'' $\translatename{\tau}$, defined in
\autoref{section:cbv-cbn}.
They hence cannot be directly compared.
Our solution is inspired by Reynolds's work relating direct and
continuation semantics of the $\lambda$-calculus
\cite{reynolds:relation-direct-continuation}.

The first step is to define a family of (set-theoretic) relations (in
the style of a logical relation) that compares the observable
behaviour of a term of call-by-value type with observable behaviour of a
term of call-by-name type.
We can then ask whether $\translatevalue{e}$ is related in this sense to
$\translatename{e}$.
This is not the case in general.
In the presence of arbitrary computational effects, we cannot expect to say
anything useful about how the behaviour of
$\translatevalue{e}$ relates to the behaviour of $\translatename{e}$.
However, under certain conditions satisfied only for certain
effects, $\translatevalue{e}$ is related to $\translatename{e}$.
These conditions say roughly that we can discard, duplicate, and reorder
effects.
The main result of the first step is a theorem relating the two
translations of $e$ when these conditions hold
(\autoref{thm:fundamental}).
This does not quite say what happens if we were to replace call-by-value
with call-by-name within some program; that is the goal of the second
step.

The second step is to identify maps between the call-by-value and
call-by-name interpretations, forming \emph{Galois connections} (one for
each source-language type) between the two interpretations.
We compose these maps with the translations of
expressions, to arrive at two terms that \emph{can} be compared directly.
For this step we assume a stronger condition on computational effects than in the
first, saying informally that effects can be thunked.
Under this condition we show that the maps between call-by-value and
call-by-name \emph{represent} the relations from the first step.
By combining this fact with \autoref{thm:fundamental} we
prove a result that directly relates the two terms we construct by
composition with the Galois connections.

We therefore arrive at a general \emph{reasoning principle}
(\autoref{theorem:cbv-cbn}) that
we use to compare call-by-value with call-by-name.
Given any preorder $\syntaxleq$ that captures the property we wish to
show about programs, our reasoning principle gives
conditions
that imply $M \syntaxleq M'$, where $M'$ is constructed as above by
replacing call-by-value with call-by-name.
We apply our reasoning principle to examples by choosing different
relations $\syntaxleq$; each of these relations indicates the extent to
which changing evaluation order affects the behaviour of the program.
In the divergence example $N \syntaxleq N'$ is defined to mean
termination of $N$ implies termination of $N'$ with the same result; in
the other examples $\syntaxleq$ similarly mirrors the
properties described informally above.

Rather than just considering some fixed collection of
(allowable)
effects, we
work abstractly and identify properties of computational effects that enable us
to relate call-by-value and call-by-name.

Our reasoning principle relies on the existence of some
denotational model $\mathcal{M}$ of the computational effects, and we reason
primarily inside $\mathcal{M}$.
In the first step we in fact relate the behaviour of the call-by-value
and call-by-name translations of terms within the given model
$\mathcal{M}$.
On the other hand, in the second step we are able to prove a result (our
reasoning principle \autoref{theorem:cbv-cbn}) in which the conclusion
is independent of $\mathcal{M}$ (though \autoref{theorem:cbv-cbn} does
assume the existence of a suitable $\mathcal{M}$, since the proof relies
on the first step).

Crucially, we use \emph{order-enriched}
models, which come with a partial order on the denotations of
terms. The ordering on denotations is necessary to obtain a general reasoning
principle. (Our example properties cannot be proved by showing that denotations are
equal, because they are not symmetric.)
Working inside the semantics rather than using syntactic logical
relations makes it easier to prove and to use our reasoning principle,
especially for the divergence example.

In \autoref{section:cbpv} we summarize the call-by-push-value calculus
(CBPV) and the
call-by-value and call-by-name translations. We then make the following
contributions:
\begin{itemize}
  \item We describe an \emph{order-enriched} categorical
    semantics for CBPV (\autoref{section:cbpv-models}).
  \item We define a family of relations for comparing the observable
    behaviours of a term of call-by-value type with a term of
    call-by-name type (\autoref{section:relation}).
    We prove that, for effects satisfying certain conditions, the
    call-by-value and call-by-name translations of expressions are
    related by these (\autoref{thm:fundamental}).
    As a corollary, we directly relate the call-by-value and
    call-by-name translations of closed expressions of type $\booltype$
    (\autoref{corollary:cbv-cbn-programs}).
  \item We define the Galois connections between the call-by-value and
    call-by-name translations (\autoref{section:galois-connection}), and
    show that they represent the relations from the first step
    (\autoref{thm:represent-relation}).
  \item We use the Galois connections to prove a novel reasoning
    principle (\autoref{theorem:cbv-cbn}) that
    relates the call-by-value and call-by-name translations of
    expressions (\autoref{section:main-theorem}).
\end{itemize}
We apply our reasoning principle to three different examples:
no effects, divergence, and nondeterminism.
In this way we establish all of three facts listed at the beginning of
this introduction.
Our motivation is partly to demonstrate the Galois
connection technique as a way of reasoning about different semantics of
a given language.
Call-by-value and call-by-name is one example of this (and
Reynolds's original application to direct and continuation semantics is
another).

This paper is a revised and extended version of
\cite{mcdermott2022galois}.
The primary difference is the addition of \autoref{section:relation},
containing the first step outlined above.
The conference version \cite{mcdermott2022galois} skips this step and
goes directly to the Galois connections.
The first step in particular enables us to prove a statement about closed terms of
type $\booltype$ (\autoref{corollary:cbv-cbn-programs}) under weaker
assumptions than in the conference version
\cite[Corollary~22]{mcdermott2022galois}.
We also add an extra example (immutable state), add products to the
source language, and include more detailed proofs than in the conference
version.

\section{Call-by-push-value, call-by-value, and call-by-name}
\label{section:cbpv}
Levy~\cite{levy:call-by-push-value,levy:call-by-push-value-decomposing}
introduced call-by-push-value (CBPV) as a calculus that captures both
call-by-value and call-by-name.
We reason about the relationship between call-by-value and call-by-name
evaluation inside CBPV.\footnotemark{}
\footnotetext{%
  Our use of CBPV should not be regarded as essential.
  We use it here because it is known to capture call-by-value and
  call-by-name in a strong sense (see \cite{levy:call-by-push-value}).
  It may be possible to replace CBPV with some other language that
  captures call-by-value and call-by-name, and obtain similar results to
  ours.
}

The syntax of CBPV terms is stratified into two kinds: \emph{values} $V,
W$ do not reduce, \emph{computations} $M, N$ might reduce (possibly with
computational effects). The syntax of
types is similarly stratified into \emph{value types} $A, B$ and
\emph{computation types} $\comp{C}, \comp{D}$.
\begin{align*}
  \text{value types} &&
  A,B~&\Coloneqq
  ~\unittype~|~A_1\times A_2~|
  ~\booltype~|~\thunktype{\comp{C}}
  \\
  \text{computation types} &&
  \comp{C},\comp{D}~&\Coloneqq
  ~\comp{C}_1\compprod\comp{C}_2~|
  ~A \to \comp{C}~|~\freetype{A}
  \\
  \text{values} &&
  V, W~&\Coloneqq~x
  ~|~\unitterm~|~\pairterm{V_1}{V_2}
  ~|~\trueterm~|~\falseterm
  ~|~\thunkterm{M}
  \\
  \text{computations} &&
  M, N~&\Coloneqq
    ~\comppairterm{M_1}{M_2}
  ~|~\pushtermtight{1}{M}~|~\pushtermtight{2}{M}
  \\&&&{\mspace{14mu}}
  ~|~{\mspace{6.8mu}}\absterm{x}[A]{M}~|~\pushtermtight{V}{M}
  ~|~\returnterm{V}~|~\toterm{M}{x}{N}
  \\&&&{\mspace{14mu}}
  ~|~{\mspace{6.8mu}}\matchpairterm{V}{x_1}{x_2}{M}
  \\&&&{\mspace{14mu}}
  ~|~{\mspace{6.8mu}}\ifterm{V}{M_1}{M_2}
  ~|~\forceterm{V}
\end{align*}
We restrict to only the subset of CBPV required for this paper.

The value type $\thunktype{\comp C}$ is the type of \emph{thunks} of
computations of type $\comp C$.
Elements of $\thunktype{\comp C}$ are introduced using
$\thunktermkeyword{}$: the value $\thunkterm M$ is the suspension of the
computation term $M$.
The corresponding eliminator is $\forcetermkeyword$, which behaves as the inverse of
$\thunktermkeyword$.
Computation types include
binary products; the pairing of two computations $M_1$ and $M_2$ is
written $\comppairterm{M_1}{M_2}$, and the first and second projections
are $\pushtermtight{1}{M}$ and $\pushtermtight{2}{M}$.
Computation types also include
function types (where functions send values to
computations).
Function application is written
$\pushtermtight{V}{M}$, where $V$ is the argument and $M$ is the
function to apply.
The \emph{returner type} $\freetype{A}$ has as elements computations
that return elements of the value type $A$; these computations may have
effects.
Elements of $\freetype A$ are introduced by $\returntermkeyword$; the
computation $\returnterm{V}$ immediately returns the value $V$ (with no
effects).
Computations can be sequenced using $\toterm{M}{x}{N}$.
This first evaluates $M$ (which is required to have returner
type), and then evaluates $N$ with $x$ bound to the result of $M$.
(It is similar to \verb+M >>= \x -> N+ in Haskell.)
The syntax we give here does not include any method of introducing
effects; we extend CBPV with some example computational effects in
\autoref{section:cbpv-examples}.

The evaluation order in CBPV is fixed for each program.  The only
primitive that causes the evaluation of two separate computations is
$\totermkeyword$, which implements eager sequencing.
Thunks give us more control
over the evaluation order: they can be arbitrarily duplicated and
discarded, and can be forced in any order chosen by the program.
This is how CBPV captures both call-by-value and call-by-name (see
\autoref{section:cbv-cbn} below).

\begin{figure}[t]
  \[\boxed{\valuetyped{\Gamma}{V}{A}}\]
  \begin{mathpar}
    \inferrule
      { }
      { \valuetyped{\Gamma}{x}{A} }
    ~\text{if}~(x : A) \in \Gamma

    \inferrule
      { }
      { \valuetyped{\Gamma}{\unitterm}{\unittype} }

    \inferrule
      { \valuetyped{\Gamma}{V_1}{A_1} \\
        \valuetyped{\Gamma}{V_2}{A_2} }
      { \valuetyped{\Gamma}{\pairterm{V_1}{V_2}}{A_1 \times A_2} }\\

    \inferrule
      { }
      { \valuetyped{\Gamma}{\trueterm}{\booltype} }

    \inferrule
      { }
      { \valuetyped{\Gamma}{\falseterm}{\booltype} }

    \inferrule
      { \comptyped{\Gamma}{M}{\comp{C}} }
      { \valuetyped{\Gamma}{\thunkterm{M}}{\thunktype{\comp{C}}} }
  \end{mathpar}
  \vspace{-0.3em}

  \[\boxed{\comptyped{\Gamma}{M}{\comp{C}}}\]
  \begin{mathpar}
    \mprset{andskip=1.5em,sep=1.4em}
    \inferrule
      { \comptyped{\Gamma}{M_1}{\comp{C}_1} \\
        \comptyped{\Gamma}{M_2}{\comp{C}_2} }
      { \comptyped{\Gamma}{\comppairterm{M_1}{M_2}}
          {\comp{C}_1 \compprod \comp{C}_2 } }

    \inferrule
      { \comptyped{\Gamma}{M}{\comp{C}_1 \compprod \comp{C}_2 } }
      { \comptyped{\Gamma}{\pushtermtight{1}{M}}{\comp{C}_1} }

    \inferrule
      { \comptyped{\Gamma}{M}{\comp{C}_1 \compprod \comp{C}_2 } }
      { \comptyped{\Gamma}{\pushtermtight{2}{M}}{\comp{C}_2} }\\

    \inferrule
      { \comptyped{\Gamma, x : A}{M}{\comp{C}} }
      { \comptyped{\Gamma}{\absterm{x}[A]{M}}{A \to \comp{C}} }

    \inferrule
      { \valuetyped{\Gamma}{V}{A} \\
        \comptyped{\Gamma}{M}{A \to \comp{C}} }
      { \comptyped{\Gamma}{\pushtermtight{V}{M}}{\comp{C}} }

    \inferrule
      { \valuetyped{\Gamma}{V}{A} }
      { \comptyped{\Gamma}{\returnterm{V}}{\freetype{A}} }

    \inferrule
      { \comptyped{\Gamma}{M}{\freetype{A}} \\
        \comptyped{\Gamma, x : A}{N}{\comp{C}} }
      { \comptyped{\Gamma}{\toterm{M}{x}{N}}{\comp{C}} }

    \inferrule
      { \valuetyped{\Gamma}{V}{A_1 \times A_2} \\
        \comptyped{\Gamma, x_1 : A_1, x_2 : A_2}{M}{\comp{C}} }
      { \comptyped{\Gamma}{\matchpairterm{V}{x_1}{x_2}{M}}{\comp{C}} }

    \inferrule
      { \valuetyped{\Gamma}{V}{\booltype} \\
        \comptyped{\Gamma}{M_1}{\comp{C}} \\
        \comptyped{\Gamma}{M_2}{\comp{C}} }
      { \comptyped{\Gamma}{\ifterm{V}{M_1}{M_2}}{\comp{C}} }

    \inferrule
      { \valuetyped{\Gamma}{V}{\thunktype{\comp{C}}} }
      { \comptyped{\Gamma}{\forceterm{V}}{\comp{C}} }
  \end{mathpar}

  \caption{CBPV typing rules}
  \label{figure:cbpv-typing}
\end{figure}
CBPV has two typing judgments: $\valuetyped{\Gamma}{V}{A}$
for values and $\comptyped{\Gamma}{M}{\comp{C}}$ for computations.
\emph{Typing contexts} $\Gamma$ are ordered lists of (variable, value
type) pairs.
We require that no variable appears more than once in any typing
context.
\autoref{figure:cbpv-typing} gives the typing rules.
Rules that add a new variable to a typing context implicitly require
that the variable is fresh.
We write $\emptycontext$ for the empty typing context, $V : A$ as an
abbreviation for $\valuetyped{\emptycontext}{V}{A}$, and $M : \comp{C}$
as an abbreviation for $\comptyped{\emptycontext}{M}{\comp{C}}$.

\begin{figure}
  \begin{mathpar}
    \inferrule
      { }
      { \comppairterm{M_1}{M_2} \reducesto \comppairterm{M_1}{M_2} }

    \inferrule
      { M \reducesto \comppairterm{N_1}{N_2} \\ N_i \reducesto R }
      { \pushtermtight{i}{M} \reducesto R }
    ~i \in \{1, 2\}

    \inferrule
      { }
      { \absterm{x}[A]{M} \reducesto \absterm{x}[A]{M} }

    \inferrule
      { M \reducesto \absterm{x}[A]{N} \\
        \substitute{N}{x\mapsto V} \reducesto R }
      { \pushtermtight{V}{M} \reducesto R } \\

    \inferrule
      { }
      { \returnterm{V} \reducesto \returnterm{V} }

    \inferrule
      { M \reducesto \returnterm{V} \\
        \substitute{N}{x\mapsto V} \reducesto R }
      { \toterm{M}{x}{N} \reducesto R }

    \inferrule
      { M_1 \reducesto R }
      { \ifterm{\trueterm}{M_1}{M_2} \reducesto R }

    \inferrule
      { M_2 \reducesto R }
      { \ifterm{\falseterm}{M_1}{M_2} \reducesto R }

    \inferrule
      { \substitute{M}{x_1 \mapsto V_1, x_2 \mapsto V_2} \reducesto R }
      { \matchpairterm{\pairterm{V_1}{V_2}}{x_1}{x_2}{M} \reducesto R }

    \inferrule
      { M \reducesto R}
      { \forceterm{(\thunkterm{M})} \reducesto R }
  \end{mathpar}
  \caption{Big-step operational semantics of CBPV}
  \label{figure:cbpv-semantics}
\end{figure}
We give an operational semantics for CBPV.
This consists of a big-step evaluation relation $M \reducesto R$, which
means the computation $M$ evaluates to $R$. 
Here $R$ ranges over \emph{terminal computations}, which are the subset
of computations with an introduction form on the outside:
\[
  R~\Coloneqq~\comppairterm{M_1}{M_2}~|~\absterm{x}[A]{M}
  ~|~\returnterm{V}
\]
We only evaluate closed, well-typed computations, so when we write $M
\reducesto R$ we assume $M : \comp{C}$ for some $\comp{C}$ (this implies
$R : \comp{C}$).
Reduction therefore cannot get stuck.
The rules defining $\reducesto$ are given in
\autoref{figure:cbpv-semantics}.
All terminal computations evaluate to themselves.
Products of computations are \emph{lazy}: to evaluate a
projection $\pushtermtight{i}{M}$, only the $i$th component of the pair
$M$ is evaluated.
Since we have not yet included any way of forming
impure computations, the semantics is deterministic and
normalizing: given any $M : \comp{C}$, there is exactly one
terminal computation $R$ such that $M \reducesto R$.
\autoref{section:cbpv-examples} extends the semantics in
ways that violate these properties.
We are primarily interested in evaluating computations of returner type.

A CBPV \emph{program} is a closed computation $M :
\freetype{\booltype}$.
The reasoning principle we give for call-by-value and call-by-name
relates open terms in program contexts.
A \emph{program relation} consists of a preorder\footnotemark{}
$\syntaxleq$ on programs.
For example, we could use
\footnotetext{%
  We do not actually need to assume that $\syntaxleq$ is reflexive or transitive
  at any point, but because of constraints we add later (such as
  existence of an adequate model), we do not expect there to be any
  interesting examples in which $\syntaxleq$ is not a preorder.
}%
\[
  M \syntaxleq M'
  \quad\text{if and only if}\quad
  \forall V\!:\!\booltype.~(M \reducesto \returnterm{V})
    ~\Rightarrow~(M' \reducesto \returnterm{V})
\]
We could also use, for example, the total relation for $\syntaxleq$
(and in this case apply our reasoning principle for call-by-value and
call-by-name even if we include e.g.\ mutable state as a side effect --
but then of course the conclusion of our reasoning principle would be
trivial).
Given any program relation $\syntaxleq$, we define a
\emph{contextual preorder} $M \ctxleq[\Gamma] M'$ on arbitrary
well-typed computations (in typing context $\Gamma$) by considering the
behaviour of $M$ and $M'$ in programs as follows.
A \emph{computation context} $\mathcal{E}$ is a computation term, with a
single hole $\hole$ where a computation term is expected.
We write $\mathcal{E}[M]$ for the computation that results from
replacing $\hole$ with $M$ (which may capture some of the free variables
of $M$).
For example, if $\mathcal{E}$ is the computation context
$\toterm{N}{x}{\hole}$ then $\mathcal{E}[\returnterm{x}]$ is the
computation $\toterm{N}{x}{\returnterm{x}}$, where $x$ is captured.
We use computation contexts to define $\ctxleq[\Gamma]$.
\begin{defi}[Contextual preorder]
  Suppose that $\syntaxleq$ is a program relation, and that
  $\comptyped{\Gamma}{M}{\comp{C}}$ and
  $\comptyped{\Gamma}{M'}{\comp{C}}$ are two computations of the same
  type.
  We write $M \ctxleq[\Gamma] M'$ if,
  for all computation contexts
  $\mathcal{E}$ such that
  $\mathcal{E}[M], \mathcal{E}[M'] : \freetype{\booltype}$, we have
  $\mathcal{E}[M] \syntaxleq \mathcal{E}[M']$.
  We write $M \ctxeq[\Gamma] M'$, and say that $M$ and $M'$ are
  \emph{contextually equivalent}, when both $M \ctxleq[\Gamma] M'$ and
  $M' \ctxleq[\Gamma] M$ hold.
\end{defi}
We sometimes omit $\Gamma$, and write just $M \ctxleq M'$ or
$M \ctxeq M'$.

\subsection{Call-by-value and call-by-name}
\label{section:cbv-cbn}
We use CBPV (instead of e.g.\ Moggi's monadic metalanguage
\cite{moggi:notions-computation-monads}) because it captures both
call-by-value and call-by-name in a strong sense (see the introduction
of \cite{levy:call-by-push-value} for a detailed discussion of
this).
Levy~\cite{levy:call-by-push-value} gives two compositional translations
from a source language into CBPV: one for call-by-value and one for
call-by-name.
We recall both translations in this section; our goal is to reason about
the relationship between them.

For the source language, we use the following syntax of types $\tau$ and
expressions $e$:
\begin{align*}
  \tau~&\Coloneqq~
  \unittype~|~\booltype~|~\tau_1 \times \tau_2
  ~|~\tau \to \tau'
  \\
  e~&\Coloneqq~x
  ~|~\unitterm
  ~|~\pairterm{e_1}{e_2}~|~\fstterm{e}~|~\sndterm{e}
  ~|~\trueterm~|~\falseterm~|~\ifterm{e_0}{e_1}{e_2}
  ~|~\absterm{x}[\tau]{e}~|~e\,e'
\end{align*}
We include two base types $\unittype$ and $\booltype$ to be used in
examples.\footnotemark{}
\footnotetext{%
  Unlike in Levy~\cite{levy:call-by-push-value}, we do not include
  general sum types, only $\booltype$.
  We expect that including arbitrary sum types would complicate
  \autoref{section:relation}, because it is difficult to extend logical
  relations of varying arity with sums.
  The difficulty, and techniques for dealing with it, are discussed
  e.g.\ in~\cite{abel2019normalization,fiore1999lambda,katsumata2008characterisation}.}
The source language has a typing judgement of the form
$\welltyped{\Gamma}{e}{\tau}$, defined by the usual rules.

\begin{figure}
  \begin{subfigure}{\linewidth}
    \begin{gather*}
    \begin{array}{r @{\hspace{1em}\mapsto\hspace{1em}} l}
\multicolumn{2}{l}{\boxed{
      \text{type}~\tau \;\mapsto\; \text{value type}~\translatevalue{\tau}
}}
      \\[1ex]
      \unittype & \unittype \\
      \tau_1 \times \tau_2 &
      \translatevalue{\tau_1} \times \translatevalue{\tau_2} \\
      \booltype & \booltype \\
      \tau \to \tau' &
      \thunktype{(\translatevalue{\tau}
        \to \freetype{(\translatevalue{\tau'})})}
    \end{array}
    \begin{array}{r @{\hspace{1em}\mapsto\hspace{1em}} l}
\multicolumn{2}{l}{\boxed{
      \text{typing context}~\Gamma \;\mapsto\;
      \text{typing context}~\translatevalue{\Gamma}
}}
      \\[1ex]
      \emptycontext & \emptycontext \\
\hspace{4.5em}
      \Gamma, x : \tau &
      \translatevalue{\Gamma}, x : \translatevalue{\tau}
    \end{array}
    \\[3ex]
    \begin{array}{r @{\hspace{1em}\mapsto\hspace{1em}} l}
\multicolumn{2}{l}{\boxed{
      \text{expression}~\welltyped{\Gamma}{e}{\tau} \;\mapsto\;
      \text{computation}~
      \comptyped{\translatevalue{\Gamma}}{\translatevalue{e}}
        {\freetype{(\translatevalue{\tau})}}
}}
      \\[1ex]
      x & \returnterm{x} \\
      \unitterm & \returnterm{\unitterm} \\
      \pairterm{e_1}{e_2} &
      \toterm{\translatevalue{e_1}}{z_1}
        {\toterm{\translatevalue{e_2}}{z_2}
          {\returnterm{\pairterm{z_1}{z_2}}}} \\
      \fstterm{e} &
      \toterm{\translatevalue{e}}{z}
        {\matchpairterm{z}{z_1}{z_2}{\returnterm{z_1}}} \\
      \sndterm{e} &
      \toterm{\translatevalue{e}}{z}
        {\matchpairterm{z}{z_1}{z_2}{\returnterm{z_2}}} \\
      \trueterm & \returnterm{\trueterm} \\
      \falseterm & \returnterm{\falseterm} \\
      \ifterm{e_0}{e_1}{e_2} &
      \toterm{\translatevalue{e_0}}{z}
        {\ifterm{z}{\translatevalue{e_1}}{\translatevalue{e_2}}} \\
      \absterm{x}[\tau]{e} &
      \returnterm{\thunkterm{
        \absterm{x}[\translatevalue{\tau}]{\translatevalue{e}}}} \\
      e\,e' &
      \toterm{\translatevalue{e}}{y}
        {\toterm{\translatevalue{e'}}{z}
          {\pushterm{z}{\forceterm{y}}}}
    \end{array}
    \end{gather*}
    \caption{Call-by-value translation $\translatevalue{{-}}$}
  \end{subfigure}

  \begin{subfigure}{\linewidth}
  \begin{gather*}\\
    \begin{array}{r @{\hspace{1em}\mapsto\hspace{1em}} l}
\multicolumn{2}{l}{\boxed{
      \text{type}~\tau \;\mapsto\; \text{computation type}~\translatename{\tau}
}}
      \\[1ex]
      \unittype & \freetype{\mspace{2mu}\unittype} \\
      \tau_1 \times \tau_2 &
      \translatename{\tau_1} \compprod \translatename{\tau_2} \\
      \booltype & \freetype{\mspace{2mu}\booltype} \\
      \tau \to \tau' &
      (\thunktype{(\translatename{\tau})}) \to \translatename{\tau'}
    \end{array}
    \begin{array}{r @{\hspace{1em}\mapsto\hspace{1em}} l}
\multicolumn{2}{l}{\boxed{
      \text{typing context}~\Gamma \;\mapsto\;
      \text{typing context}~\translatename{\Gamma}
}}
      \\[1ex]
      \emptycontext & \emptycontext \\
\hspace{4.5em}
      \Gamma, x : \tau &
      \translatename{\Gamma}, x : \thunktype{(\translatename{\tau})}
    \end{array}
    \\[3ex]
    \begin{array}{r @{\hspace{1em}\mapsto\hspace{1em}} l}
\multicolumn{2}{l}{\boxed{
      \text{expression}~\welltyped{\Gamma}{e}{\tau} \;\mapsto\;
      \text{computation}~
      \comptyped{\translatename{\Gamma}}{\translatename{e}}
        {\translatename{\tau}}
}}
      \\[1ex]
      x & \forceterm{x} \\
      \unitterm & \returnterm{\unitterm} \\
      \pairterm{e_1}{e_2} &
      \comppairterm{\translatename{e_1}}{\translatename{e_2}} \\
      \fstterm{e} &
      \pushterm{1}{\translatename{e}} \\
      \sndterm{e} &
      \pushterm{2}{\translatename{e}} \\
      \trueterm & \returnterm{\trueterm} \\
      \falseterm & \returnterm{\falseterm} \\
      \ifterm{e_0}{e_1}{e_2} &
      \toterm{\translatename{e_0}}{z}
        {\ifterm{z}{\translatename{e_1}}{\translatename{e_2}}} \\
      \absterm{x}[\tau]{e} &
      \absterm{x}[\thunktype{(\translatename{\tau})}]{\translatename{e}}\\
      e\,e' &
      \pushterm{(\thunkterm{\translatename{e'}})}{\translatename{e}}
    \end{array}
   \end{gather*}
    \caption{Call-by-name translation $\translatename{{-}}$}
  \end{subfigure}
  \caption{Translations from the source language into CBPV}
  \label{figure:translations}
\end{figure}

The two translations from the source language to CBPV are
defined in \autoref{figure:translations}. For call-by-value, each source
language type $\tau$ is mapped to a CBPV value type
$\translatevalue{\tau}$ that contains the results of call-by-value
computations. For call-by-name, $\tau$ is translated to a
computation type $\translatename{\tau}$, which contains the
computations themselves.
Products in call-by-value use the value-type
products of CBPV (which means they are necessarily \emph{strict}: both
components of a pair are always evaluated). For call-by-name we give a lazy
interpretation of binary products, using products of CBPV computation types.
(Though note that we do not interpret $\unittype$ as a nullary product
of computation types. We instead treat $\unittype$ as a base type, so
that effects can happen at type $\unittype$, which matches typical
functional languages.)
Functions under the call-by-value translation accept values of type
$\translatevalue{\tau}$ as arguments; arguments are evaluated before
being passed to the function.
Under the call-by-name translation, functions accept thunks of
computations as arguments; instead of evaluating them, arguments are
thunked before passing them to call-by-name functions.
Source-language typing contexts $\Gamma$ are translated to CBPV typing
contexts $\translatevalue{\Gamma}$ and $\translatename{\Gamma}$.
In call-by-value they contain values, in call-by-name they contain
thunks of computations.
Source-language expressions $e$ are mapped to CBPV computations
$\translatevalue{e}$ and $\translatename{e}$.
The translation uses some auxiliary program variables, which are
assumed fresh.

For call-by-value we arbitrarily choose
left-to-right evaluation for both pairing and function application.
Under the call-by-name translation, computational effects occur only at the base
types $\unittype$ and $\booltype$ (since this is where the returner
types appear).

Of course, we have to justify that these translations actually capture
call-by-value and call-by-name. There are two semantics of
interest for
the source language: a call-by-value semantics (that evaluates
left-to-right), and a call-by-name semantics (with lazy products). Since
we consider the observable behaviour of CBPV terms, the properties we
want are that if the call-by-value translations $\translatevalue{e}$ and $\translatevalue{e'}$ have
the same observable behaviour then $e$ and $e'$ have the same observable
behaviour with respect to the call-by-value semantics, and similarly
for call-by-name. Levy~\cite{levy:call-by-push-value} proves both of
these properties (though without products in the source language). We take this as the required justification, and do not
give the details.

\subsection{Examples}
\label{section:cbpv-examples}
We consider three collections of
(allowable)
effects as examples throughout the
paper.

\begin{exa}[No effects]\label{ex:no-side-effects}
  We include the simplest possible example: the case where there are no
  computational effects at all.
  For this example, call-by-value and call-by-name turn out to have
  identical behaviour.
  We define the program relation $M \syntaxleq_{\mathrm{pure}} M'$ (for closed
  computations $M, M' : \freetype{\booltype}$) as:
  \[
    M \syntaxleq_{\mathrm{pure}} M'
    \quad\text{if and only if}\quad
    \exists V\!:\!\booltype.~(M \reducesto \returnterm{V})
      ~\wedge~(M' \reducesto \returnterm{V})
  \]
  In other words, $M$ and $M'$ both evaluate to the same result $V$.
  Since evaluation is deterministic, $V$ is necessarily unique.
  The contextual preorder $M \ctxleq[\Gamma] M'$ means if we construct
  two programs by wrapping $M$ and $M'$ in the same computation context,
  then these two programs evaluate to the same result.
  This relation is symmetric.
  Our other examples use non-symmetric relations.
\end{exa}

\begin{exa}[Divergence]\label{ex:recursion}
  For our second example, the only effect is divergence (via
  recursion).
  In this case, call-by-value and call-by-name do not have identical
  behaviour (they are not related by $\ctxleq$ as it is defined in our
  no-effects example).
  We instead show that replacing
  call-by-value with call-by-name does not change a terminating program
  into a diverging one.
  
  We extend our two languages with recursion.  For CBPV we extend
  the syntax of computations with fixed points $\cbpvrecterm{x}[\comp{C}]{M}$, and
  correspondingly extend the type system and operational semantics with
  the following rules:
  \[
    \inferrule
      { \comptyped{\Gamma, x : \thunktype{\comp{C}}\,}{\,M\,}{\,\comp{C}} }
      { \comptyped{\Gamma\,}{\,\cbpvrecterm{x}[\comp{C}]{M}\,}{\,\comp{C}} }
    \qquad
    \inferrule
      { \substitute{M}{x\mapsto\thunkterm{(\cbpvrecterm{x}[\comp{C}]{M})}}
        \reducesto R }
      { \cbpvrecterm{x}[\comp{C}]{M} \reducesto R }
  \]
  The variable $x$ is bound to a thunk of the recursive computation, so
  recursion is done by forcing $x$.
  (This is not the only way to add
  recursion to CBPV
  \cite{devesas-campos-levy:syntactic-computational-adequacy}, but is the
  most convenient for our purposes.)
  Of course, by adding recursion we lose normalization (but the
  semantics is still deterministic).
  We extend the source language, and the two translations into CBPV,
  with recursive functions:
  \begin{gather*}
    \SwapAboveDisplaySkip
    e~\Coloneqq~\dots~|~\valuerecterm{f}[\tau][\tau']{x}{e}
    \qquad
    \inferrule
      { \welltyped{\Gamma, f : \tau \to \tau', x : \tau}{e}{\tau'} }
      { \welltyped{\Gamma}
          {\valuerecterm{f}[\tau][\tau']{x}{e}}
          {\tau\to\tau'} }
    \\[1ex]
    \begin{aligned}
    \translatevalue{\valuerecterm{f}[\tau][\tau']{x}{e}}
    &~=~
    \returnterm{\thunkterm{(\cbpvrecterm{f}
      [(\translatevalue{\tau}\to\freetype{(\translatevalue{\tau'})})]
      {\absterm{x}[\translatevalue{\tau}]{\translatevalue{e}}})}} \\
    \translatename{\valuerecterm{f}[\tau][\tau']{x}{e}}
    &~=~
    \cbpvrecterm{f}
      [(\thunktype{(\translatename{\tau})}\to\translatename{\tau'})]
      {\absterm{x}[\thunktype{(\translatename{\tau})}]{\translatename{e}}}
    \end{aligned}
  \end{gather*}
  Again, the translations are the same as those given by
  Levy~\cite{levy:call-by-push-value}, except that Levy has general
  fixed points for call-by-name, rather than just recursive functions.
  The expression
  $
    \Omega_\tau =
      ((\valuerecterm{f}[\booltype][\tau]{x}{f\,x})\,\falseterm)
     : \tau
  $
  enables us to distinguish between call-by-value and call-by-name:
  $(\absterm{x}[\tau]{\trueterm})\,\Omega_\tau$ diverges in
  call-by-value but not in call-by-name.
  In particular, we have
  $
    \translatename{(\absterm{x}[\tau]{\trueterm})\,\Omega_\tau}
    \reducesto
    \returnterm \trueterm
  $,
  but there is no $R$ such that
  $
    \translatevalue{(\absterm{x}[\tau]{\trueterm})\,\Omega_\tau}
    \reducesto
    R
  $.

  For this example, we define the program relation
  $\syntaxleq_{\mathrm{div}}$ by
  \[
    M \syntaxleq_{\mathrm{div}} M'
    \quad\text{if and only if}\quad
    \forall V\!:\!\booltype.~(M \reducesto \returnterm{V})
      ~\Rightarrow~(M' \reducesto \returnterm{V})
  \]
  so that $M \ctxleq[\Gamma] M'$ informally means if a program
  containing $M$ terminates with some result then the same program with
  $M'$ instead of $M$ terminates with the same result.
\end{exa}

\begin{exa}[Nondeterminism]\label{ex:nondeterminism}
  For our third example, we consider finite nondeterminism.
  Again call-by-value and call-by-name have different behaviour,
  but any result of a call-by-value execution is also a result of a
  call-by-name execution (if suitable nondeterministic choices are
  made).
  
  We consider CBPV without recursion, but augmented with
  computations $\failconstant[\comp{C}]$ for nullary nondeterministic
  choice and $\orterm{M}{N}$ for binary nondeterministic choice between
  computations;
  the typing and evaluation rules are standard:
  \[
    \inferrule
      { }
      { \comptyped{\Gamma}{\failconstant[\comp{C}]}{\comp{C}} }
    \hspace{1em}
    \inferrule
      { \comptyped{\Gamma}{M}{\comp{C}} \\
        \comptyped{\Gamma}{N}{\comp{C}} }
      { \comptyped{\Gamma}{\orterm{M}{N}}{\comp{C}} }
    \hspace{2.5em}
    \inferrule
      { M \reducesto R }
      { \orterm{M}{N} \reducesto R }
    \hspace{1em}
    \inferrule
      { N \reducesto R }
      { \orterm{M}{N} \reducesto R }
  \]
  (There is no $R$ such that $\failconstant[\comp{C}] \reducesto R$.)
  The computation $\failconstant[\comp{C}]$ is the unit for
  $\orconstant$, so $\orterm{\failconstant[\comp{C}]}{M}$ and
  $\orterm{M}{\failconstant[\comp{C}]}$ have the same behaviour as $M$.
  For each closed computation $M : \freetype{A}$ there might be zero,
  one or several values $V : A$ such that
  $M \reducesto \returnterm{V}$.

  We similarly include nullary and binary nondeterminism in the source
  language, and extend the call-by-value and call-by-name translations:
  \[
   e~\Coloneqq~\dots~|~\failconstant[\tau]~|~\orterm{e}{e'}
    \hspace{3.5em}
    \inferrule
      { }
      { \welltyped{\Gamma}{\failconstant[\tau]}{\tau} }
    \hspace{1.5em}
    \inferrule
      { \welltyped{\Gamma}{e}{\tau} \\
        \welltyped{\Gamma}{e'}{\tau} }
      { \welltyped{\Gamma}{\orterm{e}{e'}}{\tau} }
  \]
  \begin{align*}
    \translatevalue{\failconstant[\tau]}
      &= \failconstant[\freetype{(\translatevalue{\tau})}]
    &
    \translatename{\failconstant[\tau]}
      &= \failconstant[\translatename{\tau}]
    \\
    \translatevalue{\orterm{e}{e'}}
      &= \orterm{\translatevalue{e}\mspace{-1mu}}{\mspace{-1mu}\translatevalue{e'}}
    \hspace{-0.09em}
    &
    \hspace{-0.09em}
    \translatename{\orterm{e}{e'}}
      &= \orterm{\translatename{e}\mspace{-1mu}}{\mspace{-1mu}\translatename{e'}}
  \end{align*}
  As an example, evaluating the expression
  $
    e =
    (\absterm{x}{
      \ifterm{x}{x}{\trueterm}})
    (\orterm{\trueterm}{\falseterm})
  $
  under call-by-value necessarily results in $\trueterm$, but under
  call-by-name we can also get $\falseterm$.
  (We have $\translatevalue{e} \not\reducesto \returnterm \falseterm$
  but $\translatename{e} \reducesto \returnterm \falseterm$.)

  For nondeterminism, we define $\syntaxleq_{\mathrm{nd}}$ in the same way as our divergence example:
  \[
    M \syntaxleq_{\mathrm{nd}} M'
    \quad\text{if and only if}\quad
    \forall V\!:\!\booltype.~(M \reducesto \returnterm{V})
      ~\Rightarrow~(M' \reducesto \returnterm{V})
  \]
  This captures the property that any result that arises from an execution
  of $M$ (which may involve call-by-value) might arise from an execution
  of $M'$ (which may involve call-by-name).
\end{exa}

\begin{exa}[Immutable state]\label{ex:immutable}
  Finally, we consider the basic languages enriched with an extra
  construct for getting the value of a immutable state whose value is
  either $\truevalue$ or $\falsevalue$.
  Once again we do not expect there to be any difference between
  call-by-value and call-by-name, and it is indeed the case that if $e$
  is a closed expression of type $\booltype$, then call-by-value and
  call-by-name evaluations of $e$ have the same behaviour
  (this is an instance of \autoref{corollary:cbv-cbn-programs}).
  Notably however, the model we use for this example fails to satisfy
  the assumptions of our main theorem (\autoref{theorem:cbv-cbn}).

  We augment CBPV with a computation $\getconstant$.
  This gets the value of the state, producing either $\trueterm$ or
  $\falseterm$.
  \[
    \inferrule
      { }
      { \comptyped{\Gamma}{\getconstant}{\freetype{\booltype}} }
  \]
  Big-step evaluation has a slightly different form in this case.
  We write $M \reducesto_b R$ to mean $M$ evaluates to $R$ when the
  state is $b \in \{\truevalue, \falsevalue\}$.
  The rules are those of \autoref{figure:cbpv-semantics} (with the
  subscript $b$ added), plus
  \[
    \inferrule
      { }
      { \getconstant \reducesto_{\truevalue} \returnterm{\trueterm} }
    \qquad
    \inferrule
      { }
      { \getconstant \reducesto_{\falsevalue} \returnterm{\falseterm} }
  \]
  Again we extend the source language, and also the call-by-value and
  call-by-name translations:
  \[
   e~\Coloneqq~\dots~|~\getconstant
    \hspace{3.5em}
    \inferrule
      { }
      { \welltyped{\Gamma}{\getconstant}{\booltype} }
    \hspace{3.5em}
    \translatevalue{\getconstant}
    =
    \translatename{\getconstant}
    =
    \getconstant
  \]
  We define the program relation $\syntaxleq_{\mathrm{get}}$ as follows:
  \[
    M\mspace{3mu}{\syntaxleq_{\mathrm{get}}}\mspace{3mu} M'
    \;\;\text{if and only if}\;\;
    \forall b \in \{\truevalue,\falsevalue\}.\,\exists V\!:\!\booltype.\,(M \mspace{2mu}{\reducesto_b}\mspace{2mu} \returnterm{V})
       \wedge (M' \mspace{2mu}{\reducesto_b}\mspace{2mu} \returnterm{V})
  \]
\end{exa}

\section{Order-enriched denotational semantics}
\label{section:cbpv-models}
We give a denotational semantics for CBPV, which we use to prove
instances of $\ctxleq$.
Since $\ctxleq$ is not in general symmetric, we use
\emph{order-enriched} models, which come with partial orders $\pleq$
between denotations. 
In an \emph{adequate} model, $\sem M \pleq \sem N$ implies
$M \ctxleq N$.
Our semantics is based on Levy's \emph{algebra
models}~\cite{levy:call-by-push-value-decomposing} for CBPV, in which
each computation type is interpreted as a monad algebra.
(We restrict to algebra models for simplicity. Other forms of model,
such as
\emph{adjunction models}~\cite{levy:call-by-push-value-adjunctions} can
be used for the same purpose.)

\subsection{Order-enriched categories and strong monads}
We define the basic categorical notions we need for the rest of the
paper.
We assume no knowledge of enriched category theory; instead we give the
relevant order-enriched (specifically $\Poset$-enriched) definitions
here.
(We do however assume some basic ordinary category theory.)
\begin{defi}
  A \emph{$\Poset$-category} $\C$ is an ordinary category, together with
  a partial order $\pleq$ on each hom-set $\C(X, Y)$, such that
  composition is monotone.
\end{defi}

If $\C$ is a $\Poset$-category, we refer to the ordinary category as the
\emph{underlying} ordinary category, and write $\ob{\C}$ for the class
of objects.

\begin{exa}
  We use the following three $\Poset$-categories.
  \begin{center}
    \begin{tabular}{rlll}
      \toprule
      $\Poset$-category $\C$ &
      Objects $X \in \ob{\C}$ &
      Morphisms $f : X \to Y$ &
      Order $f \pleq f'$ \\
      \midrule
      $\Set$ &
      sets &
      functions &
      equality \\
      $\Poset$ &
      posets &
      monotone functions &
      pointwise \\
      $\wCpo$ &
      \wcpo{}s &
      $\omega$-continuous functions &
      pointwise \\
      \bottomrule
    \end{tabular}
  \end{center}
  In each case, composition and identities are defined in the usual way.
  For $\Set$, since the hom-posets $\Set(X, Y)$ are discrete, all of the
  $\Poset$-enriched definitions coincide with the ordinary (unenriched)
  definitions.
  The objects of $\wCpo$ are posets $(X, \pleq)$ for which $\pleq$ is
  \emph{$\omega$-complete}, i.e.\ for which every $\omega$-chain
  $x_0 \pleq x_1 \pleq \cdots$ has a least upper bound $\Lub{x}$.
  Morphisms are \emph{$\omega$-continuous} functions, i.e.\ monotone
  functions that preserve least upper bounds of $\omega$-chains.
\end{exa}

Let $\C$ be a $\Poset$-category.
We say that $\C$ is \emph{cartesian} when its underlying category has a
terminal object $1$ and binary products $X_1 \times X_2$,
such that the pairing functions
$
  \pairmorphism{{-}}{{-}}
    : \Hom\C W {X_1} \times \Hom\C W {X_2} \to \Hom\C W {X_1 \times X_2}
$
are monotone.
We write $\pi_i : X_1 \times X_2 \to X_i$ for the projections from a
product, and write $\terminalmorphism{X} : X \to 1$ for the unique map
into the terminal object.
In every cartesian category, there are canonical \emph{associativity}
isomorphisms
$
  \assoc_{X_1, X_2, X_3}
    : (X_1 \times X_2) \times X_3 \to X_1 \times (X_2 \times X_3)
$.
We say that $\C$ is \emph{cartesian closed} when it is cartesian and its
underlying category has exponentials $\exp{X}{Y}$ for which the
currying functions
$
  \curry : \Hom \C {W \times X} Y \to \Hom \C W {\exp X Y}
$
are monotone.
We write $\ev_{X, Y}$ for the \emph{evaluation} morphism
$\uncurry \id : (\exp X Y) \times X \to Y$.

Binary \emph{coproducts} in $\C$ are just binary coproducts in the
underlying ordinary category, except that the copairing functions
$
  \copairmorphism{{-}}{{-}}
    : \C(X_1, W) \times \C(X_2, W) \to \C(X_1 + X_2, W)
$
are required to be monotone.
We write $\inl : X_1 \to X_1 + X_2$ and $\inr : X_2 \to X_1 + X_2$ for
the coprojections.
The $\Poset$-categories $\Set$, $\Poset$, and $\wCpo$ are all cartesian
closed, and have binary coproducts given by disjoint union.

Above we ask for monotonicity of the bijections
\begin{align*}
  \pairmorphism{{-}}{{-}} &: \Hom \C W {X_1} \times \Hom \C W {X_2} \to \Hom \C W {X_1 \times X_2}
  \\
  \curry &: \Hom \C {W \times X} Y \to \Hom \C W {\exp X Y}
  \\
  \copairmorphism{{-}}{{-}} &: \Hom \C {X_1} W \times \Hom \C {X_2} W \to \Hom \C {X_1 + X_2} W
\end{align*}
We do not need to require monotonicity of their inverses explictly,
because this holds automatically.
In particular, the uncurrying functions
$\uncurry : \C(W, \exp X Y) \to \C(W \times X, Y)$ are monotone
because $\uncurry f = \ev_{X, Y} \compose (f \times \id_X)$,
and $\compose$ and $\times$ are both monotone.

We interpret computation types as (Eilenberg--Moore) algebras for an
order-enriched monad $\T$, which we need to be \emph{strong} (just as
models of Moggi's monadic
metalanguage~\cite{moggi:notions-computation-monads} use a strong
monad).
The definitions of strong $\Poset$-monad and of $\T$-algebra we give are
slightly non-standard, but are equivalent to the standard
ones (see for example~\cite{mcdermott2022what}).
In particular, it is more convenient for us to bake the strength into
the (Kleisli) extension of the monad instead of having a separate
strength.
\begin{defi}[Strong $\Poset$-monad]
  Let $\C$ be a cartesian $\Poset$-category.
  A \emph{strong $\Poset$-monad} $\T$ on $\C$ consists of:
  \begin{itemize}
    \item an object $TX \in \ob \C$ for each $X \in \ob{\C}$;
    \item a morphism $\eta_X : X \to TX$ for each $X \in \ob{\C}$ (the \emph{unit});
    \item a monotone function
      $
        \extendr[W]{({-})}
          : \Hom \C {W \times X} {TY} \to \Hom \C {W \times TX} {TY}
      $
      (\emph{Kleisli extension}) for each $W,X,Y \in \ob \C$.
  \end{itemize}
  These are required to satisfy the following four laws.\footnotemark{}%
  \footnotetext{%
    The conference version \cite{mcdermott2022galois} of this paper incorrectly omits naturality in $W$ from the definition of strong $\Poset$-monad and from the definition of Eilenberg--Moore algebra.
    (Naturality in $W$ is required in \cite[Definition 4.1]{mcdermott2022what}.)
  }
  \begin{itemize}
    \item Naturality of extension in $W$:
      \[
        \extendr[W]{f} \compose (w \times \id_{TX})
        ~=~
        \extendr[W']{(f \compose (w \times \id_{X}))}
      \]
      for all $f : W \times X \to TY$ and $w : W' \to W$.
    \item Left unit:
      \[
        \extendr[W]{f} \compose (\id_W \times \eta_X)
        ~=~
        f
      \]
      for all $f : W \times X \to TY$.
    \item Right unit:
      \[
        \extendr[1]{(\eta_X \compose \pi_2)} = \pi_2
      \]
      for all $X \in \ob \C$.
    \item Associativity:
      \[
        \extendr[(W' \times W)]{(\extendr[W']{g} \compose (\id_{W'}{\times} f) \compose \assoc)}
        = \extendr[W']{g} \compose (\id_{W'}{\times}\extendr[W]{f})\compose\assoc
      \]
      for all $f : W \times X \to TY$ and $g : W' \times Y \to TZ$.
  \end{itemize}
\end{defi}
Specializing the Kleisli extension of $\T$ to $W = 1$ produces a
(non-strong) extension operator
$\extend {({-})} : \Hom \C X {TY} \to \Hom \C {TX} {TY}$, satisfying the usual monad laws:
\[
  \extend{f} \compose \eta_X = f
  \qquad
  \extend{\eta_X} = \id_X
  \qquad
  \extend{(\extend g \compose f)} = \extend g \compose \extend f
\]
We use this to define, for every $f : X \to Y$, a morphism
$Tf : TX \to TY$ by $Tf = \extend{(\eta_Y \compose f)}$.
The latter definition makes $T$ into a \emph{$\Poset$-functor}: the mapping $f \mapsto Tf$ is monotone, and preserves identities and composition.
The definition of $T$ on morphisms also ensures that the unit and Kleisli extension of $\T$ satisfy the following naturality laws:
\[
  Tf \compose \eta_X = \eta_{Y} \compose f
  \quad
  \extendr[W]{(Tg \compose f)} = Tg \compose \extendr[W]{f}
  \quad
  \extendr[W]{(f \compose (\id_W \times g))} = \extendr[W]{f} \compose (\id_W \times Tg)
\]

In the notation $\extendr[W]{f} : W \times TX \to TY$, the square
$\square$ indicates the position of $T$ in the domain.
Since products are symmetric, choosing to put $T$ to the right of $W$ is
arbitrary.
We construct a Kleisli extension operator with the square to the left as
follows:
\[
  \extendl[W]{f} = \extendr[W]{(f \compose \pairmorphism{\pi_2}{\pi_1})}
  \compose \pairmorphism{\pi_2}{\pi_1}: TX \times W \to TY
  \qquad
  \text{(where}~f : X \times W \to TY\text{)}
\]
We also define two natural transformations for sequencing of computations:
$\seql$ for left-to-right and $\seqr$ for right-to-left, as follows.
\begin{align*}
  \seql_{X_1, X_2}&~=~
    \extendl[TX_2]{(\extendr[X_1]{\eta_{X_1 \times X_2}})}
  ~:~T X_1 \times T X_2 \to T (X_1 \times X_2)
  \\
  \seqr_{X_1, X_2}&~=~
    \extendr[TX_1]{(\extendl[X_2]{\eta_{X_1 \times X_2}})}
  ~:~T X_1 \times T X_2 \to T (X_1 \times X_2)
\end{align*}
We further define an \emph{effectful} pairing operation
$\vpairmorphism{{-}}{{-}}$:
\[
  \vpairmorphism{f_1}{f_2} = \seql_{X_1, X_2} \compose
  \pairmorphism{f_1}{f_2}~:~W \to T(X_1 \times X_2)
  \qquad
  \text{(where~}f_i : W \to TX_i\text{)}
\]
This evaluates from left to right; we do not need the right-to-left
version.

\begin{defi}[Eilenberg--Moore algebra]
  Let $\T$ be a strong $\Poset$-monad on a cartesian $\Poset$-category
  $\C$.
  A \emph{$\T$-algebra} $\Z = (Z, \algextend{({-})})$ is a pair of:
  \begin{itemize}
    \item an object $Z \in \ob{\C}$ (the \emph{carrier});
    \item a monotone function
       $
         \algextendr[W]{({-})}
           : \Hom \C {W \times X} {Z} \to \Hom \C {W \times TX} {Z}
       $
       (the \emph{extension} operator) for each $W, X \in \ob \C$.
  \end{itemize}
  These are required to satisfy the following three laws.
  \begin{itemize}
    \item Naturality in $W$:
      \[
        \algextendr[W]{f} \compose (w \times \id_{TX})
        ~=~
        \algextendr[W']{(f \compose (w \times \id_{X}))}
      \]
      for all $f : W \times X \to Z$ and $w : W' \to W$.
    \item Left unit:
      \[
        \algextendr[W]{f} \compose (\id_W \times \eta_X)
        ~=~
        f
      \]
      for all $f : W \times X \to Z$.
    \item Associativity:
      \[
        \algextendr[(W' \times W)]{(\algextendr[W']{g} \compose (\id_{W'}{\times} f) \compose \assoc)}
        = \algextendr[W']{g} \compose (\id_{W'}{\times}\extendr[W]{f})\compose\assoc
      \]
      for all $f : W \times X \to TY$ and $g : W' \times Y \to Z$.
  \end{itemize}

  For each $\T$-algebra $\Z$, we write $\forget \T \Z$ for the carrier
  $Z \in \ob{\C}$.
\end{defi}
Just as for the extension operator of a strong $\Poset$-monad, we
specialize the extension operator of a $\T$-algebra to
$W = 1$ and obtain a (non-strong) extension operator
$\algextend {({-})} : \Hom \C X {Z} \to \Hom \C {TX} {Z}$.
We also have an extension operator with reversed products, written
$\algextendl[W] {({-})} : \Hom \C{X \times W}{Z} \to \Hom \C {TX \times W} {Z}$.

The following constructions of algebras are standard.
\pagebreak
\begin{defi}\label{definition:algebra-constructions}
  Let $\T$ be a strong $\Poset$-monad on a cartesian closed
  $\Poset$-category $\C$.
  \begin{itemize}
    \item The \emph{free} $\T$-algebra $\free{\T}{X}$ on an object $X \in \ob{\C}$ has
      carrier $TX$; the extension operator is Kleisli extension $\extendr{({-})}$.
    \item If $\Z_1$ and $\Z_2$ are $\T$-algebras, then their
      \emph{product} $\Z_1 \times \Z_2$ is the $\T$-algebra with carrier
      $Z_1 \times Z_2$, and extension operator
      \[
        \algextendr[W]{f} = \pairmorphism
            {\algextendr[W]{(\pi_1 \compose f)}}
            {\algextendr[W]{(\pi_2 \compose f)}}
      \]
    \item If $Y \in \ob{\C}$ and $\Z$ is a $\T$-algebra, then their
      \emph{power} $\compexp{Y}{\Z}$ is the $\T$-algebra with carrier
      $\exp{Y}{Z}$ and extension operator
      \[
        \algextendr[W]{f}
          = \curry{(\algextendr[(W \times Y)]{(\uncurry f \compose \beta_{W, Y, X})} \compose
          \beta_{W, TX, Y})}
      \]
      where
      $
        \beta_{X_1, X_2, X_3} = \pairmorphism
          {\pairmorphism{\pi_1 \compose \pi_1}{\pi_2}}
          {\pi_2 \compose \pi_1}
          : (X_1 \times X_2) \times X_3 \to (X_1 \times X_3) \times X_2
      $.
  \end{itemize}
\end{defi}
We use these constructions to interpret CBPV computation types: returner
types $\freetype{A}$ are interpreted as free $\T$-algebras, product types
$\comp{C}_1 \compprod \comp{C}_2$ are interpreted as product
$\T$-algebras, and function types $A \to \comp{C}$ are interpreted as
power $\T$-algebras.

\subsection{Models of CBPV}
We define the notion of (order-enriched, algebra) \emph{model} as
follows.

\begin{defi}
  A \emph{model} $\M = (\C, \T)$ of CBPV consists of
  \begin{itemize}
    \item a cartesian closed $\Poset$-category $\C$ that admits the
      coproduct $2 = 1 + 1$;
    \item a strong $\Poset$-monad $\T$ on $\C$.
  \end{itemize}
\end{defi}

\begin{figure}
  \begin{minipage}{0.35\linewidth}
  \begin{align*}
    \shortintertext{\[\boxed{{\C}\text{-object}~\sem{A}}\]}
    \sem{\unittype} &~=~ 1 \\
    \sem{A_1 \times A_2} &~=~ \sem{A_1} \times \sem{A_2} \\
    \sem{\booltype} &~=~ 2\quad(= 1 {+} 1) \\
    \sem{\thunktype{\comp{C}}} &~=~ \forget \T \sem{\comp{C}}
    \\[4ex]
    \shortintertext{\[\boxed{{\T}\text{-algebra}~\sem{\comp{C}}}\]}
    \sem{\comp{C}_1 \compprod \comp{C}_2} &~=~
      \sem{\comp{C}_1} \times \sem{\comp{C}_2} \\
    \sem{A \to \comp{C}} &~=~ \compexp{\sem{A}}{\sem{\comp{C}}} \\
    \sem{\freetype{A}} &~=~ \free \T \sem{A}
    \\[4ex]
    \shortintertext{\[\boxed{{\C}\text{-object}~\sem{\Gamma}}\]}
    \sem{\emptycontext} &~=~ 1 \\
    \sem{\Gamma, x : A} &~=~ \sem{\Gamma} \times \sem{A} \\
  \end{align*}
  \end{minipage}%
  \begin{minipage}{0.65\linewidth}
    \begin{align*}
      \shortintertext{\[\boxed{\sem{\valuetyped{\Gamma}{V}{A}}
        : \sem{\Gamma} \to \sem{A}}\]}
      \sem{x} &~=~ \proj{x} \\
      \sem{\unitterm} &~=~ \terminalmorphism{\sem{\Gamma}} \\
      \sem{\pairterm{V_1}{V_2}} &~=~
        \pairmorphism{\sem{V_1}}{\sem{V_2}} \\
      \sem{\trueterm} &~=~
        \inl \compose \terminalmorphism{\sem{\Gamma}} \\
      \sem{\falseterm} &~=~
        \inr \compose \terminalmorphism{\sem{\Gamma}} \\
      \sem{\thunkterm{M}} &~=~ \sem{M}
      \\[2ex]
      \shortintertext{\[\boxed{\sem{\comptyped{\Gamma}{M}{\comp{C}}}
        : \sem{\Gamma} \to \forget \T \sem{\comp{C}}}\]}
      \sem{\comppairterm{M_1}{M_2}} &~=~
        \pairmorphism{\sem{M_1}}{\sem{M_2}} \\
      \sem{\pushtermtight{i}{M}} &~=~
        \proj{i} \compose \sem{M}
        \tag{$i \in \{1, 2\}$}\\
      \sem{\absterm{x}[A]{M}} &~=~
        \curry \sem{M} \\
      \sem{\pushtermtight{V}{M}} &~=~
        \uncurry \sem{M} \compose \pairmorphism{\id}{\sem{V}} \\
      \sem{\returnterm{V}} &~=~
        \unit \compose \sem{V} \\
      \sem{\toterm{M}{x}{N}} &~=~
        \algextendr[\sem{\Gamma}]{\sem{N}} \compose \pairmorphism{\id}{\sem{M}} \\
        \hspace{-1.8em}\sem{\matchpairterm{\mspace{-2mu}V\mspace{-3mu}}{x}{y}{M}} &~=~
        \sem{M} \compose \assoc^{-1}
        \compose \pairmorphism{\id}{\sem{V}} \\
      \hspace{-1.3em}
      \sem{\ifterm{V}{M_1}{M_2}} &~=~
        \copairmorphism{\sem{M_1}}{\sem{M_2}}
        \compose \dist \compose \pairmorphism{\id}{\sem{V}} \\
      \sem{\forceterm{V}} &~=~ \sem{V}
    \end{align*}
  \end{minipage}
  \caption{Denotational semantics of CBPV}
  \label{figure:cbpv-denotations}
\end{figure}

Given a model $\M = (\C, \T)$, the interpretation $\sem{{-}}$ of CBPV is defined in
\autoref{figure:cbpv-denotations}.
Value types $A$ are interpreted as objects $\sem{A} \in \ob{\C}$, while
computation types $\comp C$ are interpreted as $\T$-algebras.
The value type $\thunktype{\comp C}$ is interpreted as the carrier
$\forget{\T}{\sem{\comp C}}$ of the $\T$-algebra $\sem{\comp C}$.
Typing contexts $\Gamma$ are interpreted as objects $\sem{\Gamma} \in \C$
using the cartesian structure of $\C$; if $(x : A) \in \Gamma$ then we
write $\proj{x}$ for the corresponding projection
$\sem{\Gamma} \to \sem{A}$.
Values $\valuetyped \Gamma V A$ (respectively computations
$\comptyped \Gamma M {\comp{C}}$) are interpreted as morphisms
$\sem{\valuetyped{\Gamma}{V}{A}}$
(resp.\ $\sem{\comptyped{\Gamma}{M}{\comp{C}}}$) in $\C$;
we often omit the typing context and type when writing these.
Programs $\comptyped{\emptycontext}{M}{\booltype}$ are therefore
interpreted as morphisms $\sem M : 1 \to T 2$.
To interpret $\iftermkeyword$, we use the fact that, since $\C$ is
cartesian closed, products distribute over the coproduct
$2 = 1 + 1$.
This means that for every $W \in \ob \C$, the coproduct $W + W$ also
exists in $\C$, and the canonical morphism
\[
  W + W
    \xto{\copairmorphism
      {\pairmorphism{\id_W}{\inl \compose \terminalmorphism W}}
      {\pairmorphism{\id_W}{\inr \compose \terminalmorphism W}}}
  W \times 2
\]
has an inverse $\dist_W : W \times 2 \to W + W$.

By composing the semantics of CBPV with the two translations of the
source language, we obtain a call-by-value semantics
$\semvalue{{-}} = \sem{\translatevalue{{-}}}$ and a call-by-name
semantics $\semname{{-}} = \sem{\translatename{{-}}}$ of the source
language.
For convenience, we spell out these composed semantics in
\autoref{figure:cbv-cbn-denotations}.

\begin{figure}
  \begin{subfigure}{\linewidth}
    \begin{gather*}
    \begin{array}{r @{\hspace{1em}\mapsto\hspace{1em}} l}
\multicolumn{2}{l}{\boxed{
      \text{type}~\tau \;\mapsto\; \text{object}~\semvalue{\tau} \in
      \ob{\C}
}}
      \\[1ex]
      \unittype & 1 \\
      \tau_1 \times \tau_2 &
      \semvalue{\tau_1} \times \semvalue{\tau_2} \\
      \booltype & 2 \\
      \tau \to \tau' &
      \exp{\semvalue{\tau}}
        {T{(\semvalue{\tau'})}}
    \end{array}
    \begin{array}{r @{\hspace{1em}\mapsto\hspace{1em}} l}
\multicolumn{2}{l}{\boxed{
      \text{typing context}~\Gamma \;\mapsto\;
      \text{object}~\semvalue{\Gamma} \in \ob{\C}
}}
      \\[1ex]
      \emptycontext & 1 \\
\hspace{4.5em}
      \Gamma, x : \tau &
      \semvalue{\Gamma} \times \semvalue{\tau}
    \end{array}
    \\[3ex]
    \begin{array}{r @{\hspace{1em}\mapsto\hspace{1em}} l}
\multicolumn{2}{l}{\boxed{
      \text{expression}~\welltyped{\Gamma}{e}{\tau} \;\mapsto\;
      \text{morphism}~
      \semvalue{e} : \semvalue{\Gamma} \to T(\semvalue{\tau})
      ~\text{in}~\C
}}
      \\[1ex]
      x & \pi_x \\
      \unitterm & \eta_1 \compose \terminalmorphism{\semvalue{\Gamma}} \\
      \pairterm{e_1}{e_2} &
      \vpairmorphism{\semvalue{e_1}}{\semvalue{e_2}} \\
      \fstterm{e} &
      T \pi_1 \compose \semvalue{e} \\
      \sndterm{e} &
      T \pi_2 \compose \semvalue{e} \\
      \trueterm & \eta_2 \compose \inl \compose \terminalmorphism{\semvalue{\Gamma}} \\
      \falseterm & \eta_2 \compose \inr \compose \terminalmorphism{\semvalue{\Gamma}} \\
      \ifterm{e_0}{e_1}{e_2} &
      \extendr[\semvalue{\Gamma}]{(\copairmorphism{\semvalue{e_1}}{\semvalue{e_2}}
      \compose \dist)}
      \compose \pairmorphism{\id_{\semvalue{\Gamma}}}{\semvalue{e_0}}
      \\
      \absterm{x}[\tau]{e} &
      \eta_{\semvalue{\tau \to \tau'}}
      \compose \curry{(\semvalue{e})} \\[0.5ex]
      e\,e' &
      \extend{\ev} \compose \vpairmorphism{\semvalue{e}}{\semvalue{e'}}
    \end{array}
    \end{gather*}
    \caption{Call-by-value semantics $\semvalue{{-}}$}
  \end{subfigure}

  \begin{subfigure}{\linewidth}
  \begin{gather*}\\
    \begin{array}{r @{\hspace{1em}\mapsto\hspace{1em}} l}
      \multicolumn{2}{l}{\mspace{5mu}\boxed{
      \text{type}~\tau \;\mapsto\; \text{$\T$-algebra}~\semname{\tau}
}}
      \\[1ex]
      \unittype & \free{\T}{1} \\
      \tau_1 \times \tau_2 &
      \semname{\tau_1} \times \semname{\tau_2} \\
      \booltype & \free{\T}{2} \\
      \tau \to \tau' &
      \exp{(\forget{\T}{(\semname{\tau})})}{\semname{\tau'}}
    \end{array}
    \begin{array}{r @{\hspace{1em}\mapsto\hspace{1em}} l}
\multicolumn{2}{l}{\boxed{
      \text{typing context}~\Gamma \;\mapsto\;
      \text{object}~\semname{\Gamma} \in \ob{\C}
}}
      \\[1ex]
      \emptycontext & 1 \\
\hspace{4.5em}
      \Gamma, x : \tau &
      \semname{\Gamma} \times \forget{\T}{(\semname{\tau})}
    \end{array}
    \\[3ex]
    \begin{array}{r @{\hspace{1em}\mapsto\hspace{1em}} l}
\multicolumn{2}{l}{\boxed{
      \text{expression}~\welltyped{\Gamma}{e}{\tau} \;\mapsto\;
      \text{morphism}~
      \semname{e} : \semname{\Gamma} \to \forget{\T}{(\semname{\tau})}
      ~\text{in}~\C
}}
      \\[1ex]
      x & \pi_x \\
      \unitterm & \eta_1 \compose \terminalmorphism{\semname{\Gamma}} \\
      \pairterm{e_1}{e_2} &
      \pairmorphism{\semname{e_1}}{\semname{e_2}} \\
      \fstterm{e} &
      \pi_1 \compose \semname{e} \\
      \sndterm{e} &
      \pi_2 \compose \semname{e} \\
      \trueterm & \eta_2 \compose \inl \compose \terminalmorphism{\semname{\Gamma}} \\
      \falseterm & \eta_2 \compose \inr \compose \terminalmorphism{\semname{\Gamma}} \\
      \ifterm{e_0}{e_1}{e_2} &
      \algextendr[\semname{\Gamma}]{(\copairmorphism{\semname{e_1}}{\semname{e_2}} \compose \dist)} \compose \pairmorphism{\id_{\semname{\Gamma}}}{\semname{e_0}}
      \\
      \absterm{x}[\tau]{e} &
      \curry{(\semname{e})} \\
      e\,e' &
      \ev \compose \pairmorphism{\semname{e}}{\semname{e'}}
    \end{array}
   \end{gather*}
    \caption{Call-by-name semantics $\semname{{-}}$}
  \end{subfigure}
  \caption{Denotational semantics of call-by-value and call-by-name}
  \label{figure:cbv-cbn-denotations}
\end{figure}

We use the denotational semantics as a tool for proving instances of
contextual preorders; for this we need \emph{adequacy}.
\begin{defi}
  A model of CBPV is \emph{adequate} with respect to a given program
  relation $\syntaxleq$ if for all computations
  $\comptyped{\Gamma}{M}{\comp{C}}$ and
  $\comptyped{\Gamma}{M'}{\comp{C}}$ we have
  \[
    \sem{\comptyped{\Gamma}{M}{\comp{C}}}
    \pleq
    \sem{\comptyped{\Gamma}{M'}{\comp{C}}}
    \quad\Rightarrow\quad
    M \ctxleq[\Gamma] M'
  \]
\end{defi}

\subsection{Examples}\label{section:semantics-examples}
We give four different models, one for each of the four examples in
\autoref{section:cbpv-examples}.
Each model is adequate with respect to the corresponding definition of
$\syntaxleq$; the proof in each case is a standard \emph{logical
relations} argument (e.g.\ \cite{winskel:formal-semantics}).

\begin{exa}
  For CBPV with no effects, we use $\C = \Set$.
  The strong $\Poset$-monad $\T$ is the identity on $\Set$.
  Each $\T$-algebra $\Z$ is completely determined by its carrier $Z$;
  the extension operator
  $\algextendr[W]{({-})} : \Set(W \times X, Z) \to \Set(W \times X, Z)$ is
  necessarily the identity.
  The interpretation $\sem M$ of a closed computation $M : \freetype{\booltype}$ is just an element of
  $2$.
\end{exa}

\begin{exa}\label{ex:divergence-monad}
  For divergence, we use $\C = \wCpo$.
  The strong $\Poset$-monad $\T$ freely adjoins a least element $\bot$
  to each $\wCpo$.
  The unit $\eta_X$ is the inclusion $X \hookrightarrow TX$, while
  Kleisli extension is given by
  \[
    \extendr[W]{f} (w, x) = \begin{cases}
      \bot & \text{if}~x = \bot \\
      f (w, x) & \text{otherwise}
    \end{cases}
  \]
  A $\T$-algebra $\Z$ is equivalently an $\wCpo$ $Z$ with a least
  element $\bot \in Z$.
  The extension operator is completely determined once the carrier is
  fixed; it is analogous to $\extendr{({-})}$.

  In this case, the product $Z_1 \times Z_2$ is the set of pairs
  ordered componentwise, and the exponential $\exp Y Z$ is the set of
  set of $\omega$-continuous functions ordered pointwise.
  Hence $Z_1 \times Z_2$ has a least element $(\bot, \bot)$ (so
  forms a $\T$-algebra) whenever $Z_1$ and $Z_2$ have least elements,
  and $\exp Y Z$ has a least element (the constantly-$\bot$ function)
  whenever $Z$ has a least element.

  If $\Z$ is a $\T$-algebra, then every $\omega$-continuous function
  $f : Z \to Z$ has a least fixed point
  $\fix{f} = \Lub_{n\in\natset} f^n \bot \in Z$.
  These enable us to interpret recursive computations, by defining
  $
    \sem{\cbpvrecterm{x}[\comp C]{M}} \rho =
      \fix{\!\big(\lambda x.\, \sem{M} (\rho, x)\big)}
  $.
  The interpretation $\sem M$ of a closed computation
  $M : \freetype \booltype$ is either $\bot$ (signifying
  divergence), or one of the two elements of $2$.
\end{exa}

\begin{exa}\label{ex:nondeterminism-monad}
  For finite nondeterminism, we use $\C = \Poset$.
  The strong $\Poset$-monad $\T$ freely adds finite joins to each poset.
  It is defined by
  \[\textstyle
    TX = (\{\downclose{S'} \mid S' \in \finpowerset{X}\}, \subseteq)
    \qquad
    \unit_{\poset{X}}\,x = \downclose{\{x\}}
    \qquad
    \extendr[W]{f} (w, S) = \bigcup_{x \in S} f(w, x)
  \]
  where $\finpowerset{X}$ is the set of finite subsets of $X$, and
  $\downclose {S'} = \{x \in X \mid \exists x'\in S'.\,x \pleq x'\}$ is
  the \emph{downwards-closure} of $S' \subseteq X$.
  Each $\T$-algebra is again completely determined by its carrier;
  a $\T$-algebra $\Z$ is equivalently a poset $Z$ that has finite
  joins.
  The extension operator is necessarily given by
  $
    \algextendr[W]{f} (w, S) = \Lub_{x \in S} f(w, x)
  $.
  (The latter join exists because $S$ is the downwards-closure of a finite set,
  even though $S$ itself might not be finite.)
  The product $Z_1 \times Z_2$ is the set of pairs ordered
  componentwise, with joins given by
  $\Lub_i (z_i, z'_i) = (\Lub_i z_i, \Lub_i z'_i)$.
  The power $\exp Y Z$ is the set of monotone functions ordered
  pointwise, with joins given by $(\Lub_i f_i) x = \Lub_i (f_i x)$.

  We interpret nondeterministic computations using nullary and binary
  joins:
  \[
    \sem{\failconstant[\comp{C}]} \rho = \bot
    \qquad
    \sem{\orterm{M}{N}} \rho = \sem{M}\rho \join \sem{N}\rho
  \]
  The interpretation $\sem M$ of a closed computation $M : \freetype \booltype$ is
  one of the four subsets of $2$.
\end{exa}

\begin{exa}\label{ex:immutable-monad}
  For immutable state, we use $\C = \Set$, with the \emph{reader} monad
  \[
    TX = (\exp{2}{X})
    \qquad
    \eta_X x = \lambda b.\,x
    \qquad
    \extendr[W]{f}(w, t) = \lambda b.\,f(w, t\,b)\,b
  \]
  where $2 = \{\truevalue, \falsevalue\}$.
  The CBPV computation $\getconstant$ is interpreted as
  \[
    \sem{\getconstant} \rho = \lambda b.\,b
  \]
\end{exa}

\section{The relation between call-by-value and call-by-name}
\label{section:relation}
We now return to the main contribution of this paper: relating
call-by-value with call-by-name.
Recall the first step outlined in the introduction.
We define a family of relations $\vnrel{}{}$ (\autoref{def:vnrel})
that compare the observable behaviour of a denotation of call-by-value
type with a denotation of call-by-name type.
The main result of this section is that, under certain conditions on
computational effects, we have
\[
  \semvalue{e} \vnrel{}{} \semname{e}
\]
for all $\welltyped{\Gamma}{e}{\tau}$ (\autoref{thm:fundamental}).
Here we work with the denotational semantics, instead of with the
syntax directly, so the relations $\vnrel{}{}$ are defined with respect
to a fixed model $\M$ that we assume to be given.
There is one relation $\vnrel{}{}$ for each source-language context
$\Gamma$ and type $\tau$:
\[
  g^v \vnrel{}{} g^n~~\text{where}~~
  g^v : \semvalue{\Gamma} \to T(\semvalue{\tau}),~
  g^n : \semname{\Gamma} \to \forget{\T}{(\semname{\tau})}
\]

To define $\vnrel{}{}$, we first give a family of relations
\[
  f^v\;\semrel{\tau}[W]\;f^n
\]
that relate elements $f^v$ of $T(\semvalue{\tau})$ with elements $f^n$ of
$\forget{\T}{(\semname{\tau})}$.
Here by element we mean \emph{generalized} element, so $f^v$ and $f^n$
are morphisms
\[
  f^v : W \to T(\semvalue{\tau})
  \qquad
  f^n : W \to \forget{\T}{(\semname{\tau})}
\]
from some $W$.
The definition of $\semrel{\tau}[W]$ is in the style of a logical relation, by
induction on the type $\tau$.
The cases are listed in \autoref{fig:relation}.
Informally, we have the following.
\begin{itemize}
  \item For $\tau = \unittype$ and $\tau = \booltype$, we compare
    $f^v$ and $f^n$ directly using the order relation $\pleq$ on
    morphisms in $\C$.
    (We can do this because
    $T(\semvalue{\tau}) = \forget{\T}{(\semname{\tau})}$.)
  \item For a product type $\tau_1 \times \tau_2$, we compare the first
    components and compare the second components.
    We get these components by composing with the call-by-value and
    call-by-name interpretations of the projections
    $\fstkeyword$ and $\sndkeyword$.
  \item For a function type $\tau \to \tau'$, we relate $f^v$ to $f^n$
    when these give related results when applied to related arguments.
    Here we use the call-by-value and call-by-name interpretations of
    application.
\end{itemize}
Note that in the function case we quantify over morphisms $w : W' \to W$
to permit varying \emph{arities} $W$ (cf.\ the Kripke logical relations
of varying arity of \cite{jung1993characterization}).
(Precisely, this ensures that $\semrel{\tau}$ is closed under
precomposition with morphisms $w : W' \to W$, as in
\autoref{relation-closure}(2) below.)

\begin{figure}
  \begin{align*}
    \intertext{
      $\boxed{
        f^v~\semrel{\tau}[W]~f^n
        \text{~(where~}
        f^v : W \to T(\semvalue{\tau})
        \text{~and~}
        f^n : W \to \forget{\T}({\semname{\tau}})
        \text{)}
      }$
    }
    f^v~\semrel{\unittype}[W]~f^n
    &\;\;\text{iff}\;\;
    f^v \pleq f^n \\
    f^v~\semrel{\tau_1 \times \tau_2}[W]~f^n
    &\;\;\text{iff}\;\;
    (T\pi_1 \compose f^v)~\semrel{\tau_1}[W]~(\pi_1 \compose f^n)
    ~~\wedge~~
    (T\pi_2 \compose f^v)~\semrel{\tau_2}[W]~(\pi_2 \compose f^n)
    \\
    f^v~\semrel{\booltype}[W]~f^n
    &\;\;\text{iff}\;\;
    f^v \pleq f^n \\
    f^v~\semrel{\tau \,{\to}\, \tau'}[W]~f^n
    &\;\;\text{iff}\;\;
    \forall W', w : W' \to W, g^v, g^n.\,
    \\
      &\hspace{2.2em}
      g^v~\semrel{\tau}[W']~g^n
      ~~\Rightarrow~~
      (\extend{\ev} \compose \vpairmorphism{f^v \compose w}{g^v})
      ~\semrel{\tau'}[W']~
      (\ev \compose \pairmorphism{f^n \compose w}{g^n})
  \end{align*}
  \caption{The relation between call-by-value and call-by-name}
  \label{fig:relation}
\end{figure}

We define $\vnrel{}{}$ in terms of $\semrel{{-}}$.
To state the definition, we need some more notation.
Let
$\Gamma' = x_1 : A_1, \dots, x_k : A_k$ be a CBPV context.
Given a morphism $f_i : W \to \sem{A_i}$ for each $i \leq k$,
we obtain a morphism $\tuplemorphism{f_i}_i : W \to \sem{\Gamma}$, by
iterated pairing.
Given instead a morphism $f_i :W \to T\sem{A_i}$ for each $i$,
we obtain a morphism $\vtuplemorphism{f_i}_i : W \to T\sem{\Gamma}$ by
iterated pairing and left-to-right evaluation:
\[
  \tuplemorphism{f_i}_{i \leq k} =
    \begin{cases*}
      \terminalmorphism{W}
      &if $k = 0$
      \\
      \pairmorphism{\tuplemorphism{f_i}_{i \leq (k - 1)}}{f_k}
      &if $k > 0$
    \end{cases*}
  \qquad
  \vtuplemorphism{f_i}_{i \leq k} =
    \begin{cases*}
      \eta_1 \compose \terminalmorphism{W}
      &if $k = 0$
      \\
      \vpairmorphism{\vtuplemorphism{f_i}_{i \leq (k - 1)}}{f_k}
      &if $k > 0$
    \end{cases*}
\]

\begin{defi}\label{def:vnrel}
  Let $\M$ be a CBPV model, and let
  \[
    g^v : \semvalue{\Gamma} \to T(\semvalue{\tau'})
    \qquad
    g^n : \semname{\Gamma} \to \forget{\T}{(\semname{\tau'})}
  \]
  be morphisms,
  where $\Gamma = x_1 : \tau_1, \dots, x_k : \tau_k$.
  We write \[g^v\;\vnrel{}{}\;g^n\] when, for all objects
  $W \in \ob{\C}$ and families of morphisms
  \[
  (f_i^v : W \to T(\semvalue{\tau_i}))_i
  \qquad
  (f_i^n : W \to \forget{\T}{(\semname{\tau_i})})_i
  \]
  we have
  \[
    f_1^v~\semrel{\tau_1}[W]~f_1^n
    ~~\wedge~~
    \cdots
    ~~\wedge~~
    f_k^v~\semrel{\tau_k}[W]~f_k^n
    \quad\Rightarrow\quad
    (\extend{g^v} \compose \vtuplemorphism{f_i^v}_i)~
    ~\semrel{\tau'}[W]~
    ({g^n} \compose \tuplemorphism{f_i^n}_i)
  \]
\end{defi}


As we mention above, our goal is relate the observable behaviour of
$\semvalue{e}$ to the observable behaviour of $\semname{e}$.
Precisely, we want to prove
$\semvalue{e} \vnrel{}{} \semname{e}$.
By considering what this means for specific expressions $e$, we can see that this is
not true in general, for three reasons:
\begin{itemize}
  \item Consider the expression
    \[
      e = (\absterm{x}[\booltype]{()}) : \booltype \to \unittype
    \]
    If we apply this to an argument, then in call-by-value we
    evaluate the argument but in call-by-name we do not.
  \item
    Consider the expression
    \[
      e =
      (\absterm{x}[\booltype]{\ifterm{x}{x}{x}})
      : \booltype \to \booltype
    \]
    If we apply this, then in call-by-value the argument is evaluated
    once, but in call-by-name the argument is evaluated twice.
  \item
    Consider the expression
    \[
      e =
      (\absterm{x}[\booltype]{\absterm{y}[\booltype]{
        \ifterm{y}{x}{x}}})
      :  \booltype \to \booltype \to \booltype
    \]
    In call-by-value the argument to the outer function is evaluated
    first, and the argument to the inner function is evaluated second.
    In call-by-name the arguments are evaluated in the opposite order.
\end{itemize}

This suggests we should assume that computations can be discarded,
copied, and reordered with respect to other computations.
Precisely, we want the following properties.
\begin{defi}\label{def:first-order-semantic-properties}
  Let $\T$ be a strong $\Poset$-monad.
  A morphism $f : X \to TY$ is:
  \begin{itemize}
    \item \emph{lax discardable} when
      \[
        T\terminalmorphism{Y} \compose f
        \pleq \eta_1 \compose \terminalmorphism{X}
        ~~:X \to T1
      \]
    \item \emph{lax copyable} when
      \[
        T\pairmorphism{\id_Y}{\id_Y} \compose f
        \pleq \seql_{Y, Y} \compose \pairmorphism{f}{f}
        ~~:X \to T(Y \times Y)
      \]
      equivalently, when
      \[
        T\pairmorphism{\id_Y}{\id_Y} \compose f
        \pleq \seqr_{Y, Y} \compose \pairmorphism{f}{f}
        ~~:X \to T(Y \times Y)
      \]
    \item \emph{lax central} when
      \[
        \seql_{Y, W} \compose (f \times \id_{TW})
        \pleq
        \seqr_{Y, W} \compose (f \times \id_{TW})
        ~~:X \times TW \to T(Y \times W)
      \]
      for all $W \in \ob{\C}$.
  \end{itemize}
\end{defi}
The non-lax versions of these properties were first defined by
F{\"u}hrmann~\cite{fuhrmann:direct-models}.

\begin{exa}
  For each of our examples from \autoref{section:semantics-examples},
  every morphism $f : X \to TY$ is lax discardable, lax copyable, and lax central.
\end{exa}

Here we define these three properties for morphisms in the model $\M$,
but there are similar notions for CBPV computations, as the following
lemma shows.

\begin{lem}\label{lemma:first-order-syntactic-properties}
  Let $\comptyped{\Gamma}{M}{\freetype{A}}$ be a CBPV computation.
  The following hold for every CBPV model that is adequate with respect
  to a program relation $\syntaxleq$.
  \begin{itemize}
    \item If $\sem{M}$ is lax discardable, then
      \[
        \toterm{M}{x}{\returnterm{\unitterm}}
        ~\ctxleq[\Gamma]~
        \returnterm{\unitterm}
      \]
    \item If $\sem{M}$ is lax copyable, then
      \[
        \toterm{M}{x}{\returnterm{\pairterm{x}{x}}}
        ~\ctxleq[\Gamma]~
        \toterm{M}{x_1}{\toterm{M}{x_2}{\returnterm{\pairterm{x_1}{x_2}}}}
      \]
    \item If $\sem{M}$ is lax central, then
      \[
        \toterm{M}{x}{\toterm{\forceterm{z}}{y}{\returnterm{\pairterm{x}{y}}}}
        ~\ctxleq[\Gamma, z : \thunktype{(\freetype{B})}]~
        \toterm{\forceterm{z}}{y}{\toterm{M}{x}{\returnterm{\pairterm{x}{y}}}}
      \]
  \end{itemize}
\end{lem}
\begin{proof}
  Since we assume adequacy, in each case we can reason inside the model.
  \begin{itemize}
    \item If $\sem{M}$ is lax discardable, then
      \begin{align*}
        \sem{\toterm{M}{x}{\returnterm{\unitterm}}}
        &~=~
        T\terminalmorphism{\sem{A}} \compose \sem{M}
        \\&~\pleq~
        \eta_1 \compose \terminalmorphism{\sem{\Gamma}}
        \\&~=~
        \sem{\returnterm{\unitterm}}
      \end{align*}

    \item If $\sem{M}$ is lax copyable, then
      \begin{align*}
        \sem{\toterm{M}{x}{\returnterm{\pairterm{x}{x}}}}
        &~=~
        T\pairmorphism{\id_{\sem{A}}}{\id_{\sem{A}}} \compose \sem{M}
        \\&~\pleq~
        \seql_{\sem{A}, \sem{A}} \compose \pairmorphism{\sem{M}}{\sem{M}}
        \\&~=~
        \sem{\toterm{M}{x_1}{\toterm{M}{x_2}{\returnterm{\pairterm{x_1}{x_2}}}}}
      \end{align*}

    \item If $\sem{M}$ is lax central, then
      \begin{align*}
        \sem{\toterm{M}{x}{\toterm{\forceterm{z}}{y}{\returnterm{\pairterm{x}{y}}}}}
        &~=~
        \seql_{\sem{A}, \sem{B}} \compose (\sem{M} \times \id_{T\sem{B}})
        \\&~\pleq~
        \seqr_{\sem{A}, \sem{B}} \compose (\sem{M} \times \id_{T\sem{B}})
        \\&~=~
        \sem{\toterm{\forceterm{z}}{y}{\toterm{M}{x}{\returnterm{\pairterm{x}{y}}}}}
        \qedhere
      \end{align*}
  \end{itemize}
\end{proof}

We turn to the proof that lax discardability, lax copyability and lax
centrality are sufficient to relate call-by-value to call-by-name.
The following two lemmas are useful for this.
The first lemma says that (even without assuming these properties of
effects), the relations $\semrel{\tau}$ are closed under various
operations.

\begin{lem}\label{relation-closure}
  Let $\M$ be a CBPV model.
  For each $\tau$, the family of relations $\semrel{\tau}$ has the
  following closure properties.
  \begin{enumerate}
    \item For all $f^v, g^v, f^n, g^n$, we have
      \[
        g^v \pleq f^v
        \;\wedge\;
        f^v\;\semrel{\tau}[W]\;f^n
        \;\wedge\;
        f^n \pleq g^n
        \quad\Rightarrow\quad
        g^v~\semrel{\tau}[W]~g^n
      \]
    \item For all $w : W' \to W$, and $f^v, f^n$, we have
      \[
        f^v~\semrel{\tau}[W]~f^n
        \quad\Rightarrow\quad
        (f^v \compose w)
        ~\semrel{\tau}[W']~
        (f^n \compose w)
      \]
    \item For all $f^v, f^n$, we have
      \[
        f^v~\semrel{\tau}[W \times X]~f^n
        \quad\Rightarrow\quad
        \extendr[W]{(f^v)}
        ~\semrel{\tau}[W \times TX]~
        \algextendr[W]{(f^n)}
      \]
    \item For all $W_1, W_2$ such that the coproduct $W_1 + W_2$
      exists, and all $f_1^v, f_2^v, f_1^n, f_2^n$, we have
      \[
        f_1^v~\semrel{\tau}[W_1]~f_1^n
        ~~\wedge~~
        f_2^v~\semrel{\tau}[W_2]~f_2^n
        \quad\Rightarrow\quad
        \copairmorphism{f_1^v}{f_2^v}
        ~\semrel{\tau}[W_1+W_2]~
        \copairmorphism{f_1^n}{f_2^n}
      \]
  \end{enumerate}
\end{lem}
\begin{proof}
  The proof of each property is by induction on the type $\tau$.

  \begin{enumerate}
  \item The $\unittype$ and $\booltype$ cases are trivial, while the
  cases for product and function types follow from the inductive
  hypothesis by monotonicity of composition and pairing.

  \item The $\unittype$ and $\booltype$ cases follow from monotonicity
  of composition.
  The case for product types follows from the inductive hypothesis.
  For a function type $\tau \to \tau'$ we need to show, for every
  $w' : W'' \to W'$ and $g^v, g^n$, that
  $g^v~\semrel{\tau}[W'']~g^n$ implies
  \[
    (\extend{\ev} \compose \vpairmorphism{f^v \compose w \compose w'}{g^v})
    ~\semrel{\tau'}~
    (\ev \compose \pairmorphism{f^n \compose w \compose w'}{g^n})
  \]
  This follows immediately from the assumption $f^v~\semrel{\tau \to \tau'}[W]~f^n$,
  instantiated with the morphism $w \compose w' : W'' \to W$.

  \item The $\unittype$ and $\booltype$ cases follow from
  monotonicity of extension operators.
  The case for product types follows from the inductive hypothesis, by
  naturality of extension operators and the definition of the product
  $\T$-algebra:
  \begin{align*}
    (T \pi_i \compose \extendr[W]{(f^v)})
    =
    (\extendr[W]{(T \pi_i \compose f^v)})
    ~\semrel{\tau_i}[W \times TX]~
    (\algextendr[W]{(\pi_i \compose f^n)})
    =
    (\pi_i \compose \algextendr[W]{(f^n)})
  \end{align*}
  For a function type $\tau \to \tau'$ we show, for every
  $w : W' \to W \times TX$ and $g^v, g^n$ satisfying
  $g^v~\semrel{\tau}[W']~g^n$, that
  \[
    (\extend{\ev} \compose \vpairmorphism{\extendr[W]{(f^v)} \compose w}{g^v})
    ~\semrel{\tau'}[W']~
    (\ev \compose \pairmorphism{\algextendr[W]{(f^n)} \compose w}{g^n})
  \]
  By property (2) above, we have
  \[
    (g^v \compose \pi_2)
    ~\semrel{\tau}[(W \times X) \times W']~
    (g^n \compose \pi_2)
  \]
  so that $f^v~\semrel{\tau \to \tau'}[W \times X]~f^n$ implies
  \[
    (\extend{\ev} \compose \vpairmorphism{f^v \compose \pi_1}{g^v \compose \pi_2})
    ~\semrel{\tau'}[(W \times X) \times W']~
    (\ev \compose \pairmorphism{f^n \compose \pi_1}{g^n \compose \pi_2})
  \]
  Hence, by applying (2) and the inductive hypothesis for $\tau'$, we
  have
  $
    h^v~\semrel{\tau'}[W']~h^n
  $,
  where we define
  \begin{align*}
    h^v &~=~
    \extendr[(W \times W')]{(\extend{\ev}
      \compose \vpairmorphism{f^v \compose \pi_1}{g^v \compose \pi_2}
      \compose \beta_{W, W', X})}
    \compose \beta_{W, TX, W'} \compose \pairmorphism{w}{\id_{W'}}
    \\
    h^n &~=~ \algextendr[(W \times W')]{(\ev
      \compose \pairmorphism{f^n \compose \pi_1}{g^n \compose \pi_2}
      \compose \beta_{W, W', X})}
    \compose \beta_{W, TX, W'} \compose \pairmorphism{w}{\id_{W'}}
  \end{align*}
  with $\beta$ as in \autoref{definition:algebra-constructions}.
  It then remains to show that $h^v$ and $h^n$ are the two sides of the
  required instance of $\semrel{\tau'}[W']$, which we prove as follows.
  To prove we have the correct left-hand side, we use the associativity law, naturality of Kleisli extension, and the associativity law again, as follows.
  \[
    \begin{aligned}
      h^v
      ~=~&
      \extend{\ev} \compose \extendr[(W \times W')]{(\vpairmorphism{f^v \compose \pi_1}{g^v \compose \pi_2} \compose \beta_{W, W', X})} \compose \beta_{W, TX, W'} \compose \pairmorphism{w}{\id_{W'}}
      \\~=~&
      \extend{\ev} \compose \extendr[(W \times T(\semvalue{\tau}))]{(\seql \compose (f^v \times \id) \compose \beta_{W,  T(\semvalue{\tau}), X})} \compose \beta_{W, TX, T(\semvalue{\tau})} \compose \pairmorphism{w}{g^v}
      \\~=~&
      \extend{\ev} \compose \seql \compose (\extendr[W]{(f^v)} \times \id_{T(\semvalue{\tau})}) \compose \pairmorphism{w}{g^v}
      \\~=~&
      \extend{\ev} \compose \vpairmorphism{\extendr[W]{(f^v)} \compose w}{g^v}
    \end{aligned}
  \]
  To prove we have the correct right-hand side, we use naturality of extension, and the definition of power $\T$-algebras, as follows.
  \[
    \begin{aligned}
      h^n
      ~=~&
      \algextendr[(W \times W')]{(\ev
        \compose (f^n \times \id_{\forget{\T}{(\semname{\tau})}})
        \compose \beta_{W, \forget{\T}{(\semname{\tau})}, X})}
      \compose \beta_{W, TX, \forget{\T}{(\semname{\tau})}} \compose \pairmorphism{w}{g^n}
      \\~=~&
      \ev \compose (\algextendr[W]{(f^n)} \times \id_{\forget{\T}{(\semname{\tau})}}) \compose \pairmorphism{w}{g^n}
      \\~=~&
      \ev \compose \pairmorphism{\algextendr[W]{(f^n)} \compose w}{g^n}
    \end{aligned}
  \]

\item The $\unittype$ and $\booltype$ cases are immediate from
  monotonicity of the copairing operator $\copairmorphism{{-}}{{-}}$.
  For product types, it is enough to note that
  $
    T\pi_i \compose \copairmorphism{f_1^v}{f_2^v}
    =
    \copairmorphism{T\pi_i \compose f_1^v}{T\pi_i \compose f_2^v}
  $
  and
  $
    \pi_i \compose \copairmorphism{f_1^n}{f_2^n}
    =
    \copairmorphism{\pi_i \compose f_1^n}{\pi_i \compose f_2^n}
  $,
  and then apply the inductive hypothesis.
  For a function type $\tau \to \tau'$, we show that
  $g^v~\semrel{\tau}[W']~g^n$ implies
  \[
    (\extend{\ev} \compose \vpairmorphism{\copairmorphism{f^v_1}{f^v_2} \compose w}{g^v})
    ~\semrel{\tau'}[W']~
    (\ev \compose \pairmorphism{\copairmorphism{f^n_1}{f^n_2} \compose w}{g^n})
  \]
  where $w : W' \to W_1 + W_2$.
  To do this, consider the following objects $W'_1, W'_2$ and morphisms
  $w'_1, w'_2$.
  \[
    W'_i = W' \times W_i
    \qquad
    w'_i = \pi_2 : W'_i \to W_i
    \qquad(i \in \{1, 2\})
  \]
  Property (2) implies
  \[(g^v \compose \pi_1)~\semrel{\tau}[W'_i]~(g^n \compose \pi_1)
  \]
  so
  from the assumption
  \[f_i^v~\semrel{\tau \to \tau'}[W_i]~f_i^n
  \]
  we obtain
  $
    k^v_i
    ~\semrel{\tau'}[W'_i]~
    k^n_i
  $,
  where
  \[
    k^v_i = \extend{\ev} \compose \vpairmorphism{f^v_i \compose w'_i}{g^v \compose \pi_1}
    \qquad
    k^n_i = \ev \compose \pairmorphism{f^n_i \compose w'_i}{g^n \compose \pi_1}
  \]
  Coproducts are distributive (because we assume cartesian closure),
  so the coproduct $W'_1 + W'_2$ exists and is isomorphic to
  $W' \times (W_1 + W_2)$.
  We can therefore apply the inductive hypothesis for $\tau$, and then
  (2), to obtain
  \[
    (\copairmorphism{k^v_1}{k^v_2} \compose h)
    ~\semrel{\tau'}[W']~
    (\copairmorphism{k^n_1}{k^n_2} \compose h)
  \]
  where
  \[
    h : W'
    \xto{\pairmorphism{\id_{W'}}{w}}
    W' \times (W_1 + W_2)
    \xto{\iso}
    W'_1 + W'_2
  \]
  The result follows because
  \[
    \copairmorphism{k^v_1}{k^v_2} \compose h
    =
    \extend{\ev} \compose \vpairmorphism{\copairmorphism{f^v_1}{f^v_2} \compose w}{g^v}
    \qquad
    \copairmorphism{k^n_1}{k^n_2} \compose h
    =
    \ev \compose \pairmorphism{\copairmorphism{f^n_1}{f^n_2} \compose w}{g^n}
    \qedhere
  \]
\end{enumerate}
\end{proof}

The second lemma consists of some technical consequences of lax discardability,
lax copyability and lax centrality; we state them here for use in the
proof of \autoref{thm:fundamental} below.
For convenience, we render each of the inequalities in the statement of
the lemma in the syntax of CBPV
(we will not need the syntactic inequalities in the following, so we
omit the precise statements and proof).
\begin{lem}\label{first-order-corollaries}
  Let $\T$ be a strong $\Poset$-monad.
  \begin{enumerate}
    \item Let $f_1 : W \to TX_1$ and $f_2 : W \to TX_2$ be morphisms.
      For each $i \in \{1, 2\}$, if $f_j$ is lax discardable for
      $j \neq i$, then
      \[
        T \pi_i \compose \vpairmorphism{f_1}{f_2}
        ~\pleq~
        f_i
        ~~:W \to TX_i
      \]
      \[
        \toterm{M_1}{x_1}{\toterm{M_2}{x_2}{\returnterm{x_i}}}
        ~\ctxleq~
        M_i
        \qquad(f_k = \sem{M_k})
      \]
    \item Let $f_1 : W_1 \to TX_1$, $f_2 : W_2 \to TX_2$ and $g :
      X_1 \times X_2 \to TY$ be morphisms.
      If $f_1$ is lax central, then
      \[
        \extendl[TX_2]{(\extendr[X_1]{g})} \compose (f_1 \times f_2)
        ~\pleq~
        \extendr[TX_1]{(\extendl[X_2]{g})} \compose (f_1 \times f_2)
        ~~:W_1 \times W_2 \to TY
      \]
      \[
        \toterm{M_1}{x_1}{\toterm{M_2}{x_2}{N}}
        ~\ctxleq~
        \toterm{M_2}{x_2}{\toterm{M_1}{x_1}{N}}
        \qquad(f_i = \sem{M_i}, g = \sem{N})
      \]
      If $f_2$ is lax central, then
      \[
        \extendr[TX_1]{(\extendl[X_2]{g})} \compose (f_1 \times f_2)
        ~\pleq~
        \extendl[TX_2]{(\extendr[X_1]{g})} \compose (f_1 \times f_2)
        ~~:W_1 \times W_2 \to TY
      \]
      \[
        \toterm{M_2}{x_2}{\toterm{M_1}{x_1}{N}}
        ~\ctxleq~
        \toterm{M_1}{x_1}{\toterm{M_2}{x_2}{N}}
        \qquad(f_i = \sem{M_i}, g = \sem{N})
      \]
    \item Let $f : W \to TX$, $g : X \to TY$ and $h : X \times Y \to TZ$
      be morphisms.
      If $f$ is lax copyable and lax central, then
      \[
        \extend{(\extendr[X]{h} \compose \pairmorphism{\id_X}{g})} \compose f
        ~\pleq~
        \extendr[W]{(\extendl[Y]{h} \compose (f \times \id_Y))} \compose \pairmorphism{\id_W}{\extend{g} \compose f}
        ~~:W \to TZ
      \]
      \[
        \toterm{M}{x}{\toterm{N}{y}{N'}}
        ~\ctxleq~
        \toterm{(\toterm{M}{x}{N})}{y}{\toterm{M}{x}{N'}}
        \quad(f = \sem{M}, g = \sem{N}, h = \sem{N'})
      \]
    \item Let $f : W \to TX$, $g_1 : X \to TY_1$ and $g_2 : X \to TY_2$
      be morphisms.
      If $f$ is lax copyable and lax central, then
      \[
        \extend{\vpairmorphism{g_1}{g_2}} \compose f
        ~\pleq~
        \vpairmorphism{\extend{g_1} \compose f}{\extend{g_2} \compose f}
        ~~:W \to T(Y_1 \times Y_2)
      \]
      \[
        \begin{array}{l}
        \toterm{M}{x}{\toterm{N_1}{y_1}{\\\toterm{N_2}{y_2}{\\\returnterm{\pairterm{y_1}{y_2}}}}}
        \end{array}
        ~\ctxleq~
        \begin{array}{l}
        \toterm{(\toterm{M}{x}{N_1})}{y_1}{\\\toterm{(\toterm{M}{x}{N_2})}{y_2}{\\\returnterm{\pairterm{y_1}{y_2}}}}
        \end{array}
        \qquad(f = \sem{M}, g_i = \sem{N_i}, h = \sem{N'})
      \]
  \end{enumerate}
\end{lem}
\begin{proof}
  \begin{enumerate}
    \item The following proves the statement for $i = 1$; the proof for $i = 2$ is
  similar.
  \[
    \begin{aligned}
      T \pi_1 \compose \vpairmorphism{f_1}{f_2}
      &~=~
      T \pi_1 \compose T(\id_{X_1} \times \terminalmorphism{X_2}) \compose \vpairmorphism{f_1}{f_2}
      \\&~=~
      T \pi_1 \compose \vpairmorphism{f_1}{T\terminalmorphism{X_2} \compose f_2}
      &\quad&\text{(naturality of extension)}
      \\&~\pleq~
      T \pi_1 \compose \vpairmorphism{f_1}{\eta_{1} \compose \terminalmorphism{W}}
      &&\text{(lax discardability of $f_2$)}
      \\&~=~
      T \pi_1 \compose \extendl[1]{(\eta_{X_1 \times 1})} \compose \pairmorphism{f_1}{\terminalmorphism{W}}
      &&\text{(left unit law)}
      \\&~=~
      f_1
      &&\text{(right unit law)}
    \end{aligned}
  \]

  \item
    Lax centrality of $f_1$ implies the first inequality of (2) as follows.
    \[
      \begin{aligned}
        &\extendl[TX_2]{(\extendr[X_1]{g})} \compose (f_1 \times f_2)
        \\~=~&
        \extend{g} \compose \seql_{X_1, X_2} \compose (f_1 \times \id_{TX_2}) \compose (\id_{W_1} \times f_2)
        &\quad&\text{(associativity and left unit laws)}
        \\~\pleq~&
        \extend{g} \compose \seqr_{X_1, X_2} \compose (f_1 \times \id_{TX_2}) \compose (\id_{W_1} \times f_2)
        &&\text{(lax centrality)}
        \\~=~&
        \extendr[TX_1]{(\extendl[X_2]{g})} \compose (f_1 \times f_2)
        &&\text{(associativity and left unit laws)}
      \end{aligned}
    \]
    For the other inequality, precomposing with the isomorphisms
    $\pairmorphism{\pi_2}{\pi_1}$ swaps the roles of $f_1$ and $f_2$,
    so that we can reuse the first inequality.

  \item The following proves the result.
  \[
    \begin{aligned}
      &\extend{(\extendr[X]{h} \compose \pairmorphism{\id_X}{g})} \compose f
      \\~=~&
      \extend{(\extendr[X]{h} \compose (\id_X \times g))} \compose T\pairmorphism{\id_X}{\id_X} \compose f
      &\quad&\text{(naturality of extension)}
      \\~\pleq~&
      \extend{(\extendr[X]{h} \compose (\id_X \times g))} \compose \seqr_{X, X} \compose \pairmorphism{f}{f}
      &&\text{(lax copyability)}
      \\~=~&
      \extendr[TX]{(\extendl[X]{(\extendr[X]{h} \compose (\id_X \times g))})} \compose \pairmorphism{f}{f}
      &&\text{(associativity, left unit)}
      \\~=~&
      \extendr[W]{(\extendl[TY]{(\extendr[X]{h})} \compose (f \times g))} \compose \pairmorphism{\id_W}{f}
      &&\text{(naturality of extension)}
      \\~\pleq~&
      \extendr[W]{(\extendr[TX]{(\extendl[Y]{h})} \compose (f \times g))} \compose \pairmorphism{\id_W}{f}
      &&\text{(\autoref{first-order-corollaries}(2))}
      \\~=~&
      \extendr[W]{(\extendr[W]{(\extendl[Y]{h} \compose (f \times \id_Y))} \compose (\id_W \times g))} \compose \pairmorphism{\id_W}{f}
      &&\text{(naturality of extension)}
      \\~=~&
      \extendr[W]{(\extendl[Y]{h} \compose (f \times \id_Y))} \compose (\id_W \times \extend{g}) \compose \pairmorphism{\id_W}{f}
      &&\text{(associativity law)}
      \\~=~&
      \extendr[W]{(\extendl[Y]{h} \compose (f \times \id_Y))} \compose \pairmorphism{\id_W}{\extend{g} \compose f}
    \end{aligned}
  \]

\item The required inequality is equivalent, by postcomposing with the isomorphism $T\pairmorphism{\pi_2}{\pi_1}$, to
  \[
    \extend{(\seqr_{Y_2, Y_1} \compose \pairmorphism{g_2}{g_1})} \compose f
    ~\pleq~
    \seqr_{Y_2, Y_1} \compose \pairmorphism{\extend{g_2} \compose f}{\extend{g_1} \compose f}
  \]
  We prove this as follows, using (3) with
  $h = \extendl[Y_1]{\eta_{Y_2 \times Y_1}} \compose (g_2 \times \id_{Y_1})$.
  \[
    \begin{aligned}
      &
      \extend{(\seqr_{Y_2, Y_1} \compose \pairmorphism{g_2}{g_1})} \compose f
      \\~=~&
      \extend{(\extendr[X]{(\extendl[Y_1]{\eta} \compose (g_2 \times \id_{Y_1}))} \compose \pairmorphism{\id_{X}}{g_1})} \compose f
      &~&\text{(naturality of extension)}
      \\~\pleq~&
      \extendr[W]{(
        \extendl[Y_1]{(
          \extendl[Y_1]{\eta} \compose (g_2 \times \id_{Y_1}))}
        \!\!\!\!\compose (f \times \id_{Y_1}))}
      \!\!\!\!\compose \pairmorphism{\id_W}{\extend{g_1} \compose f}
      &&\text{(\autoref{first-order-corollaries}(3))}
      \\~=~&
      \extendr[W]{(\extendl[Y_1]{\eta} \compose (\extend{g_2} \times \id_{Y_1}) \compose (f \times \id_{Y_1}))} \mspace{-8mu}\compose \pairmorphism{\id_W}{\extend{g_1} \compose f}
      &&\text{(associativity law)}
      \\~=~&
      \seqr_{Y_2, Y_1} \compose \pairmorphism{\extend{g_2} \compose f}{\extend{g_1} \compose f}
      &&\text{(naturality of extension)} \qedhere
    \end{aligned}
  \]
\end{enumerate}
\end{proof}

We are now ready to prove the main result of this section.
\begin{thm}\label{thm:fundamental}
  Let $\M = (\C, \T)$ be a CBPV model.
  If every morphism $f : X \to TY$ is lax discardable, lax copyable, and
  lax central, then for every expression $\welltyped{\Gamma}{e}{\tau}$
  we have
  \[
    \semvalue{e}\;\vnrel{}{}\;\semname{e}
  \]
\end{thm}
\begin{proof}
  Explicitly, we are required to show, for each expression $e$, that for
  all families of morphisms
  \[
    (f_i^v : W \to T(\semvalue{\tau_i}))_i
    \qquad
    (f_i^n : W \to \forget{\T}{(\semname{\tau_i})})_i
  \]
  we have
  \[
    f_1^v~\semrel{\tau_1}[W]~f_1^n
    ~~\wedge~~
    \cdots
    ~~\wedge~~
    f_k^v~\semrel{\tau_k}[W]~f_k^n
    \quad\Rightarrow\quad
    (\extend{\semvalue{e}} \compose \vtuplemorphism{f_i^v}_i)~
    ~\semrel{\tau'}[W]~
    ({\semname{e}} \compose \tuplemorphism{f_i^n}_i)
  \]
  We prove this by induction on $e$.
  \begin{itemize}
    \item For a variable $x_j$, we are required to prove
      \[
        (T\pi_j \compose \vtuplemorphism{f^v_i}_i)
        ~\semrel{\tau_j}[W]~
        f^n_j
      \]
      A simple induction on $j$, using
      \autoref{first-order-corollaries}(1), tells us that
      \[
        T\pi_j \compose \vtuplemorphism{f^v_i}_i
        ~\pleq~
        f^v_j
      \]
      The result then follows from the assumption
      $f^v_j~\semrel{\tau_j}[W]~f^n_j$ via
      \autoref{relation-closure}(1).
    \item For the expression $\unitterm$, we are required to show
      \[
        (T\terminalmorphism{\semvalue{\Gamma}} \compose \vtuplemorphism{f^v_i}_i)
        ~\semrel{\unittype}[W]~
        (\eta_1 \compose \terminalmorphism{W})
      \]
      By definition of $\semrel{\unittype}$, this is the same as
      \[
        (T\terminalmorphism{\semvalue{\Gamma}}) \compose \vtuplemorphism{f^v_i}_i)
        \pleq
        (\eta_1 \compose \terminalmorphism{W})
      \]
      which is immediate from lax discardability.
    \item For a pair $(e_1, e_2)$, the inductive hypothesis tells us that
      \[
        (\extend{\semvalue{e_j}} \compose \vtuplemorphism{f^v_i}_i)
        ~\semrel{\tau'_j}[W]~
        (\semname{e_j} \compose \tuplemorphism{f^n_i}_i)
      \]
      for each $j \in \{1, 2\}$.
      We also have
      \[
        \begin{aligned}
          &
          T\pi_j \compose \extend{\vpairmorphism{\semvalue{e_1}}{\semvalue{e_2}}} \compose \vtuplemorphism{f^v_i}_i
          \\~\pleq~&
          T\pi_j \compose \vpairmorphism{(\extend{\semvalue{e_1}} \compose \vtuplemorphism{f^v_i}_i)}{\,(\extend{\semvalue{e_2}} \compose \vtuplemorphism{f^v_i}_i)}
          &\quad&\text{(\autoref{first-order-corollaries}(4))}
          \\~\pleq~&
          \extend{\semvalue{e_j}} \compose \vtuplemorphism{f^v_i}_i
          &&\text{(\autoref{first-order-corollaries}(1))}
        \end{aligned}
      \]
      so that \autoref{relation-closure} implies
      \[
        (T\pi_j \compose \extend{\vpairmorphism{\semvalue{e_1}}{\semvalue{e_2}}} \compose \vtuplemorphism{f^v_i}_i)
        \quad\semrel{\tau'_j}[W]\quad
        (\pi_j \compose \pairmorphism{\semname{e_1}}{\semname{e_2}} \compose \tuplemorphism{f^n_i}_i)
      \]
      Hence
      \[
        (\extend{\vpairmorphism{\semvalue{e_1}}{\semvalue{e_2}}} \compose \vtuplemorphism{f^v_i}_i)
        ~~\semrel{\tau'_1 \times \tau'_2}[W]~~
        (\pairmorphism{\semname{e_1}}{\semname{e_2}} \compose \tuplemorphism{f^n_i}_i)
      \]
      as required.
    \item For $\fstterm e$, we need to show
      \[
        (T\pi_1 \compose \extend{\semvalue{e}} \compose \vtuplemorphism{f^v_i}_i)
        ~\semrel{\tau'_1}[W]~
        (\pi_1 \compose \semname{e} \compose \tuplemorphism{f^n_i}_i)
      \]
      which is immediate from the inductive hypothesis and the definition of
      $\semrel{\tau'_1 \times \tau'_2}$.
    \item The $\sndkeyword$ case is similar to the $\fstkeyword$ case.
    \item For the expression $\trueterm$, we need to show
      \[
        (T(\inl \compose \terminalmorphism{\semvalue{\Gamma}}) \compose \vtuplemorphism{f^v_i}_i)
        ~\semrel{\booltype}[W]~
        (\eta_2 \compose \inl \compose \terminalmorphism{W})
      \]
      which means
      \[
        (T(\inl \compose \terminalmorphism{\semvalue{\Gamma}}) \compose \vtuplemorphism{f^v_i}_i)
        ~\pleq~
        (\eta_2 \compose \inl \compose \terminalmorphism{W})
      \]
      This follows from lax discardability, and naturality of $\eta$.
    \item The $\falseterm$ case is similar to the $\trueterm$ case.
    \item For $\ifterm{e_0}{e_1}{e_2}$, the inductive hypothesis gives us
      \[
        g^v_0
        ~\pleq~
        g^n_0
        \qquad
        g^v_1
        ~\semrel{\tau'}[W]~
        g^n_1
        \qquad
        g^v_2
        ~\semrel{\tau'}[W]~
        g^n_2
      \]
      where we define
      \[
        g^v_j = \extend{\semvalue{e_j}} \compose \vtuplemorphism{f^v_i}_i
        \qquad
        g^n_j = \semname{e_j} \compose \tuplemorphism{f^n_i}_i
        \qquad
        \text{($j \in \{0, 1, 2\}$)}
      \]
      By applying all of the closure properties of \autoref{relation-closure}, we therefore have
      \[
        (\extendr[W]{(\copairmorphism{g^v_1}{g^v_2} \compose \dist)}
        \compose
        (\id_W \times g^v_0)
        \compose
        \pairmorphism{\id}{\id})
        ~
        \semrel{\tau'}[W]
        ~
        (\algextendr[W]{(\copairmorphism{g^n_1}{g^n_2}\compose \dist)}
        \compose
        (\id_W \times g^n_0)
        \compose
        \pairmorphism{\id}{\id})
      \]
      This is not quite what we need, but it does imply the result via another use of \autoref{relation-closure}(1), as follows.
      We have
      \[
        \extendl[2]{(\copairmorphism{\semvalue{e_1}}{\semvalue{e_2}} \compose \dist)}
        =
        \copairmorphism{\extend{\semvalue{e_1}}}{\extend{\semvalue{e_2}}} \compose \dist
      \]
      by precomposing with $\dist^{-1}$ and using the universal property
      of the coproduct $T(\semvalue{\Gamma}) + T(\semvalue{\Gamma})$.
      It follows that
      \[
        \begin{aligned}
          &
          \extend{\semvalue{\ifterm{e_0}{e_1}{e_2}}} \compose \vtuplemorphism{f^v_i}
          \\~=~&
          \extend{(\extendr[\semvalue{\Gamma}]{(\copairmorphism{\semvalue{e_1}}{\semvalue{e_2}} \compose \dist)} \compose \pairmorphism{\id}{\semvalue{e_0}})} \compose \vtuplemorphism{f^v_i}_i
          \\~\pleq~&
          \extendr[W]{(\extendl[2]{(\copairmorphism{\semvalue{e_1}}{\semvalue{e_2}} \compose \dist)} \compose (\vtuplemorphism{f^v_i}_i \times \id_2))}
          \compose
          \pairmorphism{\id_W}{g^v_0}
          &\quad&\text{(\autoref{first-order-corollaries}(3))}
          \\~=~&
          \extendr[W]{(\copairmorphism{\extend{\semvalue{e_1}}}{\extend{\semvalue{e_2}}} \compose \dist \compose (\vtuplemorphism{f^v_i}_i \times \id_2))}
          \compose
          \pairmorphism{\id_W}{g^v_0}
          \\~=~&
          \extendr[W]{(\copairmorphism{g^v_1}{g^v_2} \compose \dist)}
          \compose
          (\id_W \times g^v_0)
          \compose
          \pairmorphism{\id}{\id}
        \end{aligned}
      \]
      We also have
      \[
        \semname{\ifterm{e_0}{e_1}{e_2}} \compose \tuplemorphism{f^n_i}_i
        ~~=~~
        \algextendr[W]{(\copairmorphism{g^n_1}{g^n_2}\compose \dist)}
        \compose
        (\id_W \times g^n_0)
        \compose
        \pairmorphism{\id}{\id}
      \]
      so that \autoref{relation-closure}(1) implies
      \[
        (\extend{\semvalue{\ifterm{e_0}{e_1}{e_2}}} \compose \vtuplemorphism{f^v_i})
        ~\semrel{\tau'}[W]~
        (\semname{\ifterm{e_0}{e_1}{e_2}} \compose \tuplemorphism{f^n_i}_i)
      \]
      as required.
    \item For a $\lambda$-abstraction $\absterm{x}[\tau]{e}$ of type $\tau \to \tau'$, consider arbitrary $w : W' \to W$ and $g^v, g^n$ such that
      \[
        g^v~\semrel{\tau}[W']~g^n
      \]
      We need to show
      \[
        (\extend{\ev} \compose \vpairmorphism{(\extend{\semvalue{\absterm{x}[\tau]{e}}} \compose \vtuplemorphism{f^v_i}_i \compose w)}{g^v})
        ~\semrel{\tau'}[W']~
        (\ev \compose \pairmorphism{\semname{\absterm{x}[\tau]{e}} \compose \tuplemorphism{f^n_i}_i \compose w}{g^n})
      \]
      Applying \autoref{relation-closure}(3) to the assumptions yields
      \[
        (f^v_i \compose w)
        ~\semrel{\tau_i}[W']~
        (f^n_i \compose w)
      \]
      so that, by the inductive hypothesis, we have
      \[
        (\extend{\semvalue{e}}
        \compose
        \vpairmorphism{\vtuplemorphism{f^v_i \compose w}_i}{g^v})
        ~\semrel{\tau'}[W']~
        (\semname{e}
        \compose
        \pairmorphism{\tuplemorphism{f^n_i \compose w}_i}{g^n})
      \]
      We show that this is the required instance of
      $\semrel{\tau'}[W']$ by rewriting both sides as follows.
      For the left-hand side, we have
      \[
        \begin{aligned}
          &
          \extend{\ev} \compose \vpairmorphism{(\extend{\semvalue{\absterm{x}[\tau]{e}}} \compose \vtuplemorphism{f^v_i}_i \compose w)}{g^v}
          \\~=~&
          \extend{\ev} \compose \vpairmorphism{T{(\curry{(\semvalue{e})})} \compose \vtuplemorphism{f^v_i \compose w}_i}{g^v}
          \\~=~&
          \extend{(\ev \compose (\curry{(\semvalue{e})} \times \id_{\semvalue{\tau}}))} \compose \vpairmorphism{\vtuplemorphism{f^v_i \compose w}_i}{g^v}
          &&\text{(naturality of extension)}
          \\~=~&
          \extend{\semvalue{e}}
          \compose
          \vpairmorphism{\vtuplemorphism{f^v_i \compose w}_i}{g^v}
        \end{aligned}
      \]
      For the right-hand side:
      \[
        \begin{aligned}
          \ev \compose \pairmorphism{\semname{\absterm{x}[\tau]{e}} \compose \tuplemorphism{f^n_i}_i \compose w}{g^n}
          ~=~&
          \ev \compose \pairmorphism{\curry{(\semname{e})} \compose \tuplemorphism{f^n_i \compose w}_i}{g^n}
          \\~=~&
          \semname{e} \compose \pairmorphism{\tuplemorphism{f^n_i \compose w}_i}{g^n}
        \end{aligned}
      \]
    \item For an application $e\,e'$, where $e$ has type $\tau \to \tau'$, the inductive hypothesis for $e'$ gives us
      \[
        (\extend{\semvalue{e'}} \compose \vtuplemorphism{f^v_i}_i)
        ~\semrel{\tau}[W]~
        (\semname{e'} \compose \tuplemorphism{f^n_i}_i)
      \]
      so that, by the inductive hypothesis for $e$ with $w = \id_W$, we have
      \[
        (\extend{\ev} \compose \vpairmorphism{(\extend{\semvalue{e}} \compose \vtuplemorphism{f^v_i}_i)}{(\extend{\semvalue{e'}} \compose \vtuplemorphism{f^v_i}_i)})
        ~\semrel{\tau'}[W]~
        (\ev \compose \pairmorphism{(\semname{e} \compose \tuplemorphism{f^n_i}_i)}{(\semname{e'} \compose \tuplemorphism{f^n_i}_i)})
      \]
      We rewrite both sides as follows.
      For the left we have
      \[
        \begin{aligned}
          &
          \extend{\semvalue{e\,e'}} \compose \vtuplemorphism{f^v_i}_i
          \\~=~&
          \extendl[T(\semvalue{\tau})]{(\extendr[\semvalue{\tau \to \tau'}]{\ev})} \compose \pairmorphism{\semvalue{e}}{\semvalue{e'}} \compose \vtuplemorphism{f^v_i}
          &\quad&\text{(naturality of extension)}
          \\~=~&
          \extend{\ev} \compose \extend{\vpairmorphism{\semvalue{e}}{\semvalue{e'}}} \compose \vtuplemorphism{f^v_i}
          &&\text{(left unit, associativity)}
          \\~\pleq~&
          \extend{\ev} \compose \vpairmorphism{(\extend{\semvalue{e}} \compose \vtuplemorphism{f^v_i}_i)}{(\extend{\semvalue{e'}} \compose \vtuplemorphism{f^v_i}_i)}
          &&\text{(\autoref{first-order-corollaries}(4))}
        \end{aligned}
      \]
      and for the right,
      \[
        \semname{e\,e'} \compose \tuplemorphism{f^n_i}_i
        ~=~
        \ev \compose \pairmorphism{(\semname{e} \compose \tuplemorphism{f^n_i}_i)}{(\semname{e'} \compose \tuplemorphism{f^n_i}_i)}
      \]
      We therefore have
      \[
        (\extend{\semvalue{e\,e'}} \compose \vtuplemorphism{f^v_i}_i)
        ~\semrel{\tau'}[W]~
        (\semname{e\,e'} \compose \tuplemorphism{f^n_i}_i)
      \]
      as required.
      \qedhere
  \end{itemize}
\end{proof}

As a corollary, we can directly compare the call-by-value and
call-by-name translations of source-language \emph{programs} (closed
expressions of type $\booltype$).

\begin{cor}\label{corollary:cbv-cbn-programs}
  Let $\M = (\C, \T)$ be a CBPV model that is adequate with respect to a program relation $\syntaxleq$.
  If every morphism $f : X \to TY$ is lax discardable, lax copyable, and lax central,
  then for every closed expression $e : \booltype$,
  we have
  \[
    \translatevalue{e} \syntaxleq \translatename{e}
  \]
\end{cor}
\begin{proof}
  By \autoref{thm:fundamental} we have
  $\semvalue{e} \vnrel{}{} \semname{e}$, so in
  particular
  $
    \semvalue{e} ~\semrel{\booltype}[1]~ \semname{e}
  $
  By definition, the latter means
  $
    \semvalue{e} \pleq \semname{e}
  $,
  which implies the result by adequacy.
\end{proof}

The conclusion of this corollary, namely
$\translatevalue{e} \syntaxleq \translatename{e}$ is independent of the
choice of model $\M$.
In contrast, the conclusion of \autoref{thm:fundamental} is not
independent of $\M$.
\autoref{thm:fundamental} should therefore be viewed as a result about
the denotations $\semvalue{e}$ and $\semname{e}$, rather than about the
translations $\translatevalue{e}$ and $\translatename{e}$.
We rectify this in \autoref{section:main-theorem} below, where the conclusion of our main
result \autoref{theorem:cbv-cbn} relates $\translatevalue{e}$ with
$\translatename{e}$ via the contextual preorder, which is
independent of $\M$.

\subsection{Examples}
To conclude this section, we discuss the consequences of the results
above for each of our examples.

We first note that we can in fact simplify the definition of
$\vnrel{}{}$ for each of these examples, by using the fact that, in each
of the three $\Poset$-categories $\Set$, $\Poset$, $\wCpo$, morphisms
are in particular functions, and are ordered pointwise.
(We treat a set as a discrete poset here.)
It follows that instead of considering generalized elements
$f : W \to X$, it is enough to consider ordinary elements $t \in X$ (which we can
identify with morphisms $t : 1 \to X$).
The simplification of $\semrel{{-}}$ we obtain is defined as follows.
\begin{align*}
  \intertext{
    $\boxed{
      t^v~\semrels{\tau}~t^n
      \text{~(where~}
      t^v \in T(\semvalue{\tau})
      \text{~and~}
      t^n \in \forget{\T}({\semname{\tau}})
      \text{)}
    }$
  }
  t^v~\semrels{\unittype}~t^n
  &\quad\text{iff}\quad
  t^v \pleq t^n \\
  t^v~\semrels{\tau_1 \times \tau_2}~t^n
  &\quad\text{iff}\quad
  T\pi_1 t^v~\semrels{\tau_1}~\pi_1 t^n
  ~~\wedge~~
  T\pi_2 t^v~\semrels{\tau_2}~\pi_2 t^n
  \\
  t^v~\semrels{\booltype}~t^n
  &\quad\text{iff}\quad
  t^v \pleq t^n
  \\
  t^v~\semrels{\tau \to \tau'}~t^n
  &\quad\text{iff}\quad
  \forall a^v, a^n.~~
    a^v \semrels{\tau} a^n
    ~\Rightarrow~
    \extend{\ev} (\seql(t^v, a^v))
    ~\semrels{\tau'}~t^na^n
\end{align*}
The precise relationship between $\semrel{{-}}$ and $\semrels{{-}}$ is
as follows.
\begin{lem}\label{lemma:semrels-works}
  For each of our four example models, we have
  \[
    f^v~\semrel{\tau}[W]~f^n
    \quad\Leftrightarrow\quad
    \forall w \in W.\, f^v w~\semrels{\tau}~f^n w
  \]
  for every $f^v : W \to T(\semvalue{\tau})$ and
  $f^n : W \to \forget{\T}{(\semname{\tau})}$.
  Hence, for each
  $g^v : \semvalue{\Gamma} \to T(\semvalue{\tau})$
  and
  $g^n : \semname{\Gamma} \to \forget{\T}{(\semname{\tau})}$,
  we have
  \[
    g^v \vnrel{}{} g^n
  \]
  exactly when, for all $(a^v_i \in T(\semvalue{\tau_i}))_i$ and
  $(a^n_i \in \forget\T(\semname{\tau_i}))_i$,
  \[
    a^v_1\,\semrels{\tau_1}\,a^n_1
    \;\wedge\;
    \cdots
    \;\wedge\;
    a^v_k\,\semrels{\tau_k}\,a^n_k
    \quad\Rightarrow\quad
    (\extend{g^v}\vtuplemorphism{a^v_i}_i)
    \;\semrels{\tau'}\;
    (g^n(a^n_i)_i)
  \]
  where $\Gamma = x_1 : \tau_1, \dots, x_k : \tau_k$.
\end{lem}
\begin{proof}
  By induction on $\tau$.
  This is trivial for $\unittype$, $\booltype$ and product types.
  For a function type $\tau \to \tau'$, the $(\Leftarrow)$ direction is
  again trivial.
  The $(\Rightarrow)$ direction follows from the fact that, identifying an
  element $w \in W$ with a morphism $w : 1 \to W$, we have
  $\semrels{\tau} = \semrel{\tau}[1]$ and
  $\semrels{\tau'} = \semrel{\tau'}[1]$ by the inductive hypothesis.
\end{proof}

We now consider each of our examples in turn.
Note that, since the proof of \autoref{thm:fundamental} is by induction on
the expression $e$, we need to extend the proof with cases for
the extra syntax we add in these examples.

\begin{cor}
  For our no effects example, we have
  \[
    \semvalue{e} \vnrel{}{} \semname{e}
  \]
  for every $\welltyped{\Gamma}{e}{\tau}$, and in particular,
  \[
    \exists V\!:\!\booltype.~
    (\translatevalue{e'} \reducesto \returnterm{V})
    ~\wedge~
    (\translatename{e'} \reducesto \returnterm{V})
  \]
  for every closed expression $e' : \booltype$.
\end{cor}
\begin{proof}
  This is immediate from \autoref{thm:fundamental} and
  \autoref{corollary:cbv-cbn-programs}, with
  $(\syntaxleq) = (\syntaxleq_{\mathrm{pure}})$.
\end{proof}

\begin{cor}
  For our divergence example, we have
  \[
    \semvalue{e} \vnrel{}{} \semname{e}
  \]
  for every $\welltyped{\Gamma}{e}{\tau}$, and in particular,
  \[
    \forall V\!:\!\booltype.~
    (\translatevalue{e'} \reducesto \returnterm{V})
    ~\Rightarrow~
    (\translatename{e'} \reducesto \returnterm{V})
  \]
  for every closed expression $e' : \booltype$.
\end{cor}
\begin{proof}
  We first extend the inductive proof of
  \autoref{thm:fundamental} with a case for recursive functions
  $\valuerecterm{f}[\tau][\tau']{x}{e}$.
  In light of \autoref{lemma:semrels-works} above, it suffices for this
  to show that, if
  \[
    h^v : \semvalue{\tau \to \tau'} \times \semvalue{\tau} \to
    T(\semvalue{\tau'})
    \qquad
    h^n : \forget{\T}{(\semname{\tau \to \tau'})} \times
    \forget{\T}{(\semname{\tau})} \to \forget{\T}{(\semname{\tau'})}
  \]
  are $\omega$-continuous functions that satisfy
  \[
    t^v~\semrels{\tau \to \tau'}~t^n
    ~\wedge~
    a^v~\semrels{\tau}~a^n
    ~\Rightarrow~
    h^v(t^v,a^v)~\semrels{\tau'}~h^n(t^n,a^n)
  \]
  when $t^v, a^v \neq \bot$, then we have
  \[
    \fix{(\lambda g^v.\, h^v(g^v, {-}))}
    ~\semrels{\tau \to \tau'}~
    \fix{(\lambda g^n.\, h^n(g^n, {-}))}
  \]
  This follows from the fact that each $\semrels{\tau''}$ relates $\bot$
  to $\bot$ and is closed under least upper bounds of $\omega$-chains,
  which can be proved by a simple induction on $\tau''$.

  The second part follows from \autoref{corollary:cbv-cbn-programs},
  by adequacy of the model with respect to the program relation
  defined in \autoref{ex:recursion}.
\end{proof}

\begin{cor}
  For our nondeterminism example, we have
  \[
    \semvalue{e} \vnrel{}{} \semname{e}
  \]
  for every $\welltyped{\Gamma}{e}{\tau}$, and in particular,
  \[
    \forall V\!:\!\booltype.~
    (\translatevalue{e'} \reducesto \returnterm{V})
    ~\Rightarrow~
    (\translatename{e'} \reducesto \returnterm{V})
  \]
  for every closed expression $e' : \booltype$.
\end{cor}
\begin{proof}
  We first extend the inductive proof of
  \autoref{thm:fundamental} with two extra cases: one for
  $\failconstant$ and one for $\orconstant$.
  Following \autoref{lemma:semrels-works} above, to prove both of these
  cases, it suffices to show that $\semrels{\tau}$ is closed under
  finite joins for each $\tau'$, i.e.\ that
  \[
    \bot\,\semrels{\tau'}\,\bot
    \qquad
    t_1^v\,\semrels{\tau'}\,t_1^n
    \;\wedge\;
    t_2^v\,\semrels{\tau'}\,t_2^n
    \;\Rightarrow\;
    (t_1^v \join t_2^v)\,\semrels{\tau'}\,(t_1^n \join t_2^n)
  \]
  This is a simple induction on $\tau'$.
  The result follows from \autoref{thm:fundamental} and
  \autoref{corollary:cbv-cbn-programs}, by adequacy of the model with
  respect to the program relation defined in
  \autoref{ex:nondeterminism}.
\end{proof}

\begin{cor}
  For our immutable state example, we have
  \[
    \semvalue{e} \vnrel{}{} \semname{e}
  \]
  for every $\welltyped{\Gamma}{e}{\tau}$, and in particular,
   \[
    \forall b \in \{\truevalue,\falsevalue\}.\,\exists
    V\!:\!\booltype.~(\translatevalue{e'} \reducesto_b \returnterm{V})
       \wedge (\translatename{e'} \reducesto_b \returnterm{V})
  \]
  for every closed expression $e' : \booltype$.
\end{cor}
\begin{proof}
  Once again, we need to add the extra case to
  \autoref{thm:fundamental}.
  It is enough to show that
  $
    (\getconstant \compose \terminalmorphism{\semvalue{\Gamma}}
    \compose \vtuplemorphism{f^v_i}_i)
    ~\semrels{\booltype}~
    (\getconstant \compose \terminalmorphism{W})
  $,
  which follows from lax discardability.
  We can then apply
  \autoref{thm:fundamental} and
  \autoref{corollary:cbv-cbn-programs} to obtain the result, using
  adequacy with respect to the program relation defined in
  \autoref{ex:immutable}.
\end{proof}

\section{A Galois connection between call-by-value and call-by-name}
\label{section:galois-connection}
We improve on the results of the previous section by showing how to
directly relate the call-by-value semantics $\semvalue{e}$ of an
expression to a morphism derived from the call-by-name semantics
$\semname{e}$.
Under a further condition on the model (which again restricts the
allowable effects), we prove a statement
(\autoref{thm:semantic-reasoning}) of the form
\[
  \semvalue{e}
  ~\pleq~
  \tovalue{\tau} \compose \semname{e} \compose \tonamehat{\Gamma}
\]
involving morphisms $\toname{\tau}$ and $\tovalue{\tau}$ for mapping
between the call-by-value and call-by-name semantics:
\[
  \begin{tikzcd}[column sep=large]
    T (\semvalue{\tau})
    \arrow[r, "\toname{\tau}", yshift=1ex] &
    \forget \T (\semname{\tau})
    \arrow[l, "\tovalue{\tau}", yshift=-1ex]
  \end{tikzcd}
\]
(In the inequality above,
$\tonamehat{\Gamma} : \semvalue{\Gamma} \to \semname{\Gamma}$
is constructed by extending the morphisms
$
  (\toname{\tau} \compose \eta)
    : \semvalue{\tau} \to \forget{\T}{(\semname{\tau})}
$
from types to contexts.)

We do not want just \emph{any} maps between call-by-value
and call-by-name.
We show (\autoref{cor:galois-connections}) that the maps we define
(precisely, the monotone functions $\toname{\tau} \compose {({-})}$ and
$\tovalue{\tau} \compose {({-})}$) form
\emph{Galois connections}~\cite{melton-schmidt-strecker:galois-connections}.
This is the crucial property that enables us to prove the inequality
above.\footnotemark{}
\footnotetext{%
  The proof of \autoref{thm:semantic-reasoning} that we give here does
  not directly use the fact that the maps are Galois connections;
  instead, it uses \autoref{thm:fundamental}.
  This is simply to avoid another induction on expressions.
  In the conference version of this paper, the corresponding fact
  \cite[Lemma~20]{mcdermott2022galois} was proved directly using the
  fact that the maps are Galois connections.}
\begin{defi}
  A \emph{Galois connection} consists of two posets $\poset X$,
  $\poset Y$ and two monotone functions
  $
    \toname{} : \poset{X} \to \poset{Y}
  $,
  $
    \tovalue{} : \poset{Y} \to \poset{X}
  $,
  such that $x \pleq \tovalue{} (\toname{}\,x)$ for all $x \in X$ and
  $\toname{} (\tovalue{}\,y) \pleq y$ for all $y \in Y$.
\end{defi}

The results of the previous section are helpful here.
We have relations $\semrel{\tau}$ that in some sense capture the
relationship between call-by-value and call-by-name.
This suggests we should look for morphisms
$\toname{\tau}$ and $\tovalue{\tau}$ that represent the relations
$\semrel{\tau}$, i.e.\ that satisfy the following equivalences.
\[
  \toname{\tau} \compose f^v~\pleq~f^n
  \quad\Leftrightarrow\quad
  f^v~\semrel{\tau}[W]~f^n
  \quad\Leftrightarrow\quad
  f^v~\pleq~\tovalue{\tau} \compose f^n
\]
These equivalences uniquely determine $\tovalue{\tau}$ and
$\toname{\tau}$, and guarantee that we have Galois connections
$(\tovalue{\tau} \compose {({-})}, \toname{\tau} \compose {({-})})$.
Furthermore, these equivalences enable us to prove
\[
  \semvalue{e}
  ~\pleq~
  \tovalue{\tau} \compose \semname{e} \compose \tonamehat{\Gamma}
\]
as a corollary of
$\semvalue{e} \vnrel{}{} \semname{e}$.
That is, the main result of this section
(\autoref{thm:semantic-reasoning}) is a corollary of the main result of
the previous section (\autoref{thm:fundamental}).

Given a CBPV model $\M = (\C, \T)$, the morphisms $\toname{\tau}$ and
$\tovalue{\tau}$ are defined in
\autoref{figure:semantic-maps}.\footnotemark{}%
\footnotetext{%
  We present the definitions in a different way to
  \cite{mcdermott2022galois}, but the morphisms are in fact the same.
  The syntactic maps in \autoref{figure:maps} below are similarly
  presented differently to \cite{mcdermott2022galois}.}
As for the relations $\semrel{\tau}$, the definition is by induction on
$\tau$.
This is mutual induction: at contravariant positions, the definition of
$\toname{}$ uses $\tovalue{}$, and vice-versa.

\begin{figure*}
  \begin{align*}
    \shortintertext{$\boxed{\toname{\tau}
      : T (\semvalue{\tau}) \to \forget \T (\semname{\tau})}$}
    \toname{\unittype}
    ~&=~
    \id_1
     ~:~T1 \to T 1 \\
    \toname{\tau_1 \times \tau_2}
    ~&=~
    \pairmorphism
      {\toname{\tau_1} \compose T\proj{1}}
      {\toname{\tau_2} \compose T\proj{2}}
     ~:~T(\semvalue{\tau_1} \times \semvalue{\tau_2})
        \to \forget \T (\semname{\tau_1}) \times \forget \T (\semname{\tau_2}) \\
    \toname{\booltype}
    ~&=~
    \id_2
     ~:~T2 \to T 2 \\
    \toname{\tau \to \tau'}
    ~&=~
    \curry{(\toname{\tau'} \compose \extend{\ev} \compose \vpairmorphism{\pi_1}{\tovalue{\tau} \compose \pi_2})}
     ~:~T(\exp{\semvalue{\tau}}{T(\semvalue{\tau'})})
       \to \exp{\forget \T (\semname{\tau})}{\forget \T (\semname{\tau'})}
    \\[2ex]
    \shortintertext{$\boxed{\tovalue{\tau}
      : \forget\T (\semname{\tau})
        \to T (\semvalue{\tau})}$}
    \tovalue{\unittype}
    ~&=~
    \id_{T1}
     ~:~T 1 \to T 1 \\
    \tovalue{\tau_1 \times \tau_2}
    ~&=~
    \vpairmorphism
      {\tovalue{\tau_1} \compose \pi_1}
      {\tovalue{\tau_2} \compose \pi_2}
     ~:~\forget \T (\semname{\tau_1}) \times \forget \T (\semname{\tau_2})
       \to T(\semvalue{\tau_1} \times \semvalue{\tau_2})\\
    \tovalue{\booltype}
    ~&=~
    \id_{T2}
     ~:~T 2 \to T 2 \\
    \tovalue{\tau \to \tau'}
    ~&=~
    \eta_{\semvalue{\tau \to \tau'}} \compose
    (\curry{(\tovalue{\tau'} \compose \ev \compose (\id \times
    (\toname{\tau} \compose \eta)))})
    \\
    &{\qquad\qquad\qquad\qquad}
    ~:~\exp{\forget\T (\semname{\tau})}{\forget\T (\semname{\tau'})}
      \to T(\exp{\semvalue{\tau}}{T(\semvalue{\tau'})})
  \end{align*}

  \caption{%
    Semantic morphisms $\toname{}$ from call-by-value to call-by-name
    and $\tovalue{}$ from call-by-name to call-by-value}
  \label{figure:semantic-maps}
\end{figure*}

Of course we do not expect to be able to prove the properties outlined
above for a general model $\M$.
To see what conditions we should require $\M$ to satisfy, suppose that
we do have Galois connections
$(\toname{\tau} \compose ({-}), \tovalue{\tau} \compose ({-}))$,
equivalently, that we have
\[
  \toname{\tau} \compose \tovalue{\tau} \pleq \id_{\forget{\T}{(\semname{\tau})}}
  \qquad
  \id_{T(\semvalue{\tau})} \pleq \tovalue{\tau} \compose \toname{\tau}
\]
Now consider what happens when we convert a lazy pair
$N = \comppairterm{N_1}{N_2}$ of type
$\translatename{\unittype\times\unittype} =
  \freetype{\unittype} \compprod \freetype{\unittype}$
into call-by-value, and then back into call-by-name:
\[
  \sem{\comppairterm*
    {\toterm{N_1}{z_1}{\toterm{N_2}{z_2}{\returnterm{z_1}}}}
    {\toterm{N_1}{z_1}{\toterm{N_2}{z_2}{\returnterm{z_2}}}}}
  ~=~
  (\toname{\unittype\times\unittype}
  \compose \tovalue{\unittype\times\unittype}
  \compose \sem{N})
  ~\pleq~
  \sem{N}
\]
The $i$th projection of the left-hand side evaluates both $N_1$ and
$N_2$, but the $i$th projection of the right is just $\sem{N_i}$.
Thus moving from left to right discards effects.
Similarly, converting a strict pair $M$ of type
$\freetype{(\translatevalue{\unittype\times\unittype})} =
  \freetype{(\unittype\times\unittype)}$
to call-by-name and back duplicates the effects of $M$:
\[
  \sem{M}
  ~\pleq~
  (\tovalue{\unittype\times\unittype}
  \compose \toname{\unittype\times\unittype}
  \compose \sem{M})
  ~=~
  \sem{\begin{aligned}
    &\toterm{M}{\phantomprimel{z}}{
     \matchpairterm{\phantomprimel{z}}{z_1}{z_2}{\\
    &\toterm{M}{z'}{
     \matchpairterm{z'}{z'_1}{z'_2}{\\
    &\returnterm{\pairterm{z_1}{z'_2}}}}}}
  \end{aligned}}
\]
These suggest that lax discardability and lax copyability will be
useful, and indeed we use both of these properties in the proof of
\autoref{thm:represent-relation} below.

For function types we need even more.
Consider what happens when we convert a CBPV computation
$
  M : \freetype{(\translatevalue{\unittype \to \unittype})}
    = \freetype{(\thunktype{(\unittype \to \freetype \unittype)})}
$
to call-by-name and then back to call-by-value.
By doing this we obtain the denotation of a computation that immediately
returns:
\[
   \sem{M}
   ~\pleq~
   (\tovalue{\unittype\to\unittype} \compose
   \toname{\unittype\to\unittype} \compose \sem{M})
   ~=~
   \sem{\returnterm{\thunkterm{\absterm{x}[\unittype]
     {\toterm{M}{z}{\pushterm{x}{\forceterm{z}}}}}}}
\]
The round-trip from call-by-value to call-by-name and back thunks the
computational effects of $M$, suspending them until the function is applied.
The property we ask for the model to satisfy in order to make this a
valid inequality is \emph{lax thunkability} of morphisms.
\begin{defi}\label{def:semantic-properties}
  Let $\T$ be a strong $\Poset$-monad on a cartesian $\Poset$-category
  $\C$.
  A morphism $f : X \to TY$ is \emph{lax thunkable} if
  $T\eta_Y \compose f \pleq \eta_{TY} \compose f$.
  If every such $f$ is lax thunkable (equivalently, if
  $T\eta_Y \pleq \eta_{TY}$ for every $Y \in \ob{\C}$), then we say that
  $\T$ is \emph{lax idempotent}.%
  \footnote{%
    Lax idempotent $\Poset$-monads are a special case of lax idempotent
    2-monads, which are well-known, and are often called
    \emph{Kock-Z{\"o}berlein} monads~\cite{kock:kock-zoberlein}.}
\end{defi}
Again this a lax version of a property defined by
F{\"u}hrmann~\cite{fuhrmann:direct-models}.
For the corresponding property in the syntax of CBPV we have the following.
\begin{lem}\label{lemma:syntactic-properties}
  Let $\comptyped{\Gamma}{M}{\freetype{A}}$ be a computation.
  For every adequate CBPV model, if $\sem{M}$ is lax thunkable, then
  \[
    \toterm{M}{x}{\returnterm{\thunkterm{\returnterm{x}}}}
    ~\ctxleq[\Gamma]~
    \returnterm{\thunkterm{M}}
  \]
\end{lem}
\begin{proof}
  We have
  \begin{align*}
    \sem{\toterm{M}{x}{\returnterm{\thunkterm{\returnterm{x}}}}}
    &~=~
    T\eta_{\sem{A}} \compose \sem{M}
    \\&~\pleq~
    \eta_{T\sem{A}} \compose \sem{M}
    \\&~=~
    \sem{\returnterm{\thunkterm{M}}}
  \end{align*}
  which implies the result by adequacy.
\end{proof}

\begin{exa}
  For three of our examples the strong $\Poset$-monad $\T$ is lax
  idempotent.
  For no effects, we use the identity monad, which is trivially lax
  idempotent because $T\eta_Y = \id_Y = \eta_{TY}$.
  For divergence, the monad (\autoref{ex:divergence-monad}) is lax
  idempotent because
  the left hand side of $T\eta_Y t \pleq \eta_{TY} t$ is $\bot$ when
  $t = \bot$ (intuitively, we can thunk diverging computations), and
  otherwise the two sides are equal.
  For nondeterminism the monad (\autoref{ex:divergence-monad}) is lax
  idempotent because,
  since $\downclose \{y\} \subseteq S$ for every $y \in S \in TY$, we have
  \[
    T\eta_Y S
    = \downclose\{\downclose \{y\} \mid y \in S\}
    \subseteq \downclose\{S\}
    = \eta_{TY} S
  \]
  (intuitively, we can postpone nondeterministic choices).

  On the other hand, the reader monad we use for immutable state
  (\autoref{ex:immutable-monad}) is \emph{not} lax idempotent.
  Indeed, a morphism $f : X \to \exp{2}{Y}$ is lax thunkable exactly
  when it satisfies $f\,x\,\truevalue = f\,x\,\falsevalue$ for all
  $x \in X$.
  In particular, $\sem{\getconstant}$ is not lax thunkable.
  As a consequence, we cannot apply the results of this section to this
  model.
  (In fact, \autoref{prop:thunkable-necessary} below implies that
  the conclusion of \autoref{thm:semantic-reasoning} is false in this
  case.)
  This does not mean that our reasoning principle
  (\autoref{theorem:cbv-cbn}) does not apply to immutable state, only
  that this model is not good enough to instantiate it.
  Indeed, it is known that this model of immutable state fails to be
  fully abstract, i.e.\ that it distinguishes between computations that
  are contextually equivalent \cite{kammar2022fully}.
  A different model, such as the identity monad on $\Set \times \Set$,
  which is lax idempotent, may enable us to apply
  \autoref{theorem:cbv-cbn}.
\end{exa}

Lax thunkability is difficult to use directly in proofs, so we establish
the following characterizations of lax thunkable morphisms.
\begin{lem}\label{lemma:thunkable-reasoning}
  Let $\T$ be a strong $\Poset$-monad and $f : X \to TY$ be a morphism.
  The following are equivalent:
  \begin{enumerate}
    \item $f$ is lax thunkable;
    \item the implication
      \[
        g_1~\pleq~\algextendr[W]{(g_2 \compose (\id_W \times \eta_Y))}
        \quad\Rightarrow\quad
        g_1 \compose (\id_W \times f)~\pleq~g_2 \compose (\id_W \times f)
      \]
      holds for all $\T$-algebras $\Z$ and morphisms
      $g_1, g_2 : W \times TY \to Z$;
    \item the implication
      \[
        g_1~\pleq~\algextendl[W]{(g_2 \compose (\eta_Y \times \id_W))}
        \quad\Rightarrow\quad
        g_1 \compose (f \times \id_W)~\pleq~g_2 \compose (f \times \id_W)
      \]
      holds for all $\T$-algebras $\Z$ and morphisms
      $g_1, g_2 : TY \times W \to Z$;
    \item the implication
      \[
        g_1~\pleq~\algextend{(g_2 \compose \eta_Y)}
        \quad\Rightarrow\quad
        g_1 \compose f~\pleq~g_2 \compose f
      \]
      holds for all $\T$-algebras $\Z$ and morphisms
      $g_1, g_2 : TY \to Z$.
  \end{enumerate}
\end{lem}
\begin{proof}
  (1) $\Rightarrow$ (2): If
  $g_1 \pleq \algextendr[W]{(g_2 \compose (\id_W \times \eta_Y))}$
  then
  \[
    \begin{aligned}
      g_1 \compose (\id_W \times f)
      &~\pleq~
      \algextendr[W]{(g_2 \compose (\id_W \times \eta_Y))} \compose (\id_W \times f)
      &\quad&\text{(assumption)}
      \\&~=~
      \algextendr[W]{g_2} \compose (\id_W \times T\eta_Y) \compose (\id_W \times f)
      &&\text{(naturality of extension)}
      \\&~\pleq~
      \algextendr[W]{g_2} \compose (\id_W \times \eta_{TY}) \compose (\id_W \times f)
      &&\text{(lax thunkability of $f$)}
      \\&~=~
      g_2 \compose (\id_W \times f)
      &&\text{(left unit law)}
    \end{aligned}
  \]

  (2) $\Rightarrow$ (4): Specializing (2) to $W = 1$ yields (4).

  (4) $\Rightarrow$ (1): Consider the $\T$-algebra
  $\Z = \free{\T}{(TY)}$ and morphisms $g_1 = T\eta_Y$ and
  $g_2 = \eta_{TY}$.
  We have
  $
    g_1 = T \eta_Y = \extend{(\eta_{TY} \compose \eta_{Y})}
    = \extend{(g_2 \compose \eta_Y)}
  $
  by the definition of $T$ on morphisms,
  so (3) gives us the required inequality
  $g_1 \compose f \pleq g_2 \compose f$.

  (1) $\Rightarrow$ (3): Similar to the proof that (1) implies (2).

  (3) $\Rightarrow$ (4): Similar to the proof that (2) implies (4).
\end{proof}

Lax thunkability is a strong property.
In particular, it implies all of the properties we assumed in the
previous section.
(The non-lax version of this fact is noted by F{\"u}hrmann in
\cite{fuhrmann:direct-models}.)

\begin{lem}\label{lemma:thunkable-consequences}
  Let $\T$ be a strong $\Poset$-monad.
  If $f : X \to TY$ is lax thunkable, then $f$ is also lax discardable,
  lax copyable, and lax central.
\end{lem}
\begin{proof}
  Lax discardability: By the definition of $T$ on morphisms, we have
  \[
    T\terminalmorphism{Y}
    ~=~\extend{(\eta_1 \compose \terminalmorphism{Y})}
    ~=~\extend{((\eta_1 \compose \terminalmorphism{TY}) \compose \eta_Y)}
  \]
  so \autoref{lemma:thunkable-reasoning}(4) implies
  \[
    T\terminalmorphism{Y} \compose f
    ~\pleq~(\eta_1 \compose \terminalmorphism{TY}) \compose f
    ~=~\eta_1 \compose \terminalmorphism{X}
  \]

  Lax copyability: By the definition of $T$ on morphisms, and the left unit
  law for $\T$, we have
  \[
    T\pairmorphism{\id_Y}{\id_Y}
    = \extend{(\eta_{Y \times Y} \compose \pairmorphism{\id_Y}{\id_Y})}
    = \extend{(\seql_{Y, Y} \compose \pairmorphism{\eta_Y}{\eta_Y})}
    = \extend{((\seql_{Y, Y} \compose \pairmorphism{\id_Y}{\id_Y}) \compose \eta_Y)}
  \]
  so \autoref{lemma:thunkable-reasoning}(4) implies
  \[
    T\pairmorphism{\id_Y}{\id_Y} \compose f
    ~\pleq~
    (\seql_{Y, Y} \compose \pairmorphism{\id_Y}{\id_Y}) \compose f
    ~=~
    \seql_{Y, Y} \compose \pairmorphism{f}{f}
  \]

  Lax centrality: We have
  \begin{align*}
    \seql_{Y, W}
    &~=~
    \extendl[TW]{(\extendr[Y]{\eta_{Y \times W}})}
    &\quad&\text{(definition of $\seql$)}
    \\&~=~
    \extendl[TW]{(\extendr[Y]{(\extendl[W]{\eta_{Y \times W}} \compose (\eta_Y \times \id_W))})}
    &\quad&\text{(left unit law)}
    \\&~=~
    \extendl[TW]{(\extendr[TY]{(\extendl[W]{\eta_{Y \times W}})} \compose (\eta_Y \times \id_{TW}))}
    &\quad&\text{(naturality of extension)}
    \\&~=~
    \extendl[TW]{(\seqr_{Y, W} \compose (\eta_Y \times \id_{TW}))}
    &\quad&\text{(definition of $\seqr$)}
  \end{align*}
  so \autoref{lemma:thunkable-reasoning}(3) implies
  \[
    \seql_{Y, W} \compose (f \times \id_{TW})
    ~\pleq~
    \seqr_{Y, W} \compose (f \times \id_{TW})
    \qedhere
  \]
\end{proof}

Our aim is now to establish the relationship between $\semrel{\tau}$ and
$(\toname{\tau}, \tovalue{\tau})$ outlined at the beginning of this
section.
For $\tau = \unittype$ and $\tau = \booltype$ this turns out to be easy.
For product types, we use lax discardablity and lax
copyability, while for function types, we use lax thunkability.
For the latter two cases the following lemma is useful.

\begin{lem}\label{lem:product-and-function-properties}
  Let $\T$ be a strong $\Poset$-monad.
  \begin{enumerate}
    \item If $f : W \to T(X_1 \times X_2)$ is lax copyable, and both
      $g_1 : W \to TX_1$ and $g_2 : W \to TX_2$ are lax discardable,
      then
      \[
        f \pleq \vpairmorphism{g_1}{g_2}
        \quad\Leftrightarrow\quad
        T\pi_1 \compose f \pleq g_1
        ~\wedge~
        T\pi_2 \compose f \pleq g_2
      \]
    \item If $f : W \to T(\exp{X}{Z})$ is lax thunkable, where $\Z$ is a
      $\T$-algebra, then for every $g : W \times X \to Z$, we have
      \[
        f~\pleq~\eta_{\exp{X}{Z}} \compose \curry{g}
        \quad\Leftrightarrow\quad
        \algextendl[X]{\ev} \compose (f \times \id_{X})~\pleq~g
      \]
  \end{enumerate}
\end{lem}
\begin{proof}
  \begin{enumerate}
    \item For the $(\Rightarrow)$ direction, we have
      \[
        \begin{aligned}
          T \pi_i \compose f
          ~\pleq~&
          T \pi_i \compose \vpairmorphism{g_1}{g_2}
          &\quad&\text{(assumption)}
          \\~\pleq~&
          g_i
          &&\text{(\autoref{first-order-corollaries}(1))}
        \end{aligned}
      \]
      for each $i \in \{1, 2\}$.
      For the $(\Leftarrow)$ direction, we have
      \[
        \begin{aligned}
          f
          ~=~&
          T (\pi_1 \times \pi_2) \compose T\pairmorphism{\id}{\id} \compose f
          \\~\pleq~&
          T (\pi_1 \times \pi_2) \compose \seql \compose \pairmorphism{f}{f}
          &\quad&\text{(lax copyability)}
          \\~=~&
          \vpairmorphism{T\pi_1 \compose f}{T\pi_2 \compose f}
          &&\text{(naturality of extension)}
          \\~\pleq~&
          \vpairmorphism{g_1}{g_2}
          &&\text{(assumption)}
        \end{aligned}
      \]
    \item For the $(\Rightarrow)$ direction, we have
      \[
        \begin{aligned}
          \algextendl[X]{\ev} \compose (f \times \id_{X})
          ~\pleq~&
          \algextendl[X]{\ev} \compose ((\eta_{\exp{X}{Z}} \compose \curry{g}) \times \id_{X})
          &\quad&\text{(assumption)}
          \\~=~&
          \ev \compose (\curry{g} \times \id_{X})
          &&\text{(left unit law)}
          \\~=~&
          g
        \end{aligned}
      \]
      For the $(\Leftarrow)$ direction, we note that
      \[
        \begin{aligned}
          \eta_{\exp{X}{Z}} \compose \curry{(\algextendl[X]{\ev})} \compose f
          ~=~&
          \eta_{\exp{X}{Z}} \compose \curry{(\algextendl[X]{\ev} \compose (f \times \id_{X}))}
          \\~\pleq~&
          \eta_{\exp{X}{Z}} \compose \curry{g}
          &\quad&\text{(assumption)}
        \end{aligned}
      \]
      so it suffices to show $f \pleq \eta_{\exp{X}{Z}} \compose \curry{(\algextendl[X]{\ev})} \compose f$.
      Since $f$ is lax thunkable, this is a consequence of \autoref{lemma:thunkable-reasoning}(4) applied to the following.
      \begin{align*}
        \id_{T(\exp{X}{Z})}
        ~=~&
        \extend{\eta_{\exp{X}{Z}}}
        &&\text{(right unit law)}
        \\~=~&
        \extend{(\eta_{\exp{X}{Z}} \compose \curry{\ev})}
        \\~=~&
        \extend{(\eta_{\exp{X}{Z}} \compose \curry{(\algextendl[X]{\ev} \compose (\eta_{\exp{X}{Z}} \times \id_X))})}
        &&\text{(left unit law)}
        \\~=~&
        \extend{(\eta_{\exp{X}{Z}} \compose \curry{(\algextendl[X]{\ev})} \compose \eta_{\exp{X}{Z}})}
        \tag*{\qedhere}
      \end{align*}
  \end{enumerate}
\end{proof}

We can now relate $\semrel{\tau}$ to the morphisms $\toname{\tau}$ and
$\tovalue{\tau}$, as follows.
\begin{lem}\label{thm:represent-relation}
  Let $\M = (\C, \T)$ be a CBPV model for which $\T$ is lax idempotent.
  For every type $\tau$, object $W \in \ob{\C}$, and pair of morphisms
  \[
    f^v : W \to T(\semvalue{\tau})
    \qquad
    f^n : W \to \forget{\T}{(\semname{\tau})}
  \]
  we have the following equivalences.
  \[
    \toname{\tau} \compose f^v~\pleq~f^n
    \quad\Leftrightarrow\quad
    f^v~\semrel{\tau}[W]~f^n
    \quad\Leftrightarrow\quad
    f^v~\pleq~\tovalue{\tau} \compose f^n
  \]
\end{lem}
\begin{proof}
  By induction on the type $\tau$.
  \begin{itemize}
    \item The $\unittype$ case is trivial.

    \item For a product type $\tau_1 \times \tau_2$, we have that
      $f^v~\semrel{\tau_1 \times \tau_2}~f^n$ is equivalent, by
      expanding out the definition and applying the inductive
      hypothesis, to each of the following properties.
      \[
        \forall i \in \{1, 2\}.~
          \toname{\tau_i} \compose T\pi_i \compose f^v
          \pleq
          \pi_i\compose f^n
        \quad\qquad
        \forall i \in \{1, 2\}.~
          T\pi_i \compose f^v
          \pleq
          \tovalue{\tau_i} \compose \pi_i \compose f^n
      \]
      It therefore suffices to show that the left is equivalent to
      $\toname{\tau_1 \times \tau_2} \compose f^v \pleq f^n$, and that
      the right is equivalent to
      $f^v \pleq \tovalue{\tau_1 \times \tau_2} \compose f^n$.
      For the left, the inequality
      $\toname{\tau_1 \times \tau_2} \compose f^v \pleq f^n$ holds
      exactly when
      $
        \pi_i \compose \toname{\tau_1 \times \tau_2} \compose f^v
        \pleq
        \pi_i \compose f^n
      $
      holds for all $i \in \{1, 2\}$.
      The required equivalence therefore follows from
      \[
        \pi_i \compose \toname{\tau_1 \times \tau_2} \compose f^v
        ~=~
        T\pi_i \compose \toname{\tau_i} \compose f^v
      \]
      which is immediate from the definition of
      $\toname{\tau_1 \times \tau_2}$.
      For the right, it is enough to apply
      \autoref{lem:product-and-function-properties}(1) to the
      inequality
      $f^v \pleq \tovalue{\tau_1 \times \tau_2} \compose f^n$.
      We can do this because $\T$ is lax idempotent, which implies lax
      discardability and lax copyability
      by \autoref{lemma:thunkable-consequences}.

    \item The $\booltype$ case is trivial.

    \item For a function type $\tau \to \tau'$, we consider the two equivalences separately.

      For the equivalence on the left, we have
      \[
        \begin{aligned}
          \toname{\tau \to \tau'} \compose f^v~\pleq~f^n
          ~\Leftrightarrow~&
          \uncurry{(\toname{\tau \to \tau'} \compose f^v)}~\pleq~\uncurry{f^n}
          \\~\Leftrightarrow~&
          \toname{\tau'} \compose \extend{\ev} \compose \vpairmorphism{f^v \compose \pi_1}{\tovalue{\tau} \compose \pi_2}~\pleq~\ev \compose (f^n \times \id)
          \\~\Leftrightarrow~&
          (\extend{\ev} \compose \vpairmorphism{f^v \compose \pi_1}{\tovalue{\tau} \compose \pi_2})
          ~\semrel{\tau'}[W \times \forget{\T}{(\semname{\tau})}]~
          (\ev \compose (f^n \times \id))
        \end{aligned}
      \]
      where the second step uses the definition of
      $\toname{\tau \to \tau'}$, and the final step uses the inductive
      hypothesis.
      Call the instance of
      $\semrel{\tau'}[W \times \forget{\T}{(\semname{\tau})}]$ above
      (\textasteriskcentered).
      To show that $f^v~\semrel{\tau \to \tau'}[W]~f^n$ follows from
      (\textasteriskcentered), consider arbitrary $w : W' \to W$
      and $g^v, g^n$ such that $g^v~\semrel{\tau}[W']~g^n$.
      By \autoref{relation-closure}(2), we can precompose both sides of
      (\textasteriskcentered) with $\pairmorphism{w}{g^n}$ to obtain
      \[
        (\extend{\ev} \compose \vpairmorphism{f^v \compose w}{\tovalue{\tau} \compose g^n})
        ~\semrel{\tau'}[W']~
        (\ev \compose \pairmorphism{f^n \compose w}{g^n})
      \]
      The inductive hypothesis implies
      $g^v \pleq \tovalue{\tau} \compose g^n$, so by
      \autoref{relation-closure}(1) it follows that
      \[
        (\extend{\ev} \compose \vpairmorphism{f^v \compose w}{g^v})
        ~\semrel{\tau'}[W']~
        (\ev \compose \pairmorphism{f^n \compose w}{g^n})
      \]
      as required.
      Conversely, $f^v~\semrel{\tau \to \tau'}[W]~f^n$ implies
      (\textasteriskcentered) by taking
      \[
        W' = W \times \forget{\T}{(\semname{\tau})}
        \qquad
        w = \pi_1 : W' \to W
        \qquad
        g^v = \tovalue{\tau} \compose \pi_2
        \qquad
        g^n = \pi_2
      \]
      and noting that
      $\tovalue{\tau} \compose \pi_2 \pleq \tovalue{\tau} \compose \pi_2$
      implies
      $(\tovalue{\tau} \compose \pi_2)~\semrel{\tau}[W']~\pi_2$
      by the inductive hypothesis.

      For the remaining equivalence, we have
      \[
        \begin{aligned}
          f^v~\pleq~\tovalue{\tau \to \tau'} \compose f^n
          ~\Leftrightarrow~&
          \extendl[\semvalue{\tau}]{\ev} \compose (f^v \times \id)
          ~\pleq~
          \tovalue{\tau'} \compose \ev \compose (f^n \times (\toname{\tau} \compose \eta))
          \\~\Leftrightarrow~&
          (\extendl[\semvalue{\tau}]{\ev} \compose (f^v \times \id))
          ~\semrel{\tau'}[W \times \semvalue{\tau}]~
          (\ev \compose (f^n \times (\toname{\tau} \compose \eta)))
        \end{aligned}
      \]
      where the first step uses
      \autoref{lem:product-and-function-properties}(2) and the second
      uses the inductive hypothesis.
      Call the instance of $\semrel{\tau'}[W \times \semvalue{\tau}]$
      above (\textasteriskcentered\textasteriskcentered).
      To show that (\textasteriskcentered\textasteriskcentered)
      implies $f^v~\semrel{\tau \to \tau'}[W]~f^n$, consider arbitrary
      $w : W' \to W$ and $g^v, g^n$ such that $g^v~\semrel{\tau}[W']~g^n$.
      By \autoref{relation-closure}(2,3), we have
      \[
        \extendr[W]{(\extendl[\semvalue{\tau}]{\ev} \compose (f^v \times
        \id))} \compose \pairmorphism{w}{g^v}
        ~\semrel{\tau'}[W']~
        \algextendr[W]{(\ev \compose (f^n \times (\toname{\tau} \compose
        \eta)))} \compose \pairmorphism{w}{g^v}
      \]
      Lax idempotence of $\T$ implies lax centrality by
      \autoref{lemma:thunkable-consequences}, so we have
      \[
        \begin{aligned}
          &
          \extendr[W]{(\extendl[\semvalue{\tau}]{\ev} \compose (f^v \times \id))} \compose \pairmorphism{w}{g^v}
          \\~=~&
          \extendr[T(\semvalue{\tau \to \tau'})]{(\extendl[\semvalue{\tau}]{\ev})} \compose \pairmorphism{f^v \compose w}{g^v}
          &&\text{(naturality of extension)}
          \\~=~&
          \extendl[T(\semvalue{\tau})]{(\extendr[\semvalue{\tau \to \tau'}]{\ev})} \compose \pairmorphism{f^v \compose w}{g^v}
          &&\text{(\autoref{first-order-corollaries}(2))}
          \\~=~&
          \extend{\ev}\compose \vpairmorphism{f^v \compose w}{g^v}
          &&\text{(left unit, associativity)}
        \end{aligned}
      \]
      Since $g^v~\semrel{\tau}[W']~g^n$, we also have
      \[
        \begin{aligned}
          \algextendr[W]{(\ev \compose (f^n \times (\toname{\tau} \compose \eta)))} \compose \pairmorphism{w}{g^v}
          &~\pleq~
          (\ev \compose (f^n \times \toname{\tau})) \compose \pairmorphism{w}{g^v}
          &&\text{(\autoref{lemma:thunkable-reasoning}(2))}
          \\&~=~
          \ev \compose \pairmorphism{f^n \compose w}{\toname{\tau} \compose g^v}
          \\&~\pleq~
          \ev \compose \pairmorphism{f^n \compose w}{g^n}
          &&\text{(inductive hypothesis)}
        \end{aligned}
      \]
      where we again use the fact that $\T$ is lax idempotent.
      It follows by \autoref{relation-closure}(1) that
      \[
        (\extend{\ev}\compose \vpairmorphism{f^v \compose w}{g^v})
        ~\semrel{\tau'}[W']~
        (\ev \compose \pairmorphism{f^n \compose w}{g^n})
      \]
      as required.
      Finally, to show that $f^v~\semrel{\tau \to \tau'}[W]~f^n$
      implies (\textasteriskcentered\textasteriskcentered), we take
      \[
        W' = W \times \semvalue{\tau}
        \qquad
        w = \pi_1
        \qquad
        g^v = \eta \compose \pi_2
        \qquad
        g^n = \toname{\tau} \compose \eta \compose \pi_2
      \]
      noting that the inductive hypothesis implies
      $(\eta \compose \pi_2)~\semrel{\tau}[W']~(\toname{\tau} \compose
      \eta \compose \pi_2)$.
      From this we obtain
      \[
        (\extend{\ev} \compose \vpairmorphism{f^v \compose \pi_1}{\eta
        \compose \pi_2})
        ~\semrel{\tau'}[W \times \semvalue{\tau}]~
        (\ev \compose (f^n \times (\toname{\tau} \compose \eta)))
      \]
      which simplifies to (\textasteriskcentered\textasteriskcentered)
      by the left unit law.
      \qedhere
  \end{itemize}
\end{proof}

An immediate corollary is that, as claimed above, the maps between
call-by-value and call-by-name form Galois connections.
\begin{cor}\label{cor:galois-connections}
  Let $\M = (\C, \T)$ be a CBPV model such that $\T$ is lax idempotent.
  The monotone functions
  \[
    \begin{tikzcd}[column sep=huge]
      \Hom{\C}{W}{T (\semvalue{\tau})}
      \arrow[r, "\toname{\tau} \compose ({-})", yshift=1ex] &
      \Hom{\C}{W}{\forget \T (\semname{\tau})}
      \arrow[l, "\tovalue{\tau} \compose ({-})", yshift=-1ex]
    \end{tikzcd}
  \]
  form a Galois connection for every source-language type $\tau$ and
  object $W \in \ob\C$.
\end{cor}
\begin{proof}
  Immediate from \autoref{thm:represent-relation}.
\end{proof}

Another corollary of \autoref{thm:represent-relation} is the following,
which is the main result of this section.
We use this result in the following section to establish our reasoning
principle.
To state it, we use morphisms $\tonamehat{\Gamma}$ for converting a
call-by-value context into a call-by-name context, defined by
\[
  \tonamehat{\Gamma}
    = \tuplemorphism{\toname{\tau_i} \compose
    \eta_{\semvalue{\tau_i}} \compose \pi_{x_i}}_i
      : \semvalue{\Gamma} \to \semname{\Gamma}
\]
where $\Gamma = x_1 : \tau_1, \dots, x_k : \tau_k$.

\begin{thm}
  \label{thm:semantic-reasoning}
  Let $\M = (\C, \T)$ be a CBPV model such that $\T$ is lax idempotent.
  For all source-language expressions $\welltyped{\Gamma}{e}{\tau}$ we
  have
  \[
    \semvalue{e}
    ~\pleq~
    \tovalue{\tau} \compose \semname{e} \compose \tonamehat{\Gamma}
  \]
\end{thm}
\begin{proof}
  The inequality we need to establish is equivalently
  \[
    \semvalue{e} \pleq \tovalue{\tau} \compose \semname{e} \compose \tuplemorphism{\toname{\tau_i} \compose \eta_{\semvalue{\tau_i}} \compose \pi_{x_i}}_i
  \]
  where $\Gamma = x_1 : \tau_1, \dots, x_k : \tau_k$.
  By \autoref{thm:represent-relation} we have
  \[
    (\eta_{\semvalue{\tau_i}} \compose \pi_{x_i})
    ~\semrel{\tau_i}[\semvalue{\Gamma}]~
    (\toname{\tau_i} \compose \eta_{\semvalue{\tau_i}} \compose \pi_{x_i})
  \]
  for each $i$.
  We invoke \autoref{thm:fundamental} (using \autoref{lemma:thunkable-consequences} to show lax discardability, lax copyability and lax centrality), to obtain
  \[
    (\extend{\semvalue{e}} \compose \vtuplemorphism{\eta_{\semvalue{\tau_i}} \compose {\pi_{x_i}}}_i)
    ~\semrel{\tau}[\semvalue{\Gamma}]~
    (\semname{e} \compose \tuplemorphism{\toname{\tau_i} \compose \eta_{\semvalue{\tau_i}} \compose {\pi_{x_i}}}_i)
  \]
  The left-hand side is equal to $\semvalue{e}$ by the left unit law of $\T$, so \autoref{thm:represent-relation} implies the required inequality.
\end{proof}

This theorem has a partial converse, as follows.
\begin{prop}\label{prop:thunkable-necessary}
  Let $\M = (\C, \T)$ be an arbitrary CBPV model.
  If
  $
    \semvalue{e}
    \pleq
    \tovalue{\tau} \compose \semname{e} \compose \tonamehat{\Gamma}
  $
  for every $\welltyped{\Gamma}{e}{\tau}$, then
  for each object $X \in \{1,2\}$ we have $T\eta_X \pleq \eta_{TX}$, and
  every morphism $Y \to TX$ is lax thunkable.
\end{prop}
\begin{proof}
  The first step is to show that
  $\id \pleq \tovalue{\tau} \compose \toname{\tau}$ for every $\tau$, by
  applying the assumption to the expression
  $
    \welltyped{x : \unittype \to \tau}
      {x\,\unitterm}{\tau}
  $.
  Indeed, we have
  \[
    \ev
    ~=~\semvalue{x\,()}
    ~\pleq~\tovalue{\tau} \compose \semname{x\,()} \compose \tonamehat{\unittype \to \tau}
    ~=~\tovalue{\tau} \compose \toname{\tau} \compose \ev
  \]
  which implies
  $\id \pleq \tovalue{\tau} \compose \toname{\tau}$ because
  $\ev : 1 \times (\exp{1}{T(\semvalue{\tau})}) \to T(\semvalue{\tau})$
  is an isomorphism.
  It follows for each $\tau' \in \{\unittype, \booltype\}$ that
  \[
    \id_{T(\exp{1}{T(\semvalue{\tau'}}))}
    \pleq \tovalue{\unittype \to \tau'}
      \compose \toname{\unittype \to \tau'}
    = \eta_{\exp 1 {T(\semvalue{\tau'})}} \compose \algextend{\id_{\exp 1
    {T(\semvalue{\tau'})}}}
  \]
  which implies
  \[
    \id_{T(T(\semvalue{\tau'}))}
    \pleq
    \eta_{T(\semvalue{\tau'})} \compose \extend{\id_{T(\semvalue{\tau'})}}
  \]
  Hence, by naturality of Kleisli extension and the right unit law, we have
  \[
    T\eta_{\semvalue{\tau'}}
    ~=~
    \id_{T(T(\semvalue{\tau'}))} \compose T\eta_{\semvalue{\tau'}}
    ~\pleq~
    \eta_{T(\semvalue{\tau'})}\compose \extend{\id_{T(\semvalue{\tau'})}}
      \compose T\eta_{\semvalue{\tau'}}
    ~=~
    \eta_{T(\semvalue{\tau'})}
  \]
  as required.
\end{proof}

In particular, it follows from this proposition that lax discardability,
lax copyability, and lax centrality are not enough.
Our immutable state example satisfies all three of those properties, but
the morphism $\sem{\getconstant} : 1 \to T2$ is not lax thunkable, so
we do not have
$
  \semvalue{e}
  \pleq
  \tovalue{\tau} \compose \semname{e} \compose \tonamehat{\Gamma}
$
for every $e$.

\section{The reasoning principle}\label{section:main-theorem}
We now use the Galois connections defined in the previous section to
relate the call-by-value and call-by-name translations of expressions, and
arrive at our main reasoning principle.

Recall that the problem with comparing $\translatevalue{e}$ with
$\translatename{e}$ directly is that they have different types.
We render the Galois connections defined in the previous section
in the syntax of CBPV, and then construct from $\translatename{e}$ a
computation that we can directly compare with $\translatevalue{e}$:
\[
  \translatevalue{e}
  \ctxleq
  \ToValue{\tau}
    \big(\substitute{\translatename{e}}{\ToNameHat{\Gamma}}\big)
\]

\begin{figure}
  \begin{align*}
    \intertext{$\boxed{
      \comptyped{\Gamma}{M}{\freetype{(\translatevalue{\tau})}}
      ~\mapsto~
      \comptyped{\Gamma}{\ToName{\tau} M}{\translatename{\tau}}
    }$}
    \ToName{\unittype} M~&=~M
    \\
    \ToName{\tau_1 \times \tau_2} M~&=~
    \comppairterm*[~]
       {\toterm{M}{x}{\matchpairterm{x}{z_1}{z_2}{\ToName{\tau_1}{(\returnterm{z_1})}}}}
       {\toterm{M}{x}{\matchpairterm{x}{z_1}{z_2}{\ToName{\tau_2}{(\returnterm{z_2})}}}}
    \\
    \ToName{\booltype} M~&=~M
    \\
    \ToName{\tau \to \tau'} M~&=~
    \absterm{x}[\thunktype{(\translatename{\tau})}]{
      \toterm{M}{f}{\,
      \toterm{\ToValue{\tau}(\forceterm{x})}{y}{\,
      \ToName{\tau'} (\pushterm{y}{\forceterm{f}})}}}
    \\[2ex]
    \intertext{$\boxed{
      \comptyped{\Gamma}{N}{\translatename{\tau}}
      ~\mapsto~
      \comptyped{\Gamma}{\ToValue{\tau} N}{\freetype{(\translatevalue{\tau}})}
    }$}
    \ToValue{\unittype} N~&=~N \\
    \ToValue{\tau_1\times\tau_2} N~&=~
    \toterm{\ToValue{\tau_1} (\pushtermtight{1}{N})}{z_1}{
    \toterm{\ToValue{\tau_2} (\pushtermtight{2}{N})}{z_2}{
    \returnterm{\pairterm{z_1}{z_2}}}} \\
    \ToValue{\booltype} N~&=~N \\
    \ToValue{\tau \to \tau'} N~&=~
    \returnterm{\thunkterm{\absterm{x}[\translatevalue{\tau}]{
      \ToValue{\tau'}\big(\pushterm{(
        \thunkterm{(\ToName{\tau} {(\returnterm{x})})})}{N}
      \big)}}}
  \end{align*}

  \caption{Syntactic maps $\ToName{}$ from call-by-value to
    call-by-name and $\ToValue{}$ from call-by-name to call-by-value}
  \label{figure:maps}
\end{figure}

More precisely, we render $\toname{\tau}$ and $\tovalue{\tau}$ in the syntax as
maps $\ToName{\tau}$ from call-by-value computations
to call-by-name computations, and $\ToValue{\tau}$ from call-by-name to
call-by-value.\footnotemark{}
These are defined, again by induction on $\tau$, in
\autoref{figure:maps}.
(We use some auxiliary variables in the definition, which are assumed
to be fresh.)%
\footnotetext{%
  We define $\Phi$ and $\Psi$ directly as maps from computations to
  computations, but we could instead have defined computations
  \[
    \comptyped
      {x : \thunktype{(\freetype{(\translatevalue{\tau})})}}
      {\Phi'_\tau}
      {\translatename{\tau}}
    \qquad
    \comptyped
      {x : \thunktype{(\translatename{\tau}})}
      {\Psi'_\tau}
      {\freetype{(\translatevalue{\tau})}}
    \]
    and then recovered $\Phi$ and $\Psi$ modulo $\beta\eta$-laws for
    thunks, by substitution.
    This definition is slightly less convenient to work with however.}
We further define, for each source-language context
$\Gamma = x_1 : \tau_1, \dots, x_k : \tau_k$, a substitution
\[
  \ToNameHat{\Gamma}
  ~=~
  x_1 \mapsto \thunkterm{\big(\ToName{\tau_1} (\returnterm{x_1})\big)}
  ,\dots,
  x_k \mapsto \thunkterm{\big(\ToName{\tau_k} (\returnterm{x_k})\big)}
\]
for converting a call-by-value context into a
call-by-name context.
This has the following typing:
\[
  \comptyped{\translatename{\Gamma}}{N}{\comp{C}}
  ~\mapsto~
  \comptyped{\translatevalue{\Gamma}}{\ToNameHat{\Gamma}{N}}{\comp{C}}
\]

The maps $\ToName{}$, $\ToValue{}$ and $\ToNameHat{}$ are syntactic
renderings of $\toname{}$, $\tovalue{}$ and $\tonamehat{}$ in the
following sense.
\begin{lem}\label{lemma:maps-interpretation}
  Given any model of CBPV, we have:
  \begin{enumerate}
    \item $\sem{\ToName{\tau} M} = \toname{\tau} \compose \sem{M}$ for
      all $\comptyped{\Gamma}{M}{\freetype{(\translatevalue{\tau}})}$;
    \item $\sem{\ToValue{\tau} N} = \tovalue{\tau} \compose \sem{N}$ for
      all $\comptyped{\Gamma}{N}{\translatename{\tau}}$;
    \item $\sem{N\smash{[\ToNameHat{\Gamma}]}} = \sem{N} \compose \tonamehat{\Gamma}$ for all
      $\comptyped{\translatename{\Gamma}}{N}{\comp{C}}$.
  \end{enumerate}
\end{lem}
\begin{proof}
  (1) and (2) are proved by mutual induction on the type $\tau$,
  with each case being an easy calculation.
  (3) then follows immediately from (1) together with the evident substitution
  lemma for the denotational semantics of CBPV.
\end{proof}

Given a source-language expression $\welltyped{\Gamma}{e}{\tau}$, the
computation we obtain by composing $\translatename{e}$ with the maps
between call-by-value and call-by-name has the same type as 
$\translatevalue{e}$:
\[
  \comptyped{\translatevalue{\Gamma}}
  {\,\ToValue{\tau}
    \big(\substitute{\translatename{e}}{\ToNameHat{\Gamma}}\big)\,}
  {\freetype{(\translatevalue{\tau})}}
\]
We can therefore compare $\translatevalue{e}$ with
$\ToValue{\tau} (\substitute{\translatename{e}}{\ToNameHat{\Gamma}})$
directly.
In particular, it makes sense to replace $\translatevalue{e}$ with
$\ToValue{\tau} (\substitute{\translatename{e}}{\ToNameHat{\Gamma}})$
within a CBPV computation, as outlined in the introduction.
Using the results of the previous section, we establish the
following result for reasoning about how replacing $\translatevalue{e}$
in this way changes the behaviour of a computation.

Recall that a program relation $\syntaxleq$ is a preorder on
CBPV programs, and that each program relation induces a contextual
preorder $\ctxleq$.
Given any program relation $\syntaxleq$, to show that the call-by-value
and call-by-name translations of source-language expressions are related
by $\ctxleq$ it is enough to find an adequate model involving a lax
idempotent $\T$:
\begin{thm}[Relationship between call-by-value and call-by-name]
  \label{theorem:cbv-cbn}
  Let $\M = (\C, \T)$ be a CBPV model that is adequate with respect to a
  given a program relation $\syntaxleq$.
  If $\T$ is lax idempotent, then for every source-language expression
  $\welltyped{\Gamma}{e}{\tau}$ we have
  \[
    \translatevalue{e}
    ~\ctxleq~
    \ToValue{\tau}
      \big(\substitute{\translatename{e}}{\ToNameHat{\Gamma}}\big)
  \]
\end{thm}
\begin{proof}
  By adequacy it suffices to show
  $
    \sem{\translatevalue{e}}
    ~\pleq~
    \sem{\ToValue{\tau}
      (\substitute{\translatename{e}}{\smash{\ToNameHat{\Gamma}}})}
  $,
  which, by \autoref{lemma:maps-interpretation}, is equivalently
  $
    \semvalue{e}
    ~\pleq~
    \tovalue{\tau}
    \compose \semname{e}
    \compose \tonamehat{\Gamma}
  $.
  The result therefore follows from \autoref{thm:semantic-reasoning}.
\end{proof}

The generality of this theorem comes from two sources.
First, we consider arbitrary program relations $\syntaxleq$.
The only requirement on these is the existence of some adequate model in
which morphisms are lax thunkable.
Second, this theorem applies to terms that are open and have higher
types, using the maps between the two evaluation orders (in contrast to
\autoref{corollary:cbv-cbn-programs} above).

For our first three examples (no effects, divergence, and
nondeterminism), the model is adequate and has a lax idempotent $\T$.
Thus in each case the assumptions of our reasoning principle are
satisfied, establishing the claims stated informally at the beginning of
the introduction.

\begin{rem}
  Given an adequate model in which $\T$ is lax idempotent, it follows
  from \autoref{cor:galois-connections} and
  \autoref{lemma:maps-interpretation} that the maps $\ToName{\tau}$ and
  $\ToValue{\tau}$ on terms form a Galois connection (with respect
  to $\ctxleq$).
  In particular, we have
  \[
    M \ctxleq \ToValue{\tau} (\ToName{\tau} M)
    \qquad
    \ToName{\tau} (\ToValue{\tau} N) \ctxleq N
  \]
  Both of these inequalities are in general proper (they are not
  contextual equivalences).
  To see this, consider our divergence example, for which the above
  inequalities hold.
  For each $\comp C$, let $\Omega_{\comp C}$ be the diverging
  computation $\cbpvrecterm{x}[\comp C]{\forceterm x}$
  (which has type $\comp C$).
  Then if $\tau = \booltype \to \booltype$ and
  $M = \Omega_{\freetype {(\translatevalue{\tau})}}$,
  we do not have
  $M \ctxgeq \ToValue{\tau} (\ToName{\tau} M)$, because for
  $\mathcal{E} = (\toterm{\hole}{f}{\returnterm \falseterm})$ the
  computation $\mathcal{E}[M]$ diverges but
  $
    \mathcal{E}[\ToValue{\tau} (\ToName{\tau} M)]
      \reducesto \returnterm \falseterm
  $.
  In this case we have
  $
    \ToValue{\tau} (\ToName{\tau} M)
      \ctxeq \returnterm{\thunkterm{\lambda
      x:\booltype.\,\Omega_{\freetype \booltype}}}
  $.
  For a counterexample to $\ToName{\tau} (\ToValue{\tau} N) \ctxgeq N$,
  let $\tau = \booltype \to \booltype$ and
  $N= \absterm{x}[\thunktype(\freetype\booltype)]{\returnterm\trueterm}$.
  Then for
  $
    \mathcal{E}'
      = (\pushterm{(\thunkterm{\Omega_{\freetype \booltype}})}{\hole})
  $,
  the computation
  $\mathcal{E}'[\ToName{\tau} (\ToValue{\tau} N)]$ diverges
  but
  $\mathcal{E}'[N] \reducesto \returnterm \trueterm$.
  Here we have
  $
    \ToName{\tau} (\ToValue{\tau} N)
    \ctxeq
    \absterm{x}[\thunktype(\freetype\booltype)]{
      \toterm{\forceterm{x}}{y}{\returnterm\trueterm}}
  $.

  In particular, our maps between call-by-value and call-by-name are
  merely Galois connections, and not sections or retractions.
  This contrasts with
  Reynolds~\cite{reynolds:relation-direct-continuation}, who obtains a
  retraction between direct and continuation semantics.
\end{rem}

\section{Related work}
\paragraph{Comparing evaluation orders}
Plotkin~\cite{plotkin:cbv-cbn-lambda-calculus} and many others
(e.g.\ \cite{inoue-taha:reasoning-multistage-programs}) relate
call-by-value and call-by-name. Crucially, they consider $\lambda$-calculi
with no effects other than divergence. This makes a significant
difference to the techniques that can be used, in particular because in
this case the
equational theory for call-by-name is strictly weaker than for
call-by-value. This is not necessarily true for other effects.  Other
evaluation orders (such as call-by-need) have also been compared
in similarly restricted settings \cite{%
  maraist-et-al:name-value-need-linear,
  mcdermott-mycroft:ecbpv,
  hackett-hutton:call-by-need-clairvoyant}. We suspect our
technique could also be adapted to these.
Here we use CBPV as a calculus in which to reason about both
call-by-value and call-by-name, but other calculi (e.g.\
the modal calculus of \cite{espiritosanto2022plotkin}) may be
suitable for this purpose.

It might also be possible to recast some of our work in terms of the
\emph{duality} between call-by-value and call-by-name
\cite{%
  filinski:declarative-continuations-duality,
  curien-herbelin:duality-computation,
  wadler:dual,
  selinger2001control},
In particular, this may shed some light on our definitions of
$\ToName{}$ and $\ToValue{}$.
It is not clear to us what the precise connection is however.
While Selinger~\cite{selinger2001control} defines translations between
call-by-value and call-by-name versions of Parigot's
$\lambda\mu$-calculus~\cite{parigot1992lambda}, these translations
behave differently to ours, in particular, they are
semantics-preserving.

\paragraph{Relating semantics of languages}
The technique we use here to relate call-by-value and call-by-name is
based on the idea used first by
Reynolds~\cite{reynolds:relation-direct-continuation} to relate direct and
continuation semantics of the $\lambda$-calculus, and later used by others
(e.g.\ \cite{%
  meyer-wand:continuation-semantics,
  kucan:retraction-approach,
  cartwright-felleisen:extensible-specifications,
  filinski:controlling-effects}).
Reynolds constructs a relation
between the two semantics, and uses this to establish a retraction
between direct and continuation semantics,
just as we construct a relation between
call-by-value and call-by-name and then use this to establish a
Galois connection.
A minor difference is that
Reynolds relies on continuations with a
large-enough domain of answers (e.g.\ a solution to a particular
recursive domain equation). Our maps exist for \emph{any} choice of
model. We are the first to use this technique to relate
call-by-value and call-by-name.
There has been some work
\cite{%
  sabry-felleisen:reasoning-about-programs-cps,
  lawall-danvy:separating-stages-cps,
  sabry-wadler:reflection-call-by-value} on soundness and completeness
properties of translations (similar to the translations into
CBPV), in particular using Galois connections (and similar structures)
for which the order is reduction of programs. Our results would fail if
we used reduction of programs directly, so we consider only the
observable behaviour of programs.

There are some similarities between our work and the work of New et al.\
\cite{new2020call,new2021gradual} on gradual typing.  In particular,
\cite{new2021gradual} has embedding-projection pairs (a special case of
Galois connections) for casting from a more dynamic type to a less
dynamic type, and vice versa.  Their application is quite different
however.  The double category perspective used in \cite{new2020call}
may also be illuminating here.

\section{Conclusions}
In this paper, we give a general reasoning principle (\autoref{theorem:cbv-cbn}) that
relates the observable behaviour of terms under call-by-value and
call-by-name. The reasoning principle works for various collections of
computational effects, in particular, it enables us to obtain
theorems about divergence and nondeterminism.
It is about open expressions, and enables us to change
evaluation order \emph{within} programs.

The technique we use involves first relating the observable behaviour of
the call-by-value and call-by-name translations of expressions via a
logical relation (\autoref{thm:fundamental}).
We obtain a result about
call-by-value and call-by-name evaluations of \emph{programs} as a
corollary (\autoref{corollary:cbv-cbn-programs}). Applying this to
divergence, we show that if the call-by-value execution terminates
with some result then the call-by-name execution terminates with the
same result. For nondeterminism, we show that all possible results of
call-by-value executions are possible results of call-by-name
executions. There may be other collections of effects we can apply
our technique to, including combinations of divergence and nondeterminism.

We expect that our technique can be applied to other evaluation orders.
Two evaluation orders can be related by giving
translations into some common language (here we use CBPV), constructing
maps between the two translations, and showing that (for some models)
these maps form Galois connections.
A major advantage of the technique is that it allows us to identify
axiomatic properties of computational effects (thunkable, etc.) that give rise to
relationships between evaluation orders.

\section*{Acknowledgments}
\noindent We thank the anonymous referees for helpful comments.
The first author was supported by an EPSRC studentship, and by Icelandic
Research Fund grants 196323-053 and 228684-052.

\bibliographystyle{alphaurl}
\bibliography{value-name}

\end{document}